\documentclass[final, twocolumn]{article}
\usepackage{graphicx}
\usepackage{epsf}
\usepackage{times}
\usepackage{float}
\usepackage{amssymb}
\usepackage{amsmath,bm }
\usepackage{slashed}
\usepackage{cite}
\usepackage{color}
\usepackage{tikz}
\usetikzlibrary{shapes, snakes, positioning,fit,calc}

\newcommand{\expd}[1]{e^{\displaystyle {#1}}}

\graphicspath{{figures/}}

\begin{document}  %

\title{\huge \bf Fundamentals of Data-Driven Approaches to Acoustic Signal  Detection,  Filtering, and Transformation}
\author{{\it  Chao Pan}
%
\thanks{C. Pan is with the Center of Intelligent Acoustics and Immersive Communications, Northwestern Polytechnical
   University, 127 Youyi West Road, Xi'an, Shaanxi 710072, China  (Email: cpan@nwpu.edu.cn). }
%
%
}

\markboth{ }{ }

\maketitle

\begin{abstract}
In recent decades, the field of signal processing has rapidly evolved due to diverse application demands, leading to a rich array of scientific questions and research areas. The forms of signals, their formation mechanisms, and the information extraction methods vary by application, resulting in diverse signal processing techniques. Common techniques can be categorized into three types: transformation, detection, and filtering. Signal transformation converts signals from their original domain to a more suitable target domain for analysis; signal detection aims to identify the existence of relevant information within a signal and its specific time and location; and signal filtering focuses on extracting or separating source signals of interest from observed signals. In acoustic signal processing, techniques include sound source localization, sound event detection, voiceprint extraction and recognition, noise reduction, and source separation, with applications in speech communication, voice interaction, smart healthcare, and industrial diagnostics. Recently, the advancement of deep learning technologies has shifted methodologies in acoustic signal processing from knowledge-driven to data-driven approaches, leading to significant research outcomes. This paper aims to systematically summarize the principles and methods of data-driven acoustic signal processing, providing a comprehensive understanding framework for academic exploration and practical applications.
 
\end{abstract} 
\tableofcontents
\section{Basic Concepts of Neural Networks} 
 
In the first section, we introduce the fundamental concepts of neural networks, laying the groundwork for subsequent sections. It covers mathematical notation, basic network structures, objective functions, parameter learning, and the overall structure of this paper.
  1) Explanation of mathematical notation and how to represent a network structure using mathematical formulas;
  2) Basic network structures, characteristics of classic architectures, and notation;
  3) Training objectives for the network, objective functions, and common methods for constructing objective functions;
  4) Parameter learning, forward data propagation, backward error propagation, derivatives of the objective function with respect to parameters, and methods for parameter learning.
 
Neural networks can be viewed as functions, which have inputs, outputs, and parameters. Given an input, the neural network processes the signal/data to produce an output. First, consider a simple function \( f(x) = ax + b \), which represents a straight line where \( x \) is the input and \( a \) and \( b \) are parameters defining the line. If we combine this function with \( g(x) = \cos(nx) \), we obtain \( f(x) = \cos[n(ax + b)] \), which results in a curve with parameters \( a, b, \) and \( n \); this curve is more complex than the linear function. In the case of two-dimensional inputs, the function becomes a surface \( f(x_1, x_2) = \cos[n(a_1 x_1 + a_2 x_2 + b)] \), illustrating an even more complex functional form; modifying parameters \( a_1, a_2, b, n \) changes the specific shape of the surface. This simple example illustrates two points: 1) By combining different functions, we can represent more complex functional forms; 2) The specific shape of the function is determined by the parameters that define it.

The process of solving practical problems using neural networks resembles the process of calculating outputs based on inputs. The design approach can be summarized as follows: first, use existing network modules to construct an overall network structure (i.e., a composite function), and then learn suitable parameters for the task from a dataset. For convenience, typical network modules will be represented as functions, such as:
\begin{itemize}
    \item \( \mathcal{C}(\cdot) \) for Convolutional Neural Networks;
    \item \( \mathcal{G}(\cdot) \) for Gated Recurrent Networks;
    \item \( \mathcal{R}(\cdot) \) for Residual Networks;
    \item \( \mathcal{U}(\cdot) \) for U-Net architectures;
    \item \( \mathcal{E}(\cdot) \) for Encoder Networks;
    \item \( \mathcal{D}(\cdot) \) for Decoder Networks;
    \item \( \mathcal{A}(\cdot) \) for Feature Fusion Networks;
    \item \( \mathcal{F}(\cdot) \) for Fully Connected Networks;
    \item \( \mathcal{T}(\cdot) \) for Transformer Networks;
    \item \( \mathcal{S}(\cdot) \) for the network output layer, typically a sigmoid function, softmax function, or linear mapping function.
\end{itemize}

\subsection{Notation in Neural Networks}

Assuming the input to the neural network is \( \bm{x} \), the function to be modeled is \( f(\bm{x}) \), and the labels for training samples are \( \bm{y} \). The set of sample pairs is denoted as \( \{(\bm{x}_n, \bm{y}_n), \forall n\} \). For instance, if a network consists of three convolutional layers, one feature fusion layer, three fully connected layers, and one softmax output layer, the function modeled by the global neural network can be expressed as:

\begin{align} \label{example-f-x}
    f(\bm{x}) &= \mathcal{S} \circ \mathcal{F}_3 \circ \mathcal{F}_2 \circ \mathcal{F}_1 \circ \mathcal{A} \circ \mathcal{C}_3 \circ \mathcal{C}_2 \circ \mathcal{C}_1(\bm{x}),
\end{align}
where \( \circ \) denotes composite  function, such that \( \mathcal{C}_2 \circ \mathcal{C}_1(x) = \mathcal{C}_2\left[\mathcal{C}_1(x)\right] \). Using the composite function notation greatly simplifies the complexity of formulas and enhances understanding.

In such a network, each module has parameters that need to be learned. Typically, the parameters corresponding to each module are not discussed in the introduction of network principles, but are learned automatically through hyperparameter settings and datasets. Therefore, in most cases, it is unnecessary to assign specific mathematical symbols for description. When necessary, the parameters will be defined accordingly. The set of parameters is uniformly denoted as \( \theta_{(\cdot)} \), where the subscript indicates the symbol of the neural network; for example, \( \theta_{\mathcal{C}_3} \) represents the set of parameters corresponding to layer \( \mathcal{C}_3(\bm{x}) \), i.e., the set of convolutional kernels.

Sometimes, we need to discuss the output layer of the network separately to construct the training objective function. To address this need, we can introduce the subscript \( (\cdot)_{\slashed{\mathcal{S}}} \) to indicate the removal of the output layer \( \mathcal{S}(\cdot) \); for example, in the neural network represented in (\ref{example-f-x}), removing the output layer can be expressed as \( f_{\slashed{\mathcal{S}}}(\bm{x}) \), and 
\begin{align}
    f_{\slashed{\mathcal{S}}}(\bm{x}) &= \mathcal{F}_3 \circ \mathcal{F}_2 \circ \mathcal{F}_1 \circ \mathcal{A} \circ \mathcal{C}_3 \circ \mathcal{C}_2 \circ \mathcal{C}_1(\bm{x}).
\end{align}
Of course, under this notation, we can also define functions that exclude any intermediate layer, such as \( f_{\slashed{\mathcal{C}}_3}(\bm{x}) \), \( f_{\slashed{\mathcal{N}}_3}(\bm{x}) \), etc.; however, such cases are rare and will not be discussed further.

\subsection{Objective Functions in Neural Networks}

Once the network is constructed, it is necessary to determine the network parameters using data/samples. The training data enables the network to achieve a specific objective. Defining the objective function is the essence of building a neural network; with a clearly defined objective, the network can optimize the parameters using the backpropagation algorithm. Common objective functions include four types: Mean Squared Error (MSE), Cross-Entropy, Negative Log-Likelihood, and Binary Cross-Entropy. The definition of Mean Squared Error is as follows:
\begin{align}
    \mathcal{J}_{\mathrm{MSE}}(\theta_f) &= \mathbb{E}_{\bm{x}} \left\| f(\bm{x}) - \bm{y} \right\|_2^2 \\
    &\approx \sum_n \left\| f(\bm{x}_n) - \bm{y}_n \right\|_2^2,
\end{align}
where \( \theta_f \) represents all parameters in the function \( f(\bm{x}) \); \( \mathbb{E}_{\bm{x}} \) denotes the mathematical expectation. The expectation can be understood as \( \mathbb{E}_{\bm{x}} f(\bm{x}) = \int f(\bm{x}) p(\bm{x}) d \bm{x} \), where \( p(\bm{x}) \) is the density function of the variable \( \bm{x} \).

Similar to norms, Cross-Entropy also measures a type of distance. Given the network output \( f(\bm{x}) \) and the labels \( \bm{y} \), the Cross-Entropy is defined as:
\begin{align}
    \mathcal{J}_{\mathrm{CE}}(\theta_f) &= -\mathbb{E}_{\bm{x}} \sum_i [\bm{y}]_i \ln [f(\bm{x})]_i \\
    &= -\mathbb{E}_{\bm{x}} \bm{y}^T \ln f(\bm{x}) \\
    &\approx -\sum_n \bm{y}_n^T \ln f(\bm{x}_n) \\
    &=-\sum_n \sum_i [\bm{y}_n]_i \ln [f(\bm{x}_n)]_i,
\end{align}
where \( [\cdot]_i \) denotes the \( i \)-th element of the vector. For multi-class tasks, the label \( \bm{y} \) is typically a one-hot vector, with only one non-zero element, which equals 1; the index of the non-zero element indicates the class to which the sample belongs. Assuming the label \( \bm{y}_n \) for the \( n \)-th sample corresponds to a non-zero index \( i_n \), the Cross-Entropy can be simplified to:
\begin{align}
    \mathcal{J}_{\mathrm{NL}}(\theta_f) &= -\mathbb{E}_{\bm{x}} \ln [f(\bm{x})]_{i_n} \\
    &\approx -\sum_n \ln [f(\bm{x}_n)]_{i_n},
\end{align}
which represents the negative log-likelihood.

For binary classification problems, the network output requires only a single output, and the label \( \bm{y} \) indicates the probability of the event/target occurring, often simplified to \( y \); equivalently, this can be represented as \( \bm{y} = [y~(1-y)]^T \). Here, \( y=1 \) signifies that the event/target occurs, while \( y=0 \) indicates its absence. Under binary classification, the Cross-Entropy can be simplified to:
\begin{align}
    \mathcal{J}_{\mathrm{BCE}}(\theta_f) &= -\mathbb{E}_{\bm{x}} \left\{ y \ln f(\bm{x}) + (1-y) \ln [1-f(\bm{x})] \right\} \\
    &\approx -\sum_n y_n \ln f(\bm{x}_n) + (1-y_n) \ln [1-f(\bm{x}_n)].
\end{align}
Binary Cross-Entropy is widely used, typically corresponding to the final layer of a neural network \( f(\bm{x}) \) being a sigmoid function, which confines the network output between 0 and 1. Under binary cross-entropy conditions, the function \( f(\bm{x}) \) describes the probability of the event/target occurring given the observation \( \bm{x}_n \), which is traditionally expressed as \( p(\mathcal{H}_1 | \bm{x}_n) \), where \( \mathcal{H}_1 \) indicates that the hypothesis is valid (i.e., the event/target occurs).

\subsection{Parameter Learning and Backpropagation Algorithm}

For each sample, the neural network learning process typically requires calculating two pieces of information: 1) During forward computation, determining the output of each layer for the given sample; 2) During backpropagation, using the sample's label and derivatives to compute the gradient of the cost function with respect to the outputs of each layer.

For convenience, let the input and output of the \( m \)-th layer be denoted as \( \bm{x}^{(m-1)} \) and \( \bm{x}^{(m)} \) (where \( m=1,2,\ldots, M \)). The input of the first layer is \( \bm{x} \), and the output of the last layer (layer \( M \)) is \( f(\bm{x}) \), thus we have:
\begin{align}
    \bm{x}^{(0)} &= \bm{x}, \\
    \bm{x}^{(M)} &= f(\bm{x}).
\end{align}
The objective of the neural network is to make all \( \bm{x}_n^{(M)} \) approximate \( \bm{y}_n \). During forward computation, the signal's flow is represented as:
\begin{align}
    f(\bm{x}_n) =& \bm{x}_n^{(M)} \leftarrow \cdots \leftarrow \bm{x}_n^{(m+1)} \leftarrow \bm{x}_n^{(m)} \leftarrow \bm{x}_n^{(m-1)} \leftarrow \nonumber
    \\& \cdots \leftarrow \bm{x}_n^{(0)} = \bm{x}_n,
\end{align}
During backpropagation, we compute:
\begin{align}
    \bm{e}_n^{(M)} \rightarrow \cdots \rightarrow \bm{e}_n^{(m+1)} \rightarrow \bm{e}_n^{(m)} \rightarrow \bm{e}_n^{(m-1)} \rightarrow \cdots \rightarrow \bm{e}_n^{(0)},
\end{align}
where \( \bm{e}_n^{(M)} \) is determined by the objective function of the network training and typically has an analytical expression; \( \bm{e}_n^{(m)} = \frac{\partial \mathcal{J}}{\partial \bm{x}_n^{(m)}} \) represents the partial derivative of the objective function with respect to the output of a specific network layer. Under the least squares criterion, \( \bm{e}_n^{(M)} \) is the error between the network output and the true label, i.e., \( \bm{e}_n^{(M)} = f(\bm{x}_n) - \bm{y}_n \).

Next, using the parameter update for a fully connected network as an example, we will illustrate how to compute the derivative of the objective function with respect to network parameters. Assume the \( m \)-th layer is a fully connected network. From previous definitions, the output of the \( m \)-th layer is \( \bm{x}_n^{(m)} \), and its input is \( \bm{x}_n^{(m-1)} \). In a fully connected network, the relationship between the input and output of the \( m \)-th layer can be described as:
\begin{align}
    \bm{x}_n^{(m)} &= \sigma\left( \bm{W}_m \bm{x}_n^{(m-1)} + \bm{b}_m \right),
\end{align}
where \( \sigma(\cdot) \) is a nonlinear activation function, and \( \bm{W}_m \) and \( \bm{b}_m \) are the parameters to be learned. When the input to the activation function \( \sigma(\cdot) \) is a vector/matrix, it equivalently applies the nonlinear mapping to each element of the vector/matrix. We can derive\cite{yu2016automatic}:
\begin{align}
    \nabla\mathcal{J}(\bm{W}_m, \bm{x}_n) &= \frac{\partial \mathcal{J}}{\partial \bm{W}_m} \\
    &= \mathrm{diag}\left[\sigma^{\prime}\left(\bm{W}_m \bm{x}_n^{(m-1)} + \bm{b}_m\right)\right] \bm{e}_n^{(m)} \bm{x}_n^{(m-1) T}, \\
    \nabla\mathcal{J}(\bm{b}_m, \bm{x}_n) &= \frac{\partial \mathcal{J}}{\partial \bm{b}_m} \\
    &= \mathrm{diag}\left[\sigma^{\prime}\left(\bm{W}_m \bm{x}_n^{(m-1)} + \bm{b}_m\right)\right] \bm{e}_n^{(m)}.
\end{align}
From the above two equations, it can be seen that the derivative of the cost function with respect to the parameters of a specific layer can be computed using the forward output of the network (the input of the current layer \( \bm{x}^{(m-1)} \)) and the backward propagated error \( \bm{e}^{(m)} \); each sample can yield such a formula, i.e., \( \bm{x}_n^{(m-1)} \) and \( \bm{e}_n^{(m)} \). For a neural network composed of multiple fully connected layers, the relationship between the propagated errors of adjacent layers is expressed as:
\begin{align}
    \bm{e}_n^{(m-1)} = \bm{W}_m^{T} \mathrm{diag}\left[\sigma^{\prime}\left(\bm{W}_m \bm{x}_n^{(m-1)} + \bm{b}_m\right)\right] \bm{e}_n^{(m)}.
\end{align}
This functional relationship intuitively provides the method for calculating propagated errors between adjacent layers. For complex networks like convolutional neural networks, recurrent neural networks, and attention-based networks, the computation relationship between propagated errors will be more complex and will not be discussed in detail here.

For neural networks constructed from basic modules as represented in (\ref{example-f-x}), we typically do not concern ourselves with the details of parameter updates for a specific layer; this work is usually handled automatically by the training platform. However, we sometimes need to compute the derivative of the objective function with respect to the output of a specific layer. Analogous to the definition of propagated error \( \bm{e}^{(m)} \), we can express the derivative of the objective function with respect to the output of module \( \square \) as:
\begin{align}
    \bm{e}^{\square}, \text{ or } \bm{e}_n^{\square}.
\end{align}
Here, \( \square \) can be replaced with the name of any module, such as \( \mathcal{F}, \mathcal{T}, \mathcal{A}, \mathcal{D}, \mathcal{S} \), etc. The previously defined derivative of the objective function with respect to the network output can also be expressed as \( \bm{e}^{(M)} \); if the output layer is defined as \( \mathcal{S}(\cdot) \), we can denote \( \bm{e}^{(M)} \) as \( \bm{e}^{(\mathcal{S})} \), meaning \( \bm{e}^{(M)} = \bm{e}^{(\mathcal{S})} \).

\subsection{Network Optimizers}

Given a neural network \( f(\bm{x}) \), a parameter set \( \theta \), and a dataset \( \{(\bm{x}_n, \bm{y}_n), \forall n\} \), a common approach is to take a small batch of data each time, compute the update direction for the network parameters, and perform parameter learning. Assuming at step \( t \), the small batch of data is \( \{(\bm{x}_n, \bm{y}_n), \forall n \in \mathbb{I}_t\} \), we can compute an average gradient based on this subset:
\begin{align}
    \varpi^{(t)} &= \frac{1}{|\mathbb{I}_t|} \sum_{n \in \mathbb{I}_t} \nabla \mathcal{J}(\theta^{(t-1)}, \bm{x}_n).
\end{align}
It can be verified that \( \varpi^{(t)} \) has the same dimensionality as the parameter set \( \theta \), with each element reflecting the update direction for the corresponding parameter. The gradient descent algorithm based on mini-batch data can be expressed as \cite{nemirovski2009robust}:

\begin{align}
    \theta^{(t)} &= \theta^{(t-1)} - \alpha \varpi^{(t)},
\end{align}
where \( \alpha > 0 \) is the learning rate, typically chosen around \( 10^{-3} \), depending on the situation. Similar to classic adaptive algorithms, a larger \( \alpha \) can enhance learning efficiency but may risk divergence; a smaller \( \alpha \) can ensure convergence but often requires more iterations. Unlike classic adaptive algorithms, a small learning rate does not guarantee good network parameter learning, as it may lead to convergence at a local optimum of the objective function. Therefore, setting the learning rate is a crucial issue in practice.

By applying a recursive smoothing operation to the average gradient of the mini-batch data, we obtain:
\begin{align}
    \varepsilon^{(t)} &= \beta_1 \varepsilon^{(t-1)} + (1-\beta_1) \varpi^{(t)},
\end{align}
where \( \beta_1 \in (0,1] \) is the smoothing factor, usually set to \( \beta_1 = 0.9 \). Since the magnitude of the gradient reflects the length of the parameter update, we need to normalize the gradients to account for the differences in the speed of updates. To achieve this, we compute:
\begin{align}
    \varrho^{(t)} &= \beta_2 \varrho^{(t-1)} + (1-\beta_2) \varpi^{(t)} \odot \varpi^{(t)},
\end{align}
where \( \beta_2 \in (0,1] \) is another smoothing factor, typically set to \( \beta_2 = 0.99 \). Given \( \theta^{(t-1)}, \varpi^{(t)}, \varepsilon^{(t)}, \) and \( \varrho^{(t)} \), the network parameters are updated as follows:

\begin{align}
    \theta^{(t)} &= \theta^{(t-1)} - \alpha \frac{1}{\sqrt{\varrho^{(t)} + \epsilon_0}} \varepsilon^{(t)}, \label{update-adam}
\end{align}
where \( \epsilon_0 > 0 \) is a small constant to prevent division by zero. Equation (\ref{update-adam}) represents the well-known Adam parameter update algorithm \cite{kingma2014adam}. The form of the parameter update formula is quite similar to the NLMS algorithm in adaptive filtering, but there are significant differences in the design of gradient computation and normalization factors, indicating they are not the same type of learning algorithm.

\subsection{Organization of This Paper}

This paper investigates the topics related to neural networks based acoustic signal processing, including the construction of objective functions, selection of training strategies, and explanations of fundamental concepts. It emphasizes basic ideas and characteristics of networks without delving into Python programming or mathematical proofs. The paper consists of a total of 9 sections, discussing data-driven acoustic signal processing, primarily focusing on detection, filtering, and transformation of acoustic signals. The content is roughly as follows: 
\begin{enumerate}
    \item This section introduces some fundamental knowledge about neural networks, including how networks express functions, understanding composite functions, objective functions, and how to use backpropagation methods to achieve parameter learning.
    \item The foundation for constructing neural networks lies in typical network modules, akin to how electronic circuits are built using capacitors, resistors, inductors, and operational amplifiers. Similarly, neural networks rely on ``basic components'', which are network structures with different characteristics. Therefore, Section 2 discusses the typical structures of neural networks, explaining the features of various architectures and how to represent them using functions, providing explanations from the perspective of digital signal processing.
    \item For a constructed network structure, it only provides a functional expression space, determining the types of data it can process. To effectively implement a specific function, network parameters must be learned from data; only when the network parameters are determined can the function's capabilities be defined. The learning of network parameters is central to the network's objective function, which dictates how the parameters should be iteratively updated. Hence, Section 3 will introduce methods for constructing objective functions. In this section, we will see that for detection, filtering, and transformation tasks, the construction of their objective functions varies significantly, as different tasks present different challenges.
    \item Sections 4, 5, 6, 7, and 8 discuss five fundamental acoustic problems, including sound source localization, sound event detection, voiceprint extraction and recognition, noise reduction, and sound source separation. Among these problems, some are signal detection issues, while others are related to filtering or transformation. Each section will describe the respective acoustic problem, its challenges, the construction of the network's objective function, and typical network structures, aiming to provide readers with focused guidance in their research or inspiration for work in other fields.
    \item Finally, Section 9 consolidates some of the author's reflections, including interpretations of Generative Adversarial Networks, intuitive understanding of optimal transport, diffusion models, modeling and optimization of the area under the ROC curve, and classical signal transformations. Through discussions of these topics, we hope to leave readers with space for imagination and inspire them to conduct valuable research in their respective fields.
\end{enumerate}
 
\section{Signal Models and Typical Frameworks}
 
In practice, the signals observed by microphone sensors represent time-varying sound pressure values in physical space. Different tasks require extracting varying information from a continuous sound pressure signal. The transformation from signal to information can be characterized as a function, constructed using classic neural network modules. This section focuses on signal models and network frameworks, covering the following main topics:
     \emph{1) Signal models: how to slice signals and construct samples;
    2) Neurons and nonlinear activation functions;
    3) Common network structures, including Convolutional Neural Networks (CNNs), Recurrent Neural Networks (RNNs), Residual Networks, and Feature Fusion Networks.}

\subsection{Signal Processing}

When a microphone sensor is placed in a physical space to observe changes in sound pressure, it captures a time-extended sound signal. In practice, useful information is extracted from numerous past segments of sound signals to analyze and process the current time segment, thus providing the desired output.

Given a finite sound signal over time \( x(t), t=0,1,\ldots,T-1 \), stacking \( L \) adjacent points results in \( \bm{x}(t) = [x(t)~ x(t-1)~\cdots~x(t-L+1)]^T, t=0,1,\ldots, T-1 \). If the starting point for stacking is not shifted point-by-point, then the vector \( \bm{x}(t) \) can be expressed as \( \bm{x}(t) = [x(tK)~ x(tK-1)~\cdots~x(tK-L+1)]^T, t=0,1,\ldots, T-1 \), where \( K \) is the step length of the shift. In sound signal processing, \( \bm{x}(t) \) is often referred to as a slice of the signal, serving as a basic unit for processing and analysis.
\begin{itemize}
    \item For time-domain filtering problems under linear systems, given the filter coefficient vector \( \bm{w} = [w(0)~ w(1)~\cdots~w(L-1)]^T \), the output of the filter can be expressed as
    \begin{align}
        y(t) &= \sum_{i=0}^{L-1} w(i)x(t-i) \\
        &= \bm{w}^T \bm{x}(t) \\
        &= \bm{w} \bullet \bm{x}(t)
    \end{align}
    where \( \bullet \) denotes the inner product operation. By designing \( \bm{w} \), the desired signal \( y(t) \) can be extracted from the observed signal \( x(t) \).

    \item For signal processing in the short-time Fourier transform domain, the time-domain slice corresponds to \( \bm{x}(t) = [x(tK)~ x(tK-1)~\cdots~x(tK-L+1)]^T \). By applying the Fourier transform, filtering, and inverse Fourier transform, the signal estimate for the current time slice can be obtained. In this framework, the filter output can be expressed as
    \begin{align}
        \bm{y}(t) = \mathrm{IFFT} \circ f \circ \mathrm{FFT}[\bm{x}(t)],
    \end{align}
    where \( f(\cdot) \) is the frequency-domain filter. Since the Fourier transform and inverse transform are fixed operations within the filtering framework, if we define the observed slice after the Fourier transform as \( \bm{x}(t) \) and the output of the frequency-domain filter as \( \bm{y}(t) \), then the filtering operation can be described as
    \begin{align}
        \bm{y}(t) = f[\bm{x}(t)].
    \end{align}
    Here, the length of \( \bm{x}(t) \) depends on the number of points in the Fourier transform, and the length of \( \bm{y}(t) \) is the same as that of \( \bm{x}(t) \).

    \item For signal detection problems, \( \bm{x}(t) \) typically represents the result after feature extraction from the time slice, usually a one-dimensional vector, such as Mel-frequency cepstral coefficients (MFCC) \cite{davis1980comparison}. By applying these features through a target detector,
    \begin{align}
        y(t) = f[\bm{x}(t)],
    \end{align}
    we can determine whether the target appears in the current frame of the signal; here, \( y(t) \) represents the probability of the target's occurrence.
\end{itemize}
In summary, sound signal processing typically describes a signal/sample using a series of time slices or features extracted from those slices. For convenience, the \( n \)-th signal in the dataset is denoted as \( \bm{x}_{n} \); the \( t \)-th slice or feature of this signal is denoted as \( \bm{x}_{n}(t) \); the number of slices for this signal is denoted as \( T_{n} \). For simplicity, we will refer to signal slices and features extracted based on slices collectively as ``slices."

\subsection{Neurons and Nonlinear Activation Functions}

A neuron acting on the signal slice \( \bm{x}(t) \) typically consists of a linear mapping and a nonlinear transformation. The output of the neuron can be expressed as
\begin{align}
    y(t) = \sigma\left[ \bm{w} \bullet \bm{x}(t) + b \right],
\end{align}
where \( \bullet \) denotes the inner product, \( \bm{w} \) is the weight coefficient vector with the same dimension as \( \bm{x}(t) \), \( b \) is the bias, and \( \sigma(\cdot) \) is the nonlinear activation function. The output \( y(t) \) is the output of the neuron. During network optimization, \( \bm{w} \) and \( b \) are the parameters that the network learns from the data.

Common activation functions include:
\begin{itemize}
    \item \emph{Sigmoid Function}:
    \begin{align}
        \sigma(z) = \frac{1}{1 + \exp(-z)}
    \end{align}
    As \( z \) approaches infinity, \( \sigma(z) \) approaches 1; as \( z \) approaches negative infinity, \( \sigma(z) \) approaches 0.

    \item \emph{Rectified Linear Unit (ReLU)} \cite{lecun2002efficient, nair2010rectified}:
    \begin{align}
        \sigma(z) = \max(0, z)
    \end{align}
    When \( z > 0 \), \( \sigma(z) = z \); when \( z \leq 0 \), \( \sigma(z) = 0 \).

    \item \emph{Leaky ReLU} \cite{maas2013rectifier}:
    \begin{align}
        \sigma(z) = \begin{cases}
            z, & z > 0; \\
            \alpha z, & z \leq 0.
        \end{cases}
    \end{align}
    Here, \( \alpha \in (0, 1) \) is a hyperparameter.

    \item \emph{Swish Function}\cite{ramachandran2017searching}:
    \begin{align}
        \sigma(z) = z \cdot \frac{1}{1 + \exp(-z)}
    \end{align}
    As \( z \) approaches infinity, \( \sigma(z) \) approaches \( z \); as \( z \) approaches negative infinity, \( \sigma(z) \) approaches 0.

    \item \emph{Hyperbolic Tangent Function, known as $\mathrm{tanh}(\cdot)$}:
    \begin{align}
        \sigma(z) = 2 \cdot \frac{1}{1 + \exp(-2z)} - 1
    \end{align}
    As \( z \) approaches infinity, \( \sigma(z) \) approaches 1; as \( z \) approaches negative infinity, \( \sigma(z) \) approaches -1. The shape of the hyperbolic tangent function is similar to that of the sigmoid function but is stretched vertically and shifted downwards.
\end{itemize}
From the properties of nonlinear activation functions, it can be observed that various nonlinear activation functions suppress outputs for inner product values \( \bm{w} \bullet \bm{x}(t) \) below certain thresholds, aiming to maintain reasonable function values.

According to the definition of the inner product, when both vectors have a norm of 1, the inner product effectively calculates the similarity between the two vectors. In other words, the more similar the two vectors, the larger the inner product; conversely, the smaller the inner product. Therefore, the function of a single neuron is essentially to perform shape matching, using \( \bm{w} \) to define a shape template and searching for slices in the data \( \bm{x}(t) \) that match this template. The output of the function describes the degree of matching. When the matching is high, the current slice can be considered to encode the shape/feature represented by \( \bm{w} \). Training the network involves finding a suitable template \( \bm{w} \) based on the dataset and objective function.

When \( K \) neurons act on the same slice, the output of that layer can be expressed as
\begin{align}
    \bm{y}(t) = \left[\begin{array}{c}
        \sigma\left[\bm{w}_{1} \bullet \bm{x}(t) + b_{1}\right] \\
        \sigma\left[\bm{w}_{2} \bullet \bm{x}(t) + b_{2}\right] \\
        \vdots \\
        \sigma\left[\bm{w}_{K} \bullet \bm{x}(t) + b_{K}\right]
    \end{array}\right] = \sigma\left[\bm{W}^{T} \bm{x}(t) + \bm{b}\right], \label{y-ful-nn}
\end{align}
where \( \bm{W} = [\bm{w}_{1}~\bm{w}_{2}~\cdots~\bm{w}_{K}] \) and \( \bm{b} = [b_{1}~b_{2}~\cdots~b_{K}]^T \). Each \( \bm{w}_{k} \) defines a template, and all \( K \) templates form a set of useful features.

\subsection{Convolutional Neural Networks}

Convolution is the fundamental operation for implementing linear system filtering. Given an observed signal \( x(t) \) and the system's impulse response \( w(t) \), the output of the filter is given by \( y(t) = \sum_{i=0}^{L-1} w(i)x(t-i), \forall t \). The length of the filter determines the number of observation points required to compute the current output. A longer filter length requires more observation points, while a shorter length requires fewer points.

For convolutional neural network operations \cite{lecun1998gradient}, we first need to express the product of weights and observations \( w(i)x(t-i) \) as the inner product of a weight vector/matrix and the observation slice, \( \bm{w}(i) \bullet \bm{x}(t-i) \). The output can then be expressed as
\begin{align}
    y(t) &= \sigma\left[ \sum_{i=0}^{L-1} \bm{w}(i) \bullet \bm{x}(t-i)\right] \label{y-cnn} \\
    &= \sigma\left[ \underline{\bm{w}} \bullet \underline{\bm{x}}(t)\right], 
\end{align}
where \( \sigma(\cdot) \) is the nonlinear activation function, \( \underline{\bm{w}} = [\bm{w}(0) ~\bm{w}(1) ~\cdots ~\bm{w}(L-1)] \), and \( \underline{\bm{x}}(t) = [\bm{x}(t)~ \bm{x}(t-1)~\cdots~\bm{x}(t-L+1)] \). When \( \bm{x}(t) \) is a vector, \( \underline{\bm{w}} \) is a matrix; when \( \bm{x}(t) \) is a tensor, \( \underline{\bm{w}} \) becomes a tensor with one additional dimension compared to \( \bm{x}(t) \).

In convolution operations, \( \underline{\bm{w}} \) is referred to as the convolution kernel, with its important parameter being the width of the kernel:
\begin{itemize}
    \item If \( \bm{x}(t) \) is treated as a multi-dimensional sequential signal, convolution must occur along the time dimension, where \( L \) in equation (\ref{y-cnn}) represents the kernel width;
    \item When \( L=1 \), the convolution operation effectively performs a merging process on the input signal across the channels.
\end{itemize}

Each convolution kernel produces one output. When multiple convolution kernels are present in the same layer, the network will have multiple outputs; the number of convolution kernels determines the depth of the next layer's signal. Given \( K \) convolution kernels, the network's output can be described as
\begin{align}
    \bm{y}(t) = \left[\begin{array}{c}
        \sigma\left[\underline{\bm{w}}_{1} \bullet \underline{\bm{x}}(t)\right] \\
        \sigma\left[\underline{\bm{w}}_{2} \bullet \underline{\bm{x}}(t)\right] \\
        \vdots \\
        \sigma\left[\underline{\bm{w}}_{K} \bullet \underline{\bm{x}}(t)\right]
    \end{array}\right]
\end{align}
In one-dimensional convolution operations, \( \bm{y}(t) \) is a \( K \times 1 \) vector. Similar to the weight coefficient vector in neural networks, each convolution kernel represents a feature template, reflecting features useful for the current task from a large amount of training data, \( \underline{\bm{x}}_{n}(t), \forall n, t \). To enhance the model's expressive ability, the number of convolution kernels is typically greater than the number of input signal channels; that is, the output signal's channel count should exceed that of the input signal. Consequently, as the network stacks layers, the amount of data output can increase dramatically. Therefore, after convolution, downsampling or pooling operations are usually required to reduce the output length from a signal of length \( T \) to a signal of length \( T/N \), where \( N \) is the stride of the downsampling. Common downsampling operations include:
\begin{itemize}
    \item \emph{Equal Interval Sampling}:
    \begin{align}
        \bm{y}^{\prime}(i) = \bm{y}(iN)
    \end{align}
    This operation is relatively simple, requiring only data extraction every \( N \) points; however, according to the basic principles of digital signal processing, direct sampling can lead to aliasing. A better approach would be to average or apply low-pass filtering to adjacent \( N \) slices directly.

    \item \emph{Max Pooling}:
    \begin{align}
        \bm{y}^{\prime}(i) = \max_{\ell=0,1,\ldots,N-1} \bm{y}(iN - \ell)
    \end{align}
    The basic principle here is to retain the most useful information among adjacent \( N \) slices, as this information encodes whether the feature represented by the current kernel appears in the current time slice, which is crucial for subsequent layer inferences.
\end{itemize}
If we combine the convolution operation, nonlinear transformation, and pooling operation into a single function \( \mathcal{C}(\cdot) \), the relationship between the output and input of the convolutional network can be expressed as
\begin{align}
    \bm{y} = \mathcal{C}(\bm{x})
\end{align}
where \( \bm{y} = [\bm{y}(0)~ \bm{y}(1)~\cdots~\bm{y}(T/N-1)] \) and \( \bm{x} = [\bm{x}(0)~\bm{x}(1)~\cdots~\bm{x}(T-1)] \) represent the output and input signals. After passing through the convolutional network, the size/length of the signal in the convolution dimension decreases due to the pooling operation, while the number of channels in the signal is determined by the number of convolution kernels. Generally, the number of channels in the output signal \( \bm{y} \) should exceed that of the input signal \( \bm{x} \); however, in specific cases where information fusion is required, a width of 1 in the network may compress the number of channels, resulting in fewer output channels than input channels.

From the operations of the convolutional network, it can be observed that each convolution kernel learns/captures patterns from the adjacent \( L \) slices of the signal \( \underline{\bm{x}}(t) \). This means that a single output point of the network is determined by the local input, and the range of input that a single output point can ``feel" is referred to as the neuron's ``receptive field." A smaller receptive field allows the neuron to capture less information, while a larger receptive field enables the neuron to capture more and more complex information. From the perspective of the receptive field, a single layer neuron’s ability to characterize input information is limited by the kernel width \( L \); however, this limitation can be overcome using multi-layer convolutional networks. A single-layer convolutional network can be seen as a filtering system, while a multi-layer convolutional network can be viewed as a cascade of multiple systems. Based on the fundamental theory of signal processing, the cascade of multiple FIR systems increases the overall impulse response length of the FIR system. Specifically, the equivalent length of the overall FIR system impulse response can be expressed as
\begin{align}
    L = \sum_{i=1}^{N} L_{i} - N + 1,
\end{align}
where \( L_{i} \) is the length of the impulse response corresponding to the \( i \)-th system, and \( N \) is the number of systems used for cascading. By extension, the cascading of multi-layer convolutional networks can significantly increase the receptive field of individual neurons. As the number of layers increases, the receptive field of the neurons also enlarges, allowing them to characterize more complex information.

The previous discussion focused on one-dimensional convolution, specifically the design of convolutional networks for multi-channel one-dimensional data. In practical applications, particularly in short-time Fourier transform domain sound signal processing, each element of the vector \( \bm{x}(t) \) represents a spectral component of the signal, where structural information exists between adjacent frequency bands. Here, \( \bm{x} = [\bm{x}(0)~\bm{x}(1)~\cdots~\bm{x}(T-1)] \) can be viewed as an image, allowing convolution operations to capture local features in both the time and frequency dimensions. In this case, \( \underline{\bm{w}} \) becomes a tensor of dimensions \( \text{length} \times \text{width} \times \text{number of channels} \); the receptive fields in both the time and frequency dimensions need to be designed separately, as do the strides in these dimensions.

\subsection{ Recurrent Neural Networks}

In classical digital signal processing, there are two typical filtering frameworks: FIR filtering and IIR filtering. FIR filtering produces the current output of a filter by weighting and summing the observed signal, akin to the operations of convolutional neural networks. In contrast, the output of an IIR filter is a linear combination of two parts: one part weights and sums the observed signal, while the other part weights the previous output of the filter. Given an input signal \( x(t) \), the output of an IIR filter is typically represented as:
\begin{align}
    y(t) = \sum_{i=0}^{L_{b}-1} b(i)x(t-i) + \sum_{j=1}^{L_{a}-1} a(j)y(t-j),
\end{align}
where \( a(j) \) and \( b(i) \) characterize the two sets of coefficients of the IIR filter, together defining the filter's properties. For a simple IIR filter given by \( y(t) = x(t) + a y(t-1) \) where both \( a(j) \) and \( b(i) \) have a single coefficient, we can rewrite the output using the iterative relation as:
\begin{align}
    y(t) = \sum_{i=0}^{\infty} a^{i} x(t-i).
\end{align}
This operation is equivalent to filtering with an FIR filter of infinite length.

From this example, we can see that: 1) IIR can be transformed into FIR, resulting in a filter of infinite length; 2) When \( |a| > 1 \), the filter output tends to infinity, leading to potential instability issues with such filters.

The neural network corresponding to FIR filters is the convolutional neural network, while the network corresponding to IIR filters is the recurrent neural network. Similar to IIR filters, recurrent neural networks have a large receptive field and face stability issues. Analogous to IIR filtering, the output of a recurrent neural network can be described as:
\begin{align}
    \bm{y}(t) = \sigma\left[ \bm{W} \bm{x}(t) + \bm{U} \bm{y}(t-1) + \bm{b} \right],
\end{align}
where \( \bm{x}(t) \) is the feature/signal of the sample at time \( t \), and \( \bm{y}(t) \) is the network's output. Assuming the length of \( \bm{x}(t) \) is \( M \) and the length of \( \bm{y}(t) \) is \( K \), the weight coefficient matrix \( \bm{W} \) has dimensions \( K \times M \), and the weight coefficient matrix \( \bm{U} \) has dimensions \( K \times K \), with the bias vector \( \bm{b} \) having length \( K \). The goal of training the recurrent neural network is to learn suitable \( \bm{W} \), \( \bm{U} \), and \( \bm{b} \) from the data, transforming features from \( \bm{x}(t) \) to \( \bm{y}(t) \).

To address stability issues in recurrent neural networks, an effective method is to introduce gating units that selectively forget past outputs. Common network structures include Long Short-Term Memory (LSTM) networks \cite{hochreiter1997long} and Gated Recurrent Units (GRU) \cite{cho2014learning}. In the construction of LSTM networks, three gating functions and an intermediate variable are used to compute the final network output, with the specific computation steps as follows:
\begin{align}
    \bm{f}(t) &= \sigma\left[ \bm{W}_{f} \bm{x}(t) + \bm{U}_{f} \bm{y}(t-1) + \bm{b}_{f} \right],\\
    \bm{i}(t) &= \sigma\left[ \bm{W}_{i} \bm{x}(t) + \bm{U}_{i} \bm{y}(t-1) + \bm{b}_{i} \right],\\
    \bm{o}(t) &= \sigma\left[ \bm{W}_{o} \bm{x}(t) + \bm{U}_{o} \bm{y}(t-1) + \bm{b}_{o} \right],\\
    \bm{c}(t) &= \bm{f}(t) \odot \bm{c}(t-1) + \nonumber\\
    &~~~~~~ \bm{i}(t) \odot \mathrm{tanh}\left[ \bm{W}_{c} \bm{x}(t) + \bm{U}_{c} \bm{y}(t-1) + \bm{b}_{c} \right],\\
    \bm{y}(t) &= \mathrm{tanh}\left[\bm{o}(t) \odot \bm{c}(t)\right],
\end{align}
where \( \bm{f}(t) \), \( \bm{i}(t) \), and \( \bm{o}(t) \) are the three gating functions, selectively forgetting past information, input information, and output information, respectively; \( \bm{c}(t) \) is an intermediate variable. Compared to LSTM networks, GRU networks require only two gating functions and one intermediate variable to complete the computations, with the specific steps as follows:
\begin{align}
    \bm{\zeta}(t) &= \sigma\left[ \bm{W}_{\zeta} \bm{x}(t) + \bm{U}_{\zeta} \bm{y}(t-1) + \bm{b}_{\zeta} \right],\\
    \bm{r}(t) &= \sigma\left[ \bm{W}_{r} \bm{x}(t) + \bm{U}_{r} \bm{y}(t-1) + \bm{b}_{r} \right],\\
    \bm{c}(t) &= \mathrm{tanh}\left\{ \bm{W}_{c} \bm{x}(t) + \bm{U}_{c} \left[\bm{r}(t) \odot \bm{y}(t-1) \right] + \bm{b}_{c} \right\},\\
    \bm{y}(t) &= \bm{\zeta}(t) \odot \bm{y}(t-1) + \left[1 - \bm{\zeta}(t)\right] \odot \bm{c}(t),
\end{align}
where \( \bm{\zeta}(t) \) and \( \bm{r}(t) \) are the two gating functions, and \( \bm{c}(t) \) is the intermediate variable. During the training of recurrent neural networks, all \( \bm{W} \), \( \bm{U} \), and \( \bm{b} \) are network parameters to be learned, determined based on data samples in conjunction with task objectives.

Compared to convolutional neural networks, a single-layer recurrent neural network can capture long-term features of signals. Such networks are typically used in time-series signal processing, serving as the output layer of the last convolutional network or the input layer of fully connected networks. In practice, due to various forgetting mechanisms, the receptive field of recurrent neural networks cannot be infinitely large; the effectiveness of these networks must be validated in the context of specific tasks.

\subsection{Residual Neural Networks}

Residual Neural Networks consist of two components \cite{he2016deep}: a nonlinear mapping \( \hbar(\cdot) \) that transforms the input into an intermediate variable, and a direct connection. Given an input signal/feature \( \bm{x} = [\bm{x}(0)~\bm{x}(1)~\cdots~\bm{x}(T-1)] \), the output of the residual network is calculated as follows:
\begin{align}
    \bm{c} &= \hbar(\bm{x}),\\
    \bm{y} &= \bm{x} + \bm{c} = \bm{x} + \hbar(\bm{x}),
\end{align}
where \( \bm{c} = \hbar(\bm{x}) = \bm{y} - \bm{x} \) is the intermediate variable, which has the same dimension as the input signal \( \bm{x} \). The addition of the direct connection is key to residual networks, as it enables training of deep neural networks by mitigating the gradient vanishing problem.

\subsection{Encoder-Decoder Networks}

Encoder-Decoder networks consist of two parts\cite{cho2014learning}: the encoder \( \mathcal{E}(\cdot) \) and the decoder \( \mathcal{D}(\cdot) \). The encoder transforms the input signal \( \bm{x} \) into a feature space to obtain \( \bm{z} \), while the decoder maps the feature \( \bm{z} \) back to the input signal space. Typically, the dimension of \( \bm{z} \) is much smaller than that of the input signal \( \bm{x} \). In the framework of fully convolutional networks, the encoder can be expressed as:
\begin{align}
    \bm{z} = \mathcal{E}(\bm{x}) = \mathcal{C}_{Q} \circ \mathcal{C}_{Q-1} \circ \cdots \circ \mathcal{C}_{1}(\bm{x}),
\end{align}
where \( \mathcal{C}_{1}(\cdot) \) is the first layer of the convolutional network, and \( Q \) is the number of layers in the convolutional network. Generally, the number of input channels at layer \( q \) is less than the number of output channels, and the number of output slices is less than the number of input slices; each slice of the subsequent convolutional network encodes specific features within a defined receptive field.

The construction of the decoder network is similar to that of the encoder network, but it utilizes transposed convolutions to increase the number of slices, thereby enhancing the temporal resolution of the signal/feature. Specifically, the decoder network can be represented as:
\begin{align}
    \hat{\bm{x}} = \mathcal{D}(\bm{z}) = \mathcal{C}_{1}^{\prime} \circ \mathcal{C}_{2}^{\prime} \circ \cdots \circ \mathcal{C}_{Q}^{\prime}(\bm{z}),
\end{align}
where \( \mathcal{C}_{q}^{\prime}(\cdot) \) denotes the transposed convolution, and its parameter dimensions correspond one-to-one with those of \( \mathcal{C}_{q}(\cdot) \); the key difference is that it performs data expansion rather than compression. From the perspective of data compression, the training of the encoder-decoder network does not require data labels; the goal is to make the decoded data approximate the original data:
\begin{align}
    \bm{x} \leftarrow \mathcal{D} \circ \mathcal{E}(\bm{x}),
\end{align}
where \( \leftarrow \) indicates that the right-hand side aims to approximate the left-hand side. In practice, constraints are often established on \( \bm{z} \) based on the distribution of the latent space or prior information, ensuring that the features obtained from encoding have specific meanings.

To enhance the robustness of the encoder features, noise is often added before decoding in the latent space (only during the training phase), which is the fundamental principle of Variational Autoencoders (VAEs) \cite{kingma2013auto}. Specifically, in the framework of VAEs, the latent variable \( \bm{z} \) consists of two parts: the current feature \( \bm{\mu} \) and another part measuring the uncertainty of this feature \( \bm{\xi} \). The encoding and decoding processes of the network can thus be expressed as follows:
\begin{align}
    \bm{z} &= \bm{\mu} \ddagger \bm{\xi}\nonumber\\
    &= \mathcal{E}(\bm{x}), \\
    \bm{c} &= \bm{\mu} + \bm{\xi} \odot \bm{\epsilon}, \\
    \hat{\bm{x}} &= \mathcal{D}(\bm{c}),
\end{align}
where \( \ddagger \) indicates concatenation along depth, \( \bm{c} \) is the intermediate variable, and \( \bm{\epsilon} \) is a sample from a zero-mean Gaussian random process. Thus, during decoding, the decoder network does not directly decode \( \bm{z} \) but rather decodes \( \bm{\mu} + \bm{\xi} \odot \bm{\epsilon} \), where \( \bm{\xi} \) can be viewed as the standard deviation of the Gaussian random process. By introducing noise into the features during training, VAEs enhance the robustness and generalization ability of the network; however, this can lead to the ``many-to-one" problem in function mapping, sometimes making the decoded signals appear ``fuzzy."

\subsection{U-Net Type Neural Networks}

U-Net type neural networks also have a similar encoder-decoder structure, utilizing feature concatenation to repeatedly leverage features of different dimensions during the ``decoding" process \cite{ronneberger2015u}. For convenience, let \( \mathcal{C}_{q}(\cdot) \) denote a convolution module composed of multiple layers of convolutions, producing output \( \bm{x}^{(q)} \); \( \mathcal{C}_{q}^{\prime}(\cdot) \) denotes the corresponding transposed convolution module, producing output \( \bm{c}^{(q)} \). In analogy to the ``encoding" part, we first have:
\begin{align}
    \bm{x}^{(q)} &= \mathcal{C}_{q}\left[\bm{x}^{(q-1)}\right] \nonumber\\
    &= \mathcal{C}_{q} \circ \mathcal{C}_{q-1} \circ \cdots \circ \mathcal{C}_{1}(\bm{x}), \forall q = 1,2,\ldots,Q.
\end{align}
For each module in the ``decoding" part, its input is formed by concatenating the output of the preceding ``decoding" module with the output of the corresponding ``encoding" module:
\begin{align}
    \bm{c}^{(q)} = \mathcal{C}_{q}^{\prime}\left[\bm{x}^{(q)} \ddagger \bm{c}^{(q+1)}\right], \forall q=Q-1, Q-2, \ldots, 1.
\end{align}
Here, \( \ddagger \) indicates concatenation of the two signals/features along depth. The literature \cite{ronneberger2015u} presents a U-Net network structure diagram for \( Q=4 \); its overall shape resembles the uppercase letter ``U," hence the name U-Net. For convenience, we define the overall function of the U-Net network as \( \mathcal{U}(\cdot) \). Given an input signal \( \bm{x} \), the network's output can be described as:
\begin{align}
    \bm{y} = \mathcal{U}(\bm{x}).
\end{align}
According to the previous definitions, we have \( \bm{y} = \bm{c}^{(1)} = \mathcal{C}_{1}^{\prime}\left[\bm{x}^{(1)} \ddagger \bm{c}^{(2)}\right] \). Since the inputs to the transposed convolution modules in the ``decoding" part do not match the depth of the corresponding ``encoding module outputs," the parameter dimensions of \( \mathcal{C}_{q}^{\prime}(\cdot) \) and \( \mathcal{C}_{q}(\cdot) \) do not correspond one-to-one.

\subsection{Feature Fusion Networks}

Feature fusion networks are typically used to convert variable-length data into fixed-length features, such as in voiceprint recognition. Given a signal/feature composed of \( T \) slices \( \bm{x} = [\bm{x}(0)~\bm{x}(1)~\cdots~\bm{x}(T-1)] \), a simple feature fusion strategy is:
\begin{align}
    \bm{y} = \frac{1}{T} \sum_{i=0}^{T-1} \bm{x}(i).
\end{align}
Here, \( \bm{y} \) has the same dimension as a single slice \( \bm{x}(i) \). Since each slice contains varying amounts of useful information, it is necessary to treat each slice differently with non-uniform weights; the corresponding feature fusion strategy can be expressed as:
\begin{align}
	\bm{y}
	&= \dfrac{1}{\sum_{j=0}^{T-1} \expd{\hbar\left[\bm{x}(j)\right]}} \sum_{i=0}^{T-1} \expd{\hbar\left[\bm{x}(i)\right]} \bm{x}(i)\\
	&= \sum_{i=0}^{T-1} \alpha_{i} \bm{x}(i)
\end{align}
where \( \hbar(\cdot) \) is a scoring function that indicates the importance of the current slice data for the task, which can be defined as the inner product between vectors or provided by another neural network directly. The coefficients are given by:
\begin{align}
    \alpha_{i}
	&= \dfrac{\expd{\hbar\left[\bm{x}(i)\right]}}{\sum_{j=0}^{T-1} \expd{\hbar\left[\bm{x}(j)\right]}},
\end{align}
which serves as the final weighting factor for the slices, satisfying 
$ \sum_{i=0}^{T-1} \alpha_{i} = 1 $.

By applying an appropriate linear transformation to the fused features, we can also derive:
\begin{align}
    \bm{y}^{\prime}
	&= \bm{V}\bm{y}\\
	&=  \sum_{i=0}^{T-1} \dfrac{\expd{\hbar\left[\bm{x}(i)\right]}}{\sum_{j=0}^{T-1} \expd{\hbar\left[\bm{x}(j)\right]}}  \bm{V}\bm{x}(i).
\end{align}
This represents a linear transformation applied to the input slices before fusion. In practice, if there are \( M \) scoring functions \( \hbar_{m}(\cdot) \) and \( M \) transformation matrices \( \bm{V}_{m} \), we can derive \( M \) fused features \( \bm{y}_{m}^{\prime} \). Finally, concatenating the \( M \) features yields the overall fused output:
\begin{align}
    \bm{y} = \bm{y}_{1}^{\prime} \ddagger \bm{y}_{2}^{\prime} \cdots \ddagger \bm{y}_{M}^{\prime}.
\end{align}

\subsection{Self-Attention Mechanism Networks}

Self-attention mechanism networks function similarly to feature fusion networks, with the key difference being that their outputs are not compressed along the time dimension; instead, their scoring functions are determined by slices at different time points. Specifically, each self-attention module comprises three learnable parameter matrices that perform different linear transformations on the slice data, resulting in three signals:
\begin{align}
    \bm{q}(t) &= \bm{T}_{q} \bm{x}(t),\\
    \bm{k}(t) &= \bm{T}_{k} \bm{x}(t),\\
    \bm{v}(t) &= \bm{T}_{v} \bm{x}(t).
\end{align}
Under single-head attention conditions, we can compute the output of the network by weighting \( \bm{v}(t) \):
\begin{align}
    \bm{y}^{\prime}(t) &= \sum_{i=0}^{T-1} \frac{\expd{\bm{k}^{T}(i) \bm{q}(t)}}{\sum_{j=0}^{T-1} \expd{\bm{k}^{T}(j) \bm{q}(t)}} \bm{v}(i)\\
    &= \sum_{i=0}^{T-1} \frac{\expd{\bm{x}^{T}(i) \bm{T}_{k}^{T} \bm{T}_{q} \bm{x}(t)}}{\sum_{j=0}^{T-1} \expd{\bm{x}^{T}(j) \bm{T}_{k}^{T} \bm{T}_{q} \bm{x}(t)}} \bm{T}_{v} \bm{x}(i).
\end{align}
Each output slice at each time point is obtained by a weighted sum of all input slices. When there are multiple \( \bm{T}_{q} \), \( \bm{T}_{k} \), and \( \bm{T}_{v} \), we can derive different output data \( \bm{y}_{m}^{\prime} \). By concatenating the outputs obtained from the attention operations, we can obtain the output slice of the multi-head attention mechanism network:
\begin{align}
    \bm{y}(t) = \bm{y}_{1}^{\prime}(t) \ddagger \bm{y}_{2}^{\prime}(t) \cdots \ddagger \bm{y}_{M}^{\prime}(t),
\end{align}
and the overall output:
\begin{align}
    \bm{y} = \bm{y}_{1}^{\prime} \ddagger \bm{y}_{2}^{\prime} \cdots \ddagger \bm{y}_{M}^{\prime}.
\end{align}
For convenience, we denote the functional relationship of the multi-head attention mechanism network as:
\begin{align}
    \bm{y} = \mathcal{M}(\bm{x}).
\end{align}
Here, \( \mathcal{M}(\cdot) \) represents the multi-head attention.

\subsection{Fully Connected Networks}

Fully connected networks facilitate nonlinear mappings of data, typically composed of multiple layers of linear transformations combined with nonlinear mappings. Given a signal/feature \( \bm{x} = [\bm{x}(0)~\bm{x}(1)~\cdots~\bm{x}(T-1)] \), if a nonlinear mapping is required for each slice, the relationship between the network input and output can be described using equation (\ref{y-ful-nn}). If a nonlinear transformation is needed for the entire signal/feature, \( \bm{x} \) must be reshaped into a vector before applying the nonlinear transformation. In this latter case, the dimensionality of the input signal/feature must remain fixed, preventing the input of signals/features of arbitrary lengths.

\subsection{Auxiliary Neural Networks}

Auxiliary neural networks generally fall into two categories: one for regularizing signals/features, such as batch normalization and layer normalization, which ensure that the regularized data conforms to a preset distribution, avoiding excessive concentration of features and enhancing the utilization efficiency of neurons; the other introduces noise to increase uncertainty, enhancing the robustness and generalization ability of the network, such as dropout operations and data augmentation.

\subsubsection{Normalization Operations}

Considering the case of multi-channel one-dimensional signals, \( \bm{x} = [\bm{x}(0)~\bm{x}(1)~\cdots~\bm{x}(T-1)] \), where each \( \bm{x}(t) \) is a \( Q \times 1 \) vector, the entire sample can be viewed as a \( Q \times T \) matrix. For convenience, we temporarily denote the \( q \)-th element of \( \bm{x}(t) \) as \( x(t,q) \). Both batch normalization and layer normalization regularize the input signal based on data samples. For each data sample \( \bm{x}_{n} \), the normalization operation for the data sample \( x_{n}(t,q) \) can be described as:
\begin{align}
    x_{n}(t,q) = \frac{x_{n}(t,q) - \mu}{\sqrt{\phi + \epsilon}} \gamma + \beta,
\end{align}
where \( \epsilon > 0 \) is a preset parameter; \( \mu \) and \( \phi \) are the mean and variance computed from the data during forward propagation based on that point, and the methods for calculating these values differ according to normalization criteria; \( \gamma \) and \( \beta \) represent scaling and shifting operations, respectively, which are the parameters for network optimization.
\begin{itemize}
    \item \emph{Batch Normalization} \cite{ioffe2015batch}: It assumes that the same batch of data shares the mean and variance at the same feature point and performs uniform data regularization. The calculations for \( \mu \) and \( \phi \) are as follows:

    \begin{align}
        \mu &= \frac{1}{N} \sum_{n=1}^{N} x_{n}(t,q),\\
        \phi &= \frac{1}{N} \sum_{n=1}^{N} |x_{n}(t,q) - \mu|^2.
    \end{align}
    From the above, it is evident that for each feature point location \( (t,q) \), corresponding \( \mu \) and \( \phi \) must be computed; naturally, different \( \gamma \) and \( \beta \) must also be learned for each feature point. Therefore, \( \mu \), \( \phi \), \( \gamma \), and \( \beta \) are functions of \( (t,q) \). Since the statistics of \( \mu \) and \( \phi \) require a batch of data, this can be implemented during training; however, calculating these values during testing and deployment is impractical. To address this, fixed values for \( \mu \) and \( \phi \) are designed based on their training phase values for network testing and deployment.

    \item \emph{Layer Normalization} \cite{ba2016layer}: It assumes that features within the same dimension/layer can share the mean and variance, performing uniform data regularization. If we assume that normalization is performed along the time dimension, the calculations for \( \mu \) and \( \phi \) are as follows:

    \begin{align}
        \mu &= \frac{1}{T} \sum_{t=0}^{T-1} x_{n}(t,q),\\
        \phi &= \frac{1}{T} \sum_{t=0}^{T-1} |x_{n}(t,q) - \mu|^2.
    \end{align}
    Here, for each data sample, the resulting \( \mu \) and \( \phi \) differ; they are functions of \( n \). We assume that normalization parameters along the time dimension can share values, meaning \( \mu \) and \( \phi \) do not depend on the time index \( t \). Thus, \( \mu \) and \( \phi \) depend on \( n \) and \( q \). Since it does not require a batch of data to calculate the mean and variance, this method can be used for both training and testing phases. For the learnable parameters \( \gamma \) and \( \beta \), they also share mechanisms across the time dimension, depending only on the feature dimension index \( q \).
\end{itemize}
In summary, both methods are point-wise normalization operations based on data samples; like nonlinear activation functions, they are computed individually. The difference lies in the assumptions about parameter sharing, which leads to different normalization effects.

\subsubsection{Noise Addition Operations}

Adding noise during network training is a common method to enhance the network's generalization ability and robustness. The rationale is straightforward: if different noise types can all be mapped to the same network output, the network will have the following advantages post-training:
\begin{itemize}
    \item From noisy signals to clean features, the network itself will exhibit a degree of denoising capability;
    \item The network will be robust against disturbances introduced by the added noise;
    \item The introduction of noise may lead to the loss or distortion of some features, allowing the final network to capture more expressive features while improving its inference ability based on missing features.
\end{itemize}
In self-supervised learning, data augmentation operations effectively function as a noise addition operation,  applying special transformations to the data, enabling the final network to cope with the challenges posed by such transformations and learn robust features. Similarly, in the previously mentioned variational autoencoders, Gaussian noise is added to the latent layer features for the same reason. In fact, early dropout operations can also be viewed as a form of noise addition \cite{srivastava2014dropout, krizhevsky2012imagenet}. Dropout occurs only during the training phase, randomly dropping features of the neural network by a certain percentage, enhancing the final network's robustness and generalization performance.

\subsection{Transformer-based Neural Networks}

Transformer-based neural networks integrate various network structures during construction \cite{vaswani2017attention, gulati2020conformer}, including fully connected networks, convolutional networks, multi-head attention networks, normalization operations, and noise addition operations. The Conformer network serves as an illustrative example. The Conformer network processes input to output through four sub-modules; let the output of the \( m \)-th module (\( m=1,2,3,4 \)) be \( \bm{x}^{(m)} \), with the network input as \( \bm{x} \) and output as \( \bm{y} \). The output computation of the Conformer network can be described as:
\begin{align}
    \bm{x}^{(1)} &= \bm{x} + \frac{1}{2} \mathcal{F}(\bm{x}),\\
    \bm{x}^{(2)} &= \bm{x}^{(1)} + \mathcal{M}(\bm{x}^{(1)}),\\
    \bm{x}^{(3)} &= \bm{x}^{(2)} + \mathcal{C}(\bm{x}^{(2)}),\\
    \bm{x}^{(4)} &= \bm{x}^{(3)} + \frac{1}{2} \mathcal{F}(\bm{x}^{(3)}),\\
    \bm{y} &= \mathcal{N}_{\mathrm{layer}}(\bm{x}^{(4)}).
\end{align}
Here, \( \mathcal{F}(\cdot) \) is a forward network module primarily composed of two fully connected layers; \( \mathcal{M}(\cdot) \) is a multi-head attention network; \( \mathcal{C}(\cdot) \) is a convolutional network module primarily composed of three convolutional layers; and \( \mathcal{N}_{\mathrm{layer}} \) is a layer normalization network. In the constructions of \( \mathcal{F}(\cdot) \), \( \mathcal{M}(\cdot) \), and \( \mathcal{C}(\cdot) \), two auxiliary neural networks are included to apply layer normalization to the signal before input and to add dropout noise after the network output to enhance robustness and generalization performance.

Specifically,
\begin{align}
    \mathcal{F}(\bm{x}) &= \mathcal{N}_{\mathrm{drop}} \circ \mathcal{L}_{2} \circ \mathcal{N}_{\mathrm{drop}} \circ \sigma_{1} \circ \mathcal{L}_{1} \circ \mathcal{N}_{\mathrm{layer}}(\bm{x}),\\
    \mathcal{M}(\bm{x}) &= \mathcal{N}_{\mathrm{drop}} \circ \mathcal{M}_{1} \circ \mathcal{N}_{\mathrm{layer}}(\bm{x}),\\
    \mathcal{C}(\bm{x}) &= \mathcal{N}_{\mathrm{drop}} \circ \mathcal{C}_{3} \circ \sigma_{2} \circ \mathcal{N}_{\mathrm{batch}} \circ \mathcal{C}_{2} \circ \sigma_{1} \circ \mathcal{C}_{1} \circ \mathcal{N}_{\mathrm{layer}}(\bm{x}),
\end{align}
where \( \sigma(\cdot) \) is the nonlinear activation function, and \( \mathcal{L}(\cdot) \) represents linear transformations applied to slices; \( \sigma \circ \mathcal{L}(\cdot) \) implements a fully connected layer. Since the final outputs of \( \mathcal{F}(\cdot) \), \( \mathcal{M}(\cdot) \), and \( \mathcal{C}(\cdot) \) require summation with their inputs, no nonlinear transformations are introduced at the last layer of each sub-module. In constructing the convolutional network, \( \mathcal{C}_{1}(\cdot) \) and \( \mathcal{C}_{3}(\cdot) \) perform convolution in the time dimension, while \( \mathcal{C}_{2}(\cdot) \) performs convolution in the channel dimension.

\section{Construction of Loss Functions}
 
The construction of loss functions is crucial for the training of deep learning networks. It is through the design of loss functions that neural networks can capture information useful for specific tasks from data. Typically, different tasks require the construction of different loss functions. This section discusses methods for constructing loss functions related to three tasks: signal detection, estimation, and transformation. 
     \emph{1) Signal detection, including binary classification problems, multi-class problems, target probability fitting, and detection probability super-resolution;
    2)  Signal estimation/filtering, including signal enhancement, denoising, and robust signal estimation;
    3)  Signal transformation, including feature learning, density modeling, and density alignment.}

Although signals may extend infinitely in time or space, in practice, signals are often sliced, obtaining finite segments for analysis and processing. Finally, merging the processed signal slices can enable complex tasks such as detection, estimation, and inference. For convenience, we treat each slice signal as a sample in deep learning; given any set of signals, we can slice them and annotate each slice, resulting in a dataset for deep learning \( \{(\bm{x}_{n}, y_{n}), \forall n\} \), where \( \bm{x}_{n} \) is the \( n \)-th sample and \( y_{n} \) is the corresponding label. When \( y_{n} \) is a vector or matrix, we denote it as \( \bm{y}_{n} \). Based on the neural network \( f(\bm{x}) \), for each sample/signal slice \( \bm{x}_{n} \), we can obtain the estimated value of the sample label \( \hat{y}_{n} = f(\bm{x}_{n}) \). The training of the neural network involves finding suitable network parameters \( \theta_{f} \) to minimize the difference between \( \hat{y}_{n} \) and \( y_{n} \). The mathematical characterization of this proximity is the loss function commonly used; different loss functions correspond to different interpretations of function approximation.

\subsection{Loss Functions in Signal Detection}

In signal processing tasks, it is typically necessary to first determine whether the expected signal is present in the observation to be processed and the probability of its presence. When multiple target signals exist, it is also essential to know the probability of each target signal's presence.

For signal detection tasks, the network mapping function usually consists of two parts: a feature extraction network \( g(\bm{x}) \) and an output layer \( S(\bm{x}) \). The feature extraction network \( g(\bm{x}) \) should be constructed based on the characteristics of the signal for different applications, which we will not discuss in detail here. For convenience, we denote the features obtained from \( g(\bm{x}) \) as \( \bm{z} \), specifically, \( \bm{z}_{n} = g(\bm{x}_{n}) \). The network's output layer typically has four forms, as follows:
\begin{itemize}
    \item \emph{Sigmoid Function}: As the network output layer, we denote this layer's function as \( S_{d}(\cdot) \):

    \begin{align}
        S_{d}(\bm{z}) = \frac{1}{1 + \expd{\bm{w}^{T} \bm{z} + b}},
    \end{align}
    where vector \( \bm{w} \) and bias \( b \) are parameters to be learned in the network output layer. This type of output layer is typically used to output a probability value between 0 and 1. Given \( S_{d}(\bm{z}) \), the overall output of the network can be described as \( \hat{y} = f(\bm{x}) = S_{d} \circ g(\bm{x}) \).

    \item \emph{Softmax Function}: As the network output layer, we denote it as \( S_{x}(\cdot) \). Its output is a vector of length \( L \), expressed as:

    \begin{align}
        S_{x}(\bm{z}) = \left[\begin{array}{c}
            \frac{\expd{\bm{w}_{0}^{T} \bm{z}}}{\sum_{\ell} \expd{\bm{w}_{\ell}^{T} \bm{z}}} \\
            \frac{\expd{\bm{w}_{1}^{T} \bm{z}}}{\sum_{\ell} \expd{\bm{w}_{\ell}^{T} \bm{z}}} \\
            \vdots \\
            \frac{\expd{\bm{w}_{L-1}^{T} \bm{z}}}{\sum_{\ell} \expd{\bm{w}_{\ell}^{T} \bm{z}}}
        \end{array}\right],
    \end{align}
    where \( \bm{w}_{\ell} \) are the parameters of the output layer to be optimized. The feature \( \bm{z} \) is the output feature of the network (with its last element being 1 to simplify the formula by incorporating the bias into the weight vector). It can be verified that the sum of all elements output by the Softmax function equals 1. When the Softmax function is used as the network output layer, the output is a vector of length \( L \), represented as \( \hat{\bm{y}} = f(\bm{x}) = S_{x} \circ g(\bm{x}) \).

    \item \emph{Multiple Sigmoid Functions}: Constructed as the network output layer, each element of the vector represents the probability of a specific target's occurrence. For convenience, we denote it as \( S_{d,\mathrm{multi}}(\cdot) \). Given the number of target types \( L \), the function \( S_{d,\mathrm{multi}}(\cdot) \) can be expressed as:

    \begin{align}
        S_{d,\mathrm{multi}}(\bm{z}) = \left[\begin{array}{c}
            \frac{1}{1 + \expd{\bm{w}^{T}_{0} \bm{z} + b_{0}}} \\
            \frac{1}{1 + \expd{\bm{w}^{T}_{1} \bm{z} + b_{1}}} \\
            \vdots \\
            \frac{1}{1 + \expd{\bm{w}^{T}_{L-1} \bm{z} + b_{L-1}}}
        \end{array}\right], \label{sigmoid-multi-def}
    \end{align}
    where the vectors \( \bm{w}_{\ell} \) and biases \( b_{\ell} \) are parameters to be optimized in the network output layer. In this case, the final output of the network can be represented as \( \hat{\bm{y}} = f(\bm{x}) = S_{d,\mathrm{multi}} \circ g(\bm{x}) \).

    \item \emph{Linear Mapping Followed by Sign Function}: As the network output layer, denoted as \( S_{s}(\cdot) \). Its output is either 1 or -1. Given the feature \( \bm{z} \), the function \( S_{s}(\cdot) \) is defined as:

    \begin{align}
        S_{s}(\bm{z}) = \mathrm{sign}\left(\bm{w}^{T} \bm{z} + b\right), \label{SVM-output}
    \end{align}
    where the vector \( \bm{w} \) and bias \( b \) are parameters to be learned in the network output layer. The term \( \bm{w}^{T} \bm{z} + b \) describes a hyperplane that maximizes the minimum distance from the positive and negative samples to the hyperplane. In this case, the network output can be represented as \( \hat{y} = f(\bm{x}) = S_{s} \circ g(\bm{x}) \).
\end{itemize}

\subsubsection{Traditional Signal Detection: Binary Classification}

Traditional signal detection can be viewed as a binary classification problem in machine learning. Given a segment of signal/sample, the task is to determine whether the expected target is present. If the target exists, the system outputs \( y = 1 \); if the target does not exist, the system outputs \( y = 0 \).

The most common loss function for binary classification is binary cross-entropy, defined as:
\begin{align}
    \mathcal{J}_{1,1}(y_{n}, \hat{y}_{n}) = - y_{n} \ln \hat{y}_{n} - (1 - y_{n}) \ln (1 - \hat{y}_{n}),
\end{align}
where \( y_{n} \) is the label of sample \( \bm{x}_{n} \); \( y_{n} = 1 \) indicates the target's presence, while \( y_{n} = 0 \) indicates its absence. \( \hat{y}_{n} \) is the output of the neural network \( f(\bm{x}_{n}) = S_{d} \circ g(\bm{x}_{n}) \), representing the probability of the target's occurrence for the given sample \( \bm{x}_{n} \). For a batch of samples during training, the corresponding loss function can be expressed as
\begin{align}
    \mathcal{J}_{1,1}^{(m)} &= \sum_{n \in \mathcal{I}_{m}} \mathcal{J}_{1,1}(y_{n}, \hat{y}_{n})\\
    & = -\sum_{n \in \mathcal{I}_{m}^{+}} \ln \hat{y}_{n} - \sum_{n \in \mathcal{I}_{m}^{-}} \ln (1 - \hat{y}_{n}), \label{J11-2cls-sum}
\end{align}
where \( \mathcal{I}_{m} \) is the index set of the samples in the \( m \)-th batch within the entire dataset, \( \mathcal{I}_{m}^{+} \) is the subset of indices corresponding to positive samples, and \( \mathcal{I}_{m}^{-} \) is the subset of indices corresponding to negative samples.

From (\ref{J11-2cls-sum}), it can be seen that each batch of samples contributes to the overall loss function through both positive and negative samples. This approach usually does not present issues. However, when the probability of the target's occurrence in the given signal sample is low, a severe ``class imbalance problem" may arise, where most of the loss returned pertains to signals without targets, ultimately causing slow convergence or even failure to converge during the training process. One solution to this problem is to adjust the loss function, as seen in the Dice loss function used in image segmentation \cite{milletari2016v}. The Dice loss function is defined as:
\begin{align}
    \mathcal{J}_{1,2}^{(m)} 
    &= 1 - \frac{2 \sum_{n \in \mathcal{I}_{m}} y_{n} \hat{y}_{n}}{\sum_{n \in \mathcal{I}_{m}} y_{n}^{2} + \sum_{n \in \mathcal{I}_{m}} \hat{y}_{n}^{2}}\\
	&= 1- \dfrac{2\sum_{n\in \mathcal{I}_{m}^{+}} \hat{y}_{n}}{N^{+} + \sum_{n\in \mathcal{I}_{m}^{-}} \hat{y}_{n}^{2}+ \sum_{n\in \mathcal{I}_{m}^{+}} \hat{y}_{n}^{2}} ,
\end{align}
where \( N^{+} \) is the number of samples in the set \( \mathcal{I}_{m}^{+} \).

\subsubsection{Mutually Exclusive Multi-target Detection: Multi-class Problems}

In many applications, it is not only necessary to know whether a target is present but also to identify which specific target or category it belongs to. For example, in the task of intelligent meeting summarization, it is crucial to determine which speaker is speaking during a segment of audio. Similarly, in image classification tasks, there are often many categories that cannot be reduced to a simple binary classification problem. In practice, when given a sample, multiple targets may need to be detected simultaneously. We will discuss a simplified case where each sample contains only one target, and multiple targets are mutually exclusive. In this scenario, the sample label \( y_{n} \) is no longer a scalar but rather a vector \( \bm{y}_{n} \), with the length corresponding to the number of target categories to be detected, denoted as \( L \). Accordingly, the network output \( \hat{\bm{y}}_{n} = f(\bm{x}_{n}) = S_{x} \circ g(\bm{x}_{n}) \) is also a vector of length \( L \).

For the loss function of multi-target detection, we can simplify the loss function of single-target detection. In a single-target detection scenario, if we define the absence of a target as a ``special target," it can be viewed as a binary detection problem with the first target's probability represented by \( y_{n} \) and the second target's probability as \( (1 - y_{n}) \). By defining \( \bm{y}_{n} = [y_{n}~(1 - y_{n})]^{T} \) and \( \hat{\bm{y}}_{n} = [\hat{y}_{n}~(1 - \hat{y}_{n})]^{T} \), the binary cross-entropy loss function can be reformulated as:
\begin{align}
    \mathcal{J}_{1,3}(\bm{y}_{n}, \hat{\bm{y}}_{n}) = -\bm{y}_{n}^{T} \ln \hat{\bm{y}}_{n}. \label{loss-cross-entropy}
\end{align}
According to the definition, the label for target detection is either \( \bm{y}_{n} = [1~0]^{T} \) (indicating the target is present) or \( \bm{y}_{n} = [0~1]^{T} \) (indicating the target is absent).

For the loss function of multi-target detection, equation (\ref{loss-cross-entropy}) remains applicable. However, at this point, both \( \bm{y}_{n} \) and \( \hat{\bm{y}}_{n} \) are vectors of length \( L \). The \( \ell \)-th element of the vector \( \hat{\bm{y}}_{n} \) represents the probability of the occurrence of the \( \ell \)-th target. The label vector \( \bm{y}_{n} \) is highly specific: it only has one element equal to 1, while all other elements are zero. When the sample corresponds to the \( \ell \)-th target, the \( \ell \)-th element of \( \bm{y}_{n} \) equals 1, e.g., \( \bm{y}_{n} = [1~0~0~\ldots~0]^{T} \) indicates that the sample corresponds to the first target. For a given batch of data, the multi-target cross-entropy loss function can be expressed as:
\begin{align}
    \mathcal{J}_{1,3}^{(m)} &= -\sum_{n \in \mathcal{I}_{m}} \bm{y}_{n}^{T} \ln \hat{\bm{y}}_{n}\\
    & = -\sum_{\ell=1}^{L} \sum_{n \in \mathcal{I}_{m}^{(\ell )}} \ln \left[\hat{\bm{y}}_{n}\right]_{\ell} , \label{J-14-batch}
\end{align}
where \( \ell \) is the index of the category, and \( [\cdot]_{\ell} \) denotes the \( \ell \)-th element of the vector. The \( \mathcal{I}_{m}^{(\ell )} \) is the index set of samples in the $m$-th batch corresponding to the $\ell$th category. The loss function given in (\ref{J-14-batch}) is a generalized form of the classic binary cross-entropy in multi-target detection. In this case, when \( \hat{y}_{n}(\ell) \) approaches 1, the target is present; when \( \hat{y}_{n}(\ell) \) approaches 0, the target is absent.

For multi-target detection, additional challenges arise due to class imbalance, where some classes are well-represented while others are not. This can lead to poor performance in detecting certain classes. Therefore, balancing sampling strategies and other techniques are often needed to improve convergence speed and network performance.

\subsubsection{Multi-Target Detection}

The aforementioned multi-target detection task addresses a specific scenario where each sample contains only one target. In practice, many tasks involve the simultaneous presence of multiple or multi-class targets. For example, when capturing an image with a camera, various objects may appear alongside the target of interest (e.g., a ``dog"), such as flowers, trees, buildings, and other animals. Similarly, in tasks like intelligent speaker diarization, multiple speakers may talk simultaneously within a given time window. In these applications, using the softmax function as the output layer of the network is not suitable, as it assumes that each sample can have only one target. To calculate the probabilities of each target's occurrence, we employ multiple sigmoid functions to construct the overall output of the network, as defined in (\ref{sigmoid-multi-def}) with \( S_{d,\mathrm{multi}}(\cdot) \). The network output can therefore be represented as: \( \hat{\bm{y}}_{n} = f(\bm{x}_{n}) = S_{d,\mathrm{multi}} \circ g(\bm{x}_{n}) \).

Under the condition of using the Sigmoid function as the network output, binary cross-entropy is typically used to construct the loss function. For an \( L \)-class multi-object detection task, given a sample \( \bm{x}_{n} \) and its label \( \bm{y}_{n} \), the loss function can be described as:
\begin{align}
    \mathcal{J}_{1,4}(\bm{y}_{n}, \hat{\bm{y}}_{n}) = -\bm{y}_{n} \ln \hat{\bm{y}}_{n} - (1-\bm{y}_{n})^{T} \ln (1-\hat{\bm{y}}_{n}).
\end{align}
For a batch of samples, the loss function is given by:
\begin{align}
    \mathcal{J}_{1,4}^{(m)} &= -\sum_{n\in \mathcal{I}_{m}}\bm{y}_{n} \ln \hat{\bm{y}}_{n} - \sum_{n\in \mathcal{I}_{m}}(1-\bm{y}_{n}) \ln (1-\hat{\bm{y}}_{n})\label{J-14-batch} \\
    &= -\sum_{\ell} \left\{\sum_{n\in \mathcal{I}_{m}^{(\ell +)}} \ln \left[ \hat{\bm{y}}_{n}\right]_{\ell} + \sum_{n\in \mathcal{I}_{m}^{(\ell -)}} \ln \left[1- \hat{\bm{y}}_{n}\right]_{\ell} \right\}, \nonumber
\end{align}
where \( \ell \) is the index of the class, and \( [\cdot]_{\ell} \) denotes the \( \ell \)-th element of the vector. The set \( \mathcal{I}_{m}^{(\ell +)} \) contains the indices of samples in the \( m \)-th batch where the \( \ell \)-th class target appears, while \( \mathcal{I}_{m}^{(\ell -)} \) contains the indices of samples where the \( \ell \)-th class target is absent. For most practical data, the number of indices in the set \( \mathcal{I}_{m}^{(\ell +)} \) is significantly smaller than that in \( \mathcal{I}_{m}^{(\ell -)} \). More critically, there can be substantial differences in the number of elements among different class index sets \( \mathcal{I}_{m}^{(\ell +)} \). For targets with fewer samples, their occurrences in the given batch may be very limited, leading to poor detection performance.

Similarly, we can also utilize the Dice loss function to mitigate this issue:
\begin{align}
    \mathcal{J}_{1,5}^{(m)} &= 1 - \frac{2\sum_{n\in \mathcal{I}_{m}}\bm{y}_{n}^{T} \hat{\bm{y}}_{n}}{\sum_{n\in \mathcal{I}_{m}}\bm{y}_{n}^{T} \bm{y}_{n} + \sum_{n\in \mathcal{I}_{m}}\hat{\bm{y}}_{n}^{T} \hat{\bm{y}}_{n}} \\
    &= 1 - \frac{2\sum_{\ell} \sum_{n\in \mathcal{I}_{m}^{(\ell +)}}\left[\hat{\bm{y}}_{n}\right]_{\ell}}{N^{+} + \sum_{\ell} \sum_{n\in \mathcal{I}_{m}^{(\ell -)}}\left[ \hat{\bm{y}}_{n} \right]_{\ell}^{2} + \sum_{\ell} \sum_{n\in \mathcal{I}_{m}^{(\ell +)}}\left[ \hat{\bm{y}}_{n} \right]_{\ell}^{2}},\nonumber
\end{align}
where \( N^{+} \) is the total occurrence count of all target classes in the batch, and \( \left[\cdot\right]_{\ell}^{2} \) denotes the square of the \( \ell \)-th element of a vector.

In practice, in addition to modifying the loss function, it is often necessary to implement strategies such as balanced sampling to accelerate convergence and improve the performance of the algorithm.

\subsubsection{Detection Probability Super-Resolution}

Neural network training typically relies on abundant labeled data. In practical applications, the cost of data labeling is often high, and in some specific applications, precise labeling may not be achievable. For example, when observing a noisy segment of audio, we might want to label the presence of specific sound sources at each time-frequency point. For simulated data, we can label signals based on the relative power of sound sources; however, for real data, we often only know which sound sources are present without being able to label every time-frequency point. Similarly, in image segmentation tasks, it is easier to label what types of targets are in an image, but manually labeling every pixel requires significant resources and is often impractical. This situation creates a fundamental need: can we allow the algorithm to complete higher-resolution labeling of targets based solely on the types of targets present in the samples?

Taking audio signal processing as an example: for a long audio signal, we typically align and frame the signal, obtaining numerous audio frames, each corresponding to a sample in deep learning. In this case, we wish to infer the label of each frame signal from the label of the entire signal. For convenience, we define:
\begin{align}
    \bm{x}_{n} &= \left[\begin{array}{cccc}
        \bm{x}_{n,0} & \bm{x}_{n,1} & \cdots & \bm{x}_{n,T-1}
    \end{array}\right],\\
    \bm{y}_{n} &= \left[\begin{array}{cccc}
        y_{n}(0) & y_{n}(1) & \cdots & y_{n}(L-1)
    \end{array}\right]^{T},\\
    \hat{\bm{y}}_{n} &= \left[\begin{array}{cccc}
        \hat{y}_{n}(0) & \hat{y}_{n}(1) & \cdots & \hat{y}_{n}(L-1)
    \end{array}\right]^{T},\\
    \hat{\bm{y}}_{n,t} &= \left[\begin{array}{cccc}
        \hat{y}_{n,t}(0) & \hat{y}_{n,t}(1) & \cdots & \hat{y}_{n,t}(L-1)
    \end{array}\right]^{T},
\end{align}
where \( \bm{x}_{n} \) is the provided signal, \( \bm{x}_{n,t} \) is the slice of the signal at time \( t \), \( \bm{y}_{n} \) is the label for the entire signal (where the value is 1 for the relevant classes), \( \hat{\bm{y}}_{n} \) is the network's output for the entire signal, \( \hat{y}_{n}(\ell) \) indicates the probability of the \( \ell \)-th target appearing in the \( n \)-th signal, and \( \hat{\bm{y}}_{n,t} \) represents the probabilities of various targets in the \( t \)-th slice of the \( n \)-th signal. The variables \( \bm{x}_{n} \) is a matrix, while \( \bm{y}_{n} \), \( \hat{\bm{y}}_{n} \), and \( \hat{\bm{y}}_{n,t} \) are vectors of length \( L \).

The task of detection probability super-resolution becomes: given the sample/signal \( \bm{x}_{n} \) and label \( \bm{y}_{n} \), can we train the network so that \( \hat{\bm{y}}_{n} \) approximates \( \bm{y}_{n} \), while automatically capturing all \( \hat{\bm{y}}_{n,t} \) and \( \hat{y}_{n,t}(\ell), \forall t, \ell \)?

Since \( \bm{y}_{n} \) is the only label information available, we can only approximate \( \bm{y}_{n} \) through \( \hat{\bm{y}}_{n} \) to facilitate the network's learning. To capture the probabilities of various targets appearing in each slice \( \hat{\bm{y}}_{n,t} \), one method is to embed the estimates of \( \hat{\bm{y}}_{n,t} \) into the construction of \( \hat{\bm{y}}_{n} \). According to the method reported in the literature \cite{wang2019comparison}, the network output and loss function can be constructed as follows:
\begin{align}
    \hat{y}_{n}(\ell) &= \sum_{t=0}^{T-1} \frac{\hat{y}_{n,t}(\ell)}{\sum_{i=0}^{T-1} \hat{y}_{n,i}(\ell)} \hat{y}_{n,t}(\ell) \label{spp-nonuniform-J16} \\
    &= \frac{\sum_{t=0}^{T-1} \hat{y}_{n,t}^{2}(\ell)}{\sum_{t=0}^{T-1} \hat{y}_{n,t}(\ell)},
\end{align}
The loss function can be defined as:
\begin{align}
    &\mathcal{J}_{1,6}(\bm{y}_{n}, \hat{\bm{y}}_{n})\nonumber \\ &= -\bm{y}_{n} \ln \hat{\bm{y}}_{n} - (1 - \bm{y}_{n})^{T} \ln (1 - \hat{\bm{y}}_{n}) \label{J-16-cross-entrop} \\
    &= -\left\{\sum_{\ell} y_{n}(\ell) \ln \hat{y}_{n}(\ell) + \left[1 - y_{n}(\ell)\right] \ln \left[1 - y_{n}(\ell)\right]\right\} \\
    &= - \sum_{\ell \in \mathcal{I}_{n}^{+}} \ln \hat{y}_{n}(\ell) - \sum_{\ell \in \mathcal{I}_{n}^{-}} \ln \left[1 - y_{n}(\ell)\right],
\end{align}
where \( \mathcal{I}_{n}^{+} \) is the set of indices for all targets present in the \( n \)-th sample, and \( \mathcal{I}_{n}^{-} \) is the set of indices for targets not present in the \( n \)-th sample. The definitions of \( \mathcal{I}_{m}^{(\ell +)} \) and \( \mathcal{I}_{m}^{(\ell -)} \) are consistent with those in (\ref{J-14-batch}).  

To analyze the advantages of this non-uniform weighted fusion, we consider a scenario where each signal has only two slices, i.e., \( T=2 \). By simplifying the variables, we can abstract a key function:
\begin{align}
    h(y_{0}, y_{1}) &= \frac{y_{0}^{2} + y_{1}^{2}}{y_{0} + y_{1}} \\
    &= \sum_{t=0}^{1}\frac{y_{t}}{\sum_{i=0}^{1} y_{i}} y_{t}, \quad \forall y_{0}, y_{1} \in [0,1].
\end{align}
If we plot the surface of \( h(y_{0}, y_{1}) \) as \( y_{0} \) and \( y_{1} \) vary, we observe that the valley of the surface occurs at \( y_{0} = y_{1} \), while peaks appear along the axes where \( y_{0} = 0 \) and \( y_{1} = 0 \). We can verify this by calculating \( h(y_{0}, y_{1}) \) for the points \( (y_{0},y_{1})=(1,0), (0,1), \) and \( (0.5,0.5) \).
Maximizing \( h(y_{0}, y_{1}) \) leads \( y_{0} \) and \( y_{1} \) to become sparse, with larger values increasing and smaller values decreasing. Conversely, minimizing \( h(y_{0}, y_{1}) \) results in \( y_{0} \) and \( y_{1} \) tending towards equality, with both values approaching smaller magnitudes. 
When the target is present, we aim to minimize \( -\ln h(y_{0}, y_{1}) \), which is equivalent to maximizing \( h(y_{0}, y_{1}) \). This results in detection probabilities at various times converging towards 0 and 1, respectively. When the target is absent, we need to minimize \( -\ln\left[1- h(y_{0}, y_{1})\right] \), which is equivalent to minimizing \( h(y_{0}, y_{1}) \). The outcome is that detection probabilities approach equality, with all values converging towards 0.
This phenomenon also holds true when generalized to multi-object detection. After the neural network is trained, the intermediate inference results \( \hat{y}_{n,t}(\ell) \) can be interpreted as the probability of the \( \ell \)-th class target appearing at time \( t \) for the \( n \)-th data sample.

\subsubsection{Target Probability Fitting}

In the previously discussed tasks, the presence or absence of targets in the sample is definitive: when the target is present, \( y_{n} = 1 \); when absent, \( y_{n} = 0 \). In some applications, the target label is a continuous value between 0 and 1: when the signal-to-noise ratio (SNR) of the sample is high, the label \( y_{n} \) approaches 1; when the SNR is low, \( y_{n} \) approaches 0; and for other ranges, the labels are continuous values between 0 and 1, monotonically increasing functions of the sample's SNR. For instance, in the framework of time-frequency audio signal processing, multiplying each time-frequency point by a mask value between 0 and 1 allows for signal enhancement; the mask value reflects the probability of the expected sound source signal being present at the given time-frequency point.

In the corresponding signal processing, the core function of the neural network is to output the probabilities of the expected sound source signal being present at each time-frequency point when given an input signal/sample. The corresponding labels are constructed based on the ratio of the expected signal power to the observed signal power. For labels \( y_{n} \) that are continuous values between 0 and 1, the network's output layer typically uses a Sigmoid function, denoted as \( S_{d}(\cdot) \); the network's output can be expressed as \( y_{n} = f(\bm{x}_{n}) = S_{d} \circ g(\bm{x}_{n}) \). For this type of problem, both binary cross-entropy and Dice loss can serve as loss functions:
\begin{align}
    \mathcal{J}_{1,7}(y_{n}, \hat{y}_{n}) &= - y_{n} \ln \hat{y}_{n} - (1 - y_{n}) \ln (1 - \hat{y}_{n}), \\
    \mathcal{J}_{1,7}^{(m)} &= \sum_{n \in \mathcal{I}_{m}} \mathcal{J}_{1,7}(y_{n}, \hat{y}_{n}) \\
    &= -\sum_{n \in \mathcal{I}_{m}} \left[y_{n} \ln \hat{y}_{n} + (1 - y_{n}) \ln (1 - \hat{y}_{n})\right], \\
    \mathcal{J}_{1,8}^{(m)} &= 1 - \frac{2 \sum_{n \in \mathcal{I}_{m}} y_{n} \hat{y}_{n}}{\sum_{n \in \mathcal{I}_{m}} y_{n}^{2} + \sum_{n \in \mathcal{I}_{m}} \hat{y}_{n}^{2}},
\end{align}
where \( \mathcal{I}_{m} \) is the index set of the samples in the \( m \)-th batch in the entire dataset; \( \mathcal{J}_{1,7}(y_{n}, \hat{y}_{n}) \) is the binary cross-entropy loss function for a single sample, \( \mathcal{J}_{1,7}^{(m)} \) is the binary cross-entropy loss function for a batch of data, and \( \mathcal{J}_{1,8}^{(m)} \) is the Dice loss function for the batch.

\subsubsection{SVM Framework for Target Detection}

The loss functions introduced earlier output values between 0 and 1, representing the probabilities of a specific target appearing in the sample. For single-target detection and binary classification problems, Support Vector Machines (SVM) are classical detection methods \cite{cortes1995support} with well-established mathematical theories and analytical processes. Since SVM typically requires linearly separable features, the features can be extracted using a neural network \( \bm{z}_{n} = g(\bm{x}_{n}) \). Through linear transformation and sign determination, we can express the network's final output as \( \hat{y}_{n} = f(\bm{x}_{n}) = S_{s} \circ g(\bm{x}_{n}) \), where the function \( S_{s}(\cdot) \) is defined in (\ref{SVM-output}). By finding the hyperplane \( \bm{w}^{T} \bm{z} + b = 0 \) that maximizes the minimum distance from positive and negative samples, the optimization problem for SVM can be described as:
\begin{align}
    \mathcal{J}_{1,9}^{(m)} &= \sum_{n \in \mathcal{I}_{m}} \max\left[0, 1 - y_{n} \left(\bm{w}^{T} \bm{z}_{n} + b\right)\right] + \lambda \left\|\bm{w}\right\|_{2}^{2} \\
    &= \sum_{n \in \mathcal{I}_{m}} \mathrm{ReLU}\left[ 1 - y_{n} \left(\bm{w}^{T} \bm{z}_{n} + b\right)\right] + \lambda \left\|\bm{w}\right\|_{2}^{2} \\
    &= \sum_{n \in \mathcal{I}_{m}} \mathrm{ReLU}\left\{ 1 - y_{n} \left[\bm{w}^{T} g(\bm{x}_{n}) + b\right]\right\} + \lambda \left\|\bm{w}\right\|_{2}^{2},\nonumber
\end{align}
where \( \lambda > 0 \) controls the capacity to tolerate erroneous labels. A larger \( \lambda \) increases tolerance, while a smaller \( \lambda \) decreases tolerance. The loss function's construction unit \( \max[0, 1 - h(\bm{x}, y)] \) is commonly referred to as the hinge loss function \cite{chapelle2007training, wu2007robust}, which has shown good performance in many neural network learning tasks. In practical terms, the overall goal is to maximize \( h(\bm{x}, y) \); when \( h(\bm{x}, y) \) outputs a large value for certain samples, we consider that the network is not gaining informative guidance from these samples and thus discard their contributions to the loss function. For samples that are still difficult to handle, a maximization of \( \max[0, 1 - h(\bm{x}, y)] \) can gradually increase the value of \( h(\bm{x}, y) \). For the loss function characterized by SVM, the \( h(\bm{x}, y) \) can be expressed as \( h(\bm{x}, y) = y_{n} \left[\bm{w}^{T} g(\bm{x}_{n}) + b\right] \). In other words, for positive samples \( y_{n} = 1 \), we wish to maximize \( \bm{w}^{T} g(\bm{x}_{n}) + b \); for negative samples \( y_{n} = -1 \), we wish to minimize \( \bm{w}^{T} g(\bm{x}_{n}) + b \). When the value of \( |\bm{w}^{T} g(\bm{x}_{n}) + b| \) is too large or too small, we aim to reduce the influence of those samples during training. The hinge loss has many applications and will be involved in other tasks later.

\subsubsection{Important Issues in Signal Detection}

Overall, signal detection is relatively straightforward compared to other tasks in neural network training. Although the training methods for such networks are well-established, certain issues still require significant research efforts, such as the design and optimization of network structures for specific data, construction of loss functions, class imbalance problems, erroneous labels, ambiguous labels, and weak label issues.
 
\subsection{Loss Functions in Signal Estimation/Filtering}

Signal estimation tasks can typically be divided into three categories: source estimation, channel estimation, and parameter estimation. The goal of source estimation is to estimate the desired source signal from noisy samples, which is often needed in tasks such as signal enhancement and denoising. Channel estimation aims to estimate the channel between the signal source and the observation point from the samples, allowing for the extraction of spatial information about the source and some characteristics of the signal propagation environment. Parameter estimation is broad, often including statistical information about source signals, parameters of source signal generation models, noise parameters, interference parameters, channel parameters, and parameters with semantic information in the propagation environment. Unlike signal detection tasks, the labels in signal estimation tasks can be relatively complex, with a wide dynamic range, and may even be complex signals.

Given the sample/observation \( \bm{x} \) and the label \( \bm{y} \), the sample in signal/parameter estimation can be modeled as:
\begin{align}
    \bm{x} = \hbar(\bm{y}) + \bm{v},
\end{align}
where \( \hbar(\cdot) \) is a function representing the part of the observed signal determined by the unknown signal/parameter, and \( \bm{v} \) represents noise and interference within the observed signal. The task of signal estimation is to estimate \( \bm{y} \) given the observation \( \bm{x} \) while both \( \hbar(\cdot) \) and \( \bm{v} \) are unknown. For convenience, we describe the network that infers \( \bm{y} \) based on \( \bm{x} \) as:
\begin{align}
    \hat{y} = f(\bm{x}),
\end{align}
where \( f(\cdot) \) is the network to be learned, and \( \theta_{f} \) are the parameters of the network. The goal of training \( f(\cdot) \) is to make \( \hat{\bm{y}} \) approximate \( \bm{y} \) according to some metric. Since the output layer in signal estimation is often a linear layer, i.e., \( \bm{w}^{T} \bm{z} + b \), we will not separate the overall network \( f(\cdot) \) into the feature extraction network \( g(\cdot) \) and the output layer \( S(\cdot) \).

The values of \( \bm{y} \) in signal estimation are typically diverse and infinite, making it challenging to manually label real measured samples. This is one of the key challenges that signal estimation faces compared to signal detection. In such cases, to obtain sufficient labeled data, signal simulation methods become crucial. In practice, to train the network, we typically need to construct the function \( \hbar(\cdot) \) to measure pure source signals \( \bm{y} \) for building the source signal dataset, and measure noise signals \( \bm{v} \) for constructing the noise signal dataset. Then, based on the characteristics of the signal model, we generate observed signals that closely resemble real observations, constructing the training set \( \{\bm{x}_{n}, \bm{y}_{n}, \forall n\} \). The quality of the model largely determines whether the final trained neural network can generalize to real data: under accurate modeling conditions, networks trained on simulated data can maintain consistent performance on real data; however, if the model deviates significantly from real data, the network may perform well on simulated data but poorly on real data. There has been extensive research on signal simulation for different applications, which exceeds the scope of this paper. We will briefly discuss methods for constructing loss functions and important issues in signal estimation tasks.

\subsubsection{Minimum Mean Square Error Loss}

Minimum mean square error (MSE) is often abbreviated as MSE. Given the observed signal/sample \( \bm{x} \) and the label \( \bm{y} \), we aim to obtain the estimated label \( \hat{\bm{y}} \) through the neural network \( f(\bm{x}) \). To achieve network training, it is necessary to define the distance between the true label \( \bm{y} \) and the network output \( \hat{\bm{y}} \) and construct a loss function based on this distance. Under the minimum mean square error criterion, the loss function is defined as:
\begin{align}
    \mathcal{J}_{2,1}(\bm{y}_{n}, \hat{\bm{y}}_{n}) 
    &= \left\|\bm{y}_{n} - \hat{\bm{y}}_{n}\right\|_{2}^{2} \\
    &= \left\|\bm{y}_{n} - f(\bm{x}_{n})\right\|_{2}^{2} = \left\|\bm{\epsilon}_{n}\right\|_{2}^{2},
\end{align}
where \( \|\cdot\|_{2} \) denotes the Euclidean norm of vectors/matrices, and \( \bm{\epsilon}_{n} = \bm{y}_{n} - \hat{\bm{y}}_{n} \) is the prediction error of the network. For a batch of data, the network's cost function can be expressed as:
\begin{align}
    \mathcal{J}_{2,1}^{(m)} 
    &= \sum_{n \in \mathcal{I}_{m}} \mathcal{J}_{2,1}(\bm{y}_{n}, \hat{\bm{y}}_{n})\\
    & = \sum_{n \in \mathcal{I}_{m}} \left\|\bm{y}_{n} - \hat{\bm{y}}_{n}\right\|_{2}^{2} = \sum_{n \in \mathcal{I}_{m}} \left\|\bm{\epsilon}_{n}\right\|_{2}^{2},
\end{align}
where \( \mathcal{I}_{m}\) is the index set of the samples corresponding to the \( m \)-th batch. In theory, if the network's capacity is sufficient and there is no ``one-to-many" situation for the observed signal and the expected signal (i.e., the same observed signal corresponds to the same expected signal), we can train the network such that \( \hat{\bm{y}}_{n} = f(\bm{x}_{n}) = \bm{y}_{n}, \forall n \); in other words, all samples in the training set can achieve accurate fitting. However, in neural network training, achieving perfect fitting often implies limited generalization ability, resulting in poor performance on the test set, and the training process should avoid overfitting. Additionally, the model's capacity is often limited, making it challenging to achieve perfect fitting for a given dataset. Therefore, in general, the error \( \bm{\epsilon}_{n} \) is not zero, and different training strategies lead to \( \bm{\epsilon}_{n} \) following specific distributions. Under the minimum mean square error criterion, \( \bm{\epsilon}_{n} \) often follows a zero-mean Gaussian distribution. When \( \bm{y}_{n} \) is a single value, the density function of the prediction error can be expressed as:
\begin{align}
    p(\bm{\epsilon}) = \frac{1}{\sigma \sqrt{2\pi}} \expd{-\frac{\epsilon^{2}}{2\sigma^{2}}},
\end{align}
where \( \sigma^{2} \) is the variance of the prediction error, and the prediction errors of samples in the training set can be considered as samples from this Gaussian distribution. A characteristic of the Gaussian distribution is that most samples cluster around the mean, and very few samples deviate significantly from it. In other words, when we want the training errors of the neural network to approach a Gaussian distribution, the model will strive to avoid large errors. When outliers appear in the samples, fitting these outliers will increase the overall error of most samples in the training set. When the dynamic range of the variable \( \bm{y} \) is large, assuming the error follows a Gaussian distribution implies that both large and small values of \( y_{n} \) must be fitted with the same precision. However, in practice, achieving good fitting for these two types of samples differs significantly in difficulty. When the network attempts to closely approximate the smaller values, it may lead to overfitting. Thus, the model's approximation may become inadequate when overfitting prevention mechanisms interrupt the training early. These various factors make minimum mean square error less effective in achieving good training results for certain specific applications. Past researchers have explored many alternatives, and we will briefly discuss several variant loss functions.

\subsubsection{$\ell_{1}$-Norm Loss}

In comparison to the mean square error loss function, the \( \ell_{1} \)-norm loss has a very similar form. For a given sample \( \bm{x}_{n} \), label \( \bm{y}_{n} \), and network prediction \( \hat{\bm{y}}_{n} = f(\bm{x}_{n}) \), the \( \ell_{1} \)-norm loss can be expressed as:
\begin{align}
    \mathcal{J}_{2,2}(\bm{y}_{n}, \hat{\bm{y}}_{n}) 
    &= \left\|\bm{y}_{n} - \hat{\bm{y}}_{n}\right\|_{1}\\ 
    &= \left\|\bm{y}_{n} - f(\bm{x}_{n})\right\|_{1} = \left\|\bm{\epsilon}_{n}\right\|_{1},
\end{align}
where \( \|\cdot\|_{1} \) denotes the \( \ell_{1} \)-norm of vectors/matrices. For a given vector \( \bm{y} = [y_{0}~y_{1}~\cdots~y_{L-1}]^{T} \), its \( \ell_{1} \)-norm is calculated as \( \left\|\bm{y}\right\|_{1} = \sum_{\ell} |y_{\ell}| \).

By minimizing the \( \ell_{1} \)-norm loss, the prediction error \( \bm{\epsilon}_{n} \) tends to follow a Laplace distribution. Compared to the zero-mean Gaussian distribution, the zero-mean Laplace distribution allows for larger errors. Under the same variance, the Laplace distribution has a higher peak at the mean than the Gaussian distribution. Therefore, optimizing the \( \ell_{1} \)-norm loss ultimately leads to a scenario where it discards some outliers and does not overly approximate their labels. Simultaneously, it aims to achieve accurate approximation for as many samples as possible. For this reason, the \( \ell_{1} \)-norm loss often demonstrates greater robustness in practical applications.

\subsubsection{Huber Loss}

Huber loss is a combination of MSE and \( \ell_{1} \)-norm loss \cite{huber1992robust, buchner2006robust}, and it has found excellent applications in echo cancellation algorithms. In applications involving Huber loss, the label's dimension is usually 1; if the label's dimension exceeds 1, the Huber loss can be computed for each dimension separately. Given a sample \( \bm{x}_{n} \), label \( y_{n} \), and network prediction \( \hat{y}_{n} = f(\bm{x}_{n}) \), the Huber loss between the label and the network prediction is defined as:
\begin{align}
    \mathcal{J}_{2,3}(\bm{y}_{n}, \hat{y}_{n}) &= \left\{\begin{array}{ll}
        \frac{1}{2} (y_{n} - \hat{y}_{n})^{2}, & \left|y_{n} - \hat{y}_{n}\right| \leq \delta;\\
        \delta \left|y_{n} - \hat{y}_{n}\right| - \frac{1}{2}\delta^{2}, & \text{otherwise}.
    \end{array}\right. \\
    &= \left\{\begin{array}{ll}
        \frac{1}{2} [y_{n} - f(\bm{x}_{n})]^{2}, & \left|y_{n} - f(\bm{x}_{n})\right| \leq \delta;\\
        \delta \left|y_{n} - f(\bm{x}_{n})\right| - \frac{1}{2}\delta^{2}, & \text{otherwise}.
    \end{array}\right.\nonumber
\end{align}
Here, \( \delta > 0 \) is a preset parameter. Clearly, when the distance between the prediction and the true label is below a certain threshold (i.e., \( \delta \)), we use the square of the error as the loss function to guide network training. When the distance exceeds this threshold, we use the absolute value of the error as the loss function to guide training. Compared to the squared error, the absolute error changes more slowly with increasing error, making Huber loss more robust against outliers than MSE and easier to converge in practice. From an optimization perspective, incorporating the absolute error as the loss function allows the training process to tolerate larger errors for certain samples; simultaneously, for samples with small errors, the training goal is to make their labels approximate as closely as possible. Therefore, for datasets with a large dynamic range of labels, Huber loss is expected to achieve good convergence performance and faster convergence speed.

\subsubsection{Feature Constraint Loss}

In most applications related to signal estimation, the label \( \bm{y}_{n} \) is the desired signal in \( \bm{x}_{n} \) and the learning objective of the network \( f(\bm{x}_{n}) \). Typically, the dimensions of \( \bm{y}_{n} \) and \( \bm{x}_{n} \) are the same, and the loss function can be easily characterized using norm distances. However, in some applications, the network's capacity is limited, and we cannot directly minimize the norm distance between \( \bm{y}_{n} \) and \( \hat{\bm{y}}_{n} = f(\bm{x}_{n}) \). In such cases, one alternative approach is to bring the layer features of \( \bm{y}_{n} \) and \( \hat{\bm{y}}_{n} \) as close as possible. Given the feature extractor:
\begin{align}
    g(\bm{y}_{n}) = w_{K} \circ w_{K-1} \circ \cdots \circ w_{1}(\bm{y}_{n}),
\end{align}
where \( w_{k}(\cdot) \) is the feature mapping function, typically composed of multiple layers of neural networks. For convenience, we define the output features of layer \( w_{k} \) as \( \bm{z}_{n}^{(k)} \); based on this definition, \( \bm{z}_{n}^{(K)} = g(\bm{y}_{n}) \). Similarly, when inputting \( \hat{\bm{y}}_{n} \) into the feature extractor, we can obtain the features at each layer as \( \hat{\bm{z}}_{n}^{(K)} \), \( \hat{\bm{z}}_{n}^{(K-1)} \), \ldots, \( \hat{\bm{z}}_{n}^{(1)} \); here, \( \hat{\bm{z}}_{n}^{(K)} = g(\hat{\bm{y}}_{n}) = g \circ f(\bm{x}_{n}) \). Under this series of constraints, the feature constraint loss function can be defined as:
\begin{align}
    \mathcal{J}_{2.4}(\bm{y}_{n}, \hat{\bm{y}}_{n}) = \sum_{k=1}^{K} \left\|\bm{z}_{n}^{(k)} - \hat{\bm{z}}_{n}^{(k)}\right\|_{2}^{2}, \label{feature-loss}
\end{align}
where \( \bm{z}_{n}^{(k)} \) is the \( k \)-th layer feature of the label \( \bm{y}_{n} \), and \( \hat{\bm{z}}_{n}^{(k)} \) is the \( k \)-th layer feature of the network output \( \hat{\bm{y}}_{n} \). The gradient of \( \mathcal{J}_{2.4}(\bm{y}_{n}, \hat{\bm{y}}_{n}) \) with respect to \( \hat{\bm{y}}_{n} \) indicates that the gradients propagated during backpropagation are jointly determined by the differences between the features at each layer.

In practice, we can also introduce weighting coefficients for each summation term in (\ref{feature-loss}) to balance the distances between features at different layers. The feature constraint loss function can also be used as a penalty term to ensure that the output signal \( f(\cdot) \) exhibits consistency in a certain feature space when estimating signals.

\subsubsection{Important Issues in Signal Estimation}

In signal estimation tasks, the robustness of the network is a very important issue. Similar to the robustness issues of filters in traditional linear filtering frameworks, pursuing extreme denoising performance often leads to instability in the filters: for certain samples, they may effectively recover the expected signal; for difficult samples, they may produce outputs that do not match the expected signal. When designing cost functions in practice, three main aspects should be considered:
\begin{itemize}
    \item Can the network output parameters of the optimal filter instead of directly outputting the signal? This is particularly important when the dynamic range of the signal is large; training a neural network to approximate functions across a wide dynamic range is often quite challenging.
    \item When constructing the loss function, can the network automatically exclude the contributions of difficult samples (or erroneous label samples) from the loss function during distance calculations? The modeling capability of neural networks is often limited; if the network is required to approximate all samples, difficult samples can lead to problems such as overfitting, non-convergence, or poor approximation for normal samples.
    \item Is it possible to introduce prior information to add penalty terms to the loss function? Prior information can effectively constrain the search space of the network function; generally, the more penalty terms there are, the smaller the search space, and the better the robustness of the trained network.
\end{itemize}

\subsection{Loss Functions in Signal Transformation}

Given a one-dimensional vector \( \bm{x} \), applying a matrix \( \bm{A} \) to this vector, i.e., \( \bm{A}\bm{x} \), completes a linear transformation of the vector. The properties of matrix \( \bm{A} \) determine the characteristics of the transformed vector. For more complex transformation methods, the essence is to transform samples in one space \( \{\bm{x}_{n}\} \) to another space \( \{\bm{y}_{n}\} \), transitioning from one domain to another. The distribution of \( \bm{y}_{n} \) adheres to certain characteristics. Typically, two types of spatial properties are described:
\begin{itemize}
    \item \emph{Clustering Properties}: Specifically, we hope that after transformation, samples of the same class in the space are as close as possible (i.e., \( \bm{y}_{n} \) is close for the same class), while samples of different classes are as far apart as possible, i.e., minimizing intra-class distances and maximizing inter-class distances. It is generally believed that after such a transformation, the samples have better representation.

    \item \emph{Density Properties}: The density function can be characterized by a function or directly by the dataset. Given the dataset in the target domain \( \{\ddot{\bm{y}}_{n}\} \), the purpose of the signal transformation is to make the density function described by \( \{\hat{\bm{y}}_{n} = f(\bm{x}_{n})\} \) approximate the density function described by \( \{\ddot{\bm{y}}_{n}\} \) as closely as possible. It is essential to note that at this point, \( \ddot{\bm{y}}_{n} \) and \( \bm{x}_{n} \) often do not have a one-to-one correspondence; they each characterize a density function.
\end{itemize}

\subsubsection{Classical Linear Discriminant Analysis}

Linear Discriminant Analysis (LDA) is a classical signal transformation method that maximizes the distance between class means while minimizing the variance among samples within the same class \cite{fisher1936use}. Given a sample set \( \{\bm{x}_{n}\} \), assuming there are \( K \) classes, and the index set corresponding to the \( k \)-th class is \( \mathcal{I}_{k} \), with the number of samples \( N_{k} = |\mathcal{I}_{k}| \); for convenience, we define the mean as \( \bm{\mu}_{k} = \frac{1}{N_{k}} \sum_{n \in \mathcal{I}_{k}} \bm{x}_{n} \). Assuming there exists an optimal transformation matrix \( \bm{A} \), the transformed samples and means can be expressed as:
\begin{align}
    \bm{y}_{n} &= \bm{A} \bm{x}_{n},\\
    \bm{\mu}_{k}^{\prime} &= \bm{A} \bm{\mu}_{k}.
\end{align}

After linear transformation, the objective function that characterizes the ``intra-class distance" can be expressed as:
\begin{align}
    \mathcal{J}_{3,1} &= \sum_{k=1}^{K} \frac{1}{N_{k}} \sum_{n \in \mathcal{I}_{k}} \left\|\bm{y}_{n} - \bm{\mu}_{k}^{\prime}\right\|_{2}^{2} \\
    &= \sum_{k=1}^{K} \frac{1}{N_{k}} \sum_{n \in \mathcal{I}_{k}} \mathrm{tr}\left[\left(\bm{y}_{n} - \bm{\mu}_{k}^{\prime}\right)\left(\bm{y}_{n} - \bm{\mu}_{k}^{\prime}\right)^{H}\right] \\
    &= \mathrm{tr}\left(\bm{A} \bm{\Phi}_{w} \bm{A}^{H}\right),
\end{align}
where
\begin{align}
    \bm{\Phi}_{w} = \sum_{k=1}^{K} \frac{1}{N_{k}} \sum_{n \in \mathcal{I}_{k}} \left(\bm{x}_{n} - \bm{\mu}_{k}\right)\left(\bm{x}_{n} - \bm{\mu}_{k}\right)^{H}
\end{align}
is the sum of the covariance matrices of each class sample. Similarly, the objective function that characterizes the distance between class means can be expressed as:
\begin{align}
    \mathcal{J}_{3,2} &= \sum_{k=1}^{K} \sum_{\ell=1}^{K} \left\|\bm{\mu}_{k}^{\prime} - \bm{\mu}_{\ell}^{\prime}\right\|_{2}^{2} \\
    &= \sum_{k=1}^{K} \sum_{\ell=1}^{K} \mathrm{tr}\left[ \left(\bm{\mu}_{k}^{\prime} - \bm{\mu}_{\ell}^{\prime}\right)\left(\bm{\mu}_{k}^{\prime} - \bm{\mu}_{\ell}^{\prime}\right)^{H}\right] \\
    &= \mathrm{tr}\left(\bm{A} \bm{\Phi}_{b} \bm{A}^{H}\right),
\end{align}
where
\begin{align}
    \bm{\Phi}_{b} = \sum_{k=1}^{K} \sum_{\ell=1}^{K} \left(\bm{\mu}_{k} - \bm{\mu}_{\ell}\right)\left(\bm{\mu}_{k} - \bm{\mu}_{\ell}\right)^{H}.
\end{align}
Under the criterion of ``maximizing inter-class sample mean distances and minimizing intra-class variance distances," the cost function for training the optimal transformation matrix \( \bm{A} \) can be expressed as:
\begin{align}
    \max_{\bm{A}} \frac{\mathcal{J}_{3,1}}{\mathcal{J}_{3,2}} = \frac{\mathrm{tr}\left(\bm{A} \bm{\Phi}_{b} \bm{A}^{H}\right)}{\mathrm{tr}\left(\bm{A} \bm{\Phi}_{w} \bm{A}^{H}\right)}.
\end{align}
The solution to this optimization problem corresponds to the eigenvectors of the matrix \( \bm{\Phi}_{w}^{-1} \bm{\Phi}_{b} \) associated with the largest eigenvalues. If the number of rows in matrix \( \bm{A} \) is \( M \), all \( M \) rows correspond to the eigenvectors of \( \bm{\Phi}_{w}^{-1} \bm{\Phi}_{b} \) corresponding to the top \( M \) largest eigenvalues.

In practical applications, given a sample set and class label information, we can first compute the class means, then calculate \( \bm{\Phi}_{w} \) and \( \bm{\Phi}_{b} \), and finally perform eigenvalue decomposition on the matrix \( \bm{\Phi}_{w}^{-1} \bm{\Phi}_{b} \) to obtain the transformation matrix \( \bm{A} \). Applying the matrix \( \bm{A} \) to the samples achieves the desired signal transformation, resulting in samples that are more compact within classes and more distant between classes.

\subsubsection{Cluster-Principle-Inspired  Loss}
\label{subsect-triple-cluster}

For deep learning, it is often possible to avoid direct solutions to matrix inversions and complex optimization problems through equivalent methods. For objectives such as minimizing intra-class distances and maximizing inter-class mean distances, the current sample's objective function can be constructed by sampling from the dataset. This section introduces two typical objective functions: contrastive loss and triplet margin loss \cite{bertinetto2016fully, schroff2015facenet}.

In constructing the contrastive loss objective function, given any sample \( \bm{x}_{n} \) and network output \( \bm{y}_{n} = f(\bm{x}_{n}) \), we also need to sample another sample \( \bm{x}_{i} \) from the dataset. For convenience, we define an additional label \( p_{n,i} \in \{0, 1\} \), with the following assignment rule: if \( \bm{x}_{n} \) and \( \bm{x}_{i} \) are of the same class, then \( p_{n,i} = 1 \); otherwise, \( p_{n,i} = 0 \).

Given \( \bm{x}_{n} \), \( \bm{x}_{i} \), and \( p_{n,i} \), the objective function for sample \( \bm{x}_{n} \) can be described as:
\begin{align}
    \mathcal{J}_{3,3} (\bm{y}_{n}, \bm{y}_{i}) &= p_{n,i} \cdot \left\|\bm{y}_{n} - \bm{y}_{i}\right\|_{2}^{2} +\nonumber\\
    & (1 - p_{n,i}) \cdot \max\left(0, \zeta_{0}
     - \left\|\bm{y}_{n} - \bm{y}_{i}\right\|_{2}^{2} \right).
\end{align}
\begin{itemize}
    \item When \( \bm{x}_{n} \) and \( \bm{x}_{i} \) are from different classes, \( p_{n,i} = 0 \), and the objective function simplifies to:

    \begin{align}
        \mathcal{J}_{3,3} (\bm{y}_{n}, \bm{y}_{i}) = \max\left(0, \zeta_{0} - \left\|\bm{y}_{n} - \bm{y}_{i}\right\|_{2}^{2}\right).
    \end{align}

    In this case, we consider two scenarios: 1) For sample pairs where \( \left\|\bm{y}_{n} - \bm{y}_{i}\right\|_{2}^{2} > \zeta_{0} \), these pairs are already distinguishable and do not contribute to the network training; 2) For pairs where \( \left\|\bm{y}_{n} - \bm{y}_{i}\right\|_{2}^{2} < \zeta_{0} \), minimizing the objective function will push \( \left\|\bm{y}_{n} - \bm{y}_{i}\right\|_{2}^{2} \) closer to \( \zeta_{0} \), thereby increasing the distance between samples of different classes.
    
    \item When \( \bm{x}_{n} \) and \( \bm{x}_{i} \) are from the same class, \( p_{n,i} = 1 \), and the objective function simplifies to:

    \begin{align}
        \mathcal{J}_{3,3} (\bm{y}_{n}, \bm{y}_{i}) = \left\|\bm{y}_{n} - \bm{y}_{i}\right\|_{2}^{2}.
    \end{align}

    Minimizing this objective function will reduce the distance between intra-class samples.
\end{itemize}
Thus, by minimizing \( \mathcal{J}_{3,3} (\bm{y}_{n}, \bm{y}_{i}) \), we can decrease the distances between intra-class samples while maximizing the distances between inter-class samples.

In the construction of the triplet loss function, for each sample \( \bm{x}_{n} \), we also need to sample two auxiliary samples: a positive sample \( \bm{x}_{n+} \) (from the same class) and a negative sample \( \bm{x}_{n-} \) (from a different class). Similar to the construction method for contrastive loss, the objective function for triplet loss is defined as: 
\begin{align}
    &\mathcal{J}_{3,4}(\bm{y}_{n}, \bm{y}_{n+}, \bm{y}_{n-}) \nonumber \\
    &= \max\left(0, \zeta_{0} + \left\|\bm{y}_{n+} - \bm{y}_{n}\right\|_{2} - \left\|\bm{y}_{n-} - \bm{y}_{n}\right\|_{2}\right).
\end{align} 
For convenience, we temporarily define the distance of the positive sample pair \( D_{n+} = \left\|\bm{y}_{n+} - \bm{y}_{n}\right\|_{2} \) and the distance of the negative sample pair \( D_{n-} = \left\|\bm{y}_{n-} - \bm{y}_{n}\right\|_{2} \); thus, the original objective function can be expressed as:
\begin{align}
    \mathcal{J}_{3,4}(\bm{y}_{n}, \bm{y}_{n+}, \bm{y}_{n-}) = \max\left(0, \zeta_{0} + D_{n+} - D_{n-}\right).
\end{align}
Through minimizing \( \mathcal{J}_{3,4}(\bm{y}_{n}, \bm{y}_{n+}, \bm{y}_{n-}) \), we can achieve the following:
\begin{itemize}
    \item When \( D_{n-} - D_{n+} > \zeta_{0} \), i.e., the distance of the negative sample pair is significantly larger than that of the positive sample pair, \( \mathcal{J}_{3,4}(\bm{y}_{n}, \bm{y}_{n+}, \bm{y}_{n-}) = 0 \). These sample pairs can already be easily distinguished and do not participate in network optimization and learning.
    
    \item When \( 0 > D_{n-} - D_{n+} > \zeta_{0} \), minimizing \( \mathcal{J}_{3,4}(\bm{y}_{n}, \bm{y}_{n+}, \bm{y}_{n-}) \) will cause \( D_{n-} - D_{n+} \) to approach \( \zeta_{0} \), effectively increasing the margin.
    
    \item When \( D_{n-} - D_{n+} < 0 \), minimizing \( \mathcal{J}_{3,4}(\bm{y}_{n}, \bm{y}_{n+}, \bm{y}_{n-}) \) will likely reduce \( D_{n+} \) or increase \( D_{n-} \), effectively reducing the distance between intra-class samples and increasing the distance between inter-class samples.
\end{itemize}
Thus, by minimizing the triplet loss, we can also ensure that the transformed samples exhibit clustering characteristics: increasing inter-class margins while reducing distances between intra-class samples.

\subsubsection{Label-Free Auto-Clustering Transformation Loss}

In some applications, we may not know the class information between samples, and sometimes it is unnecessary to know this information at all. Under such conditions, if we want to achieve signal transformations based on clustering characteristics, a straightforward approach is to assume that each sample is a class and define the objective function based on the similarities between each batch of data. Given a dataset \( \{\bm{x}_{n}\} \) and the network framework \( f(\cdot) \); for each data \( \bm{x}_{n} \), we can calculate its value in the target domain \( \bm{y}_{n} = f(\bm{x}_{n}) \). To construct sample pairs, we can augment each data sample in various ways, such as adding noise or removing segments. For convenience, we describe the augmentation operation as:
\begin{align}
    \tilde{\bm{x}}_{n} = \hbar(\bm{x}_{n}),
\end{align}
where the function \( \hbar(\cdot) \) represents the augmentation operation. Similarly, feeding the augmented samples into the neural network yields the values of the augmented samples in the target domain, denoted as \( \tilde{\bm{y}}_{n} = f(\tilde{\bm{x}}_{n}) = f \circ \hbar(\bm{x}_{n}) \).

Under the data augmentation operation, each sample will have a positive sample and multiple negative samples as references, allowing us to design a loss function for these samples. Assuming the index set for the \( m \)-th batch of samples is \( \mathcal{I}_{m} \), the target function for the \( n \)-th sample in the batch can be defined under the framework of cross-entropy loss combined with the softmax function as:
\begin{align}
    \mathcal{J}_{3,5}(\bm{y}_{n}) &= -\ln \frac{\expd{\alpha \cdot (S_{\bm{y}_{n}, \tilde{\bm{y}}_{n}} - \zeta_{0})}}{\expd{\alpha \cdot (S_{\bm{y}_{n}, \tilde{\bm{y}}_{n}} - \zeta_{0})} + \sum_{i \neq n} \expd{\alpha \cdot S_{\bm{y}_{n}, \tilde{\bm{y}}_{i}}}} \\
    &= -\ln \frac{\expd{\alpha S_{\bm{y}_{n}, \tilde{\bm{y}}_{n}}}}{\expd{\alpha S_{\bm{y}_{n}, \tilde{\bm{y}}_{n}}} + \expd{\alpha \zeta_{0}} \sum_{i \neq n} \expd{\alpha S_{\bm{y}_{n}, \tilde{\bm{y}}_{i}}}}.
\end{align}
Here, \( \alpha > 0 \) and \( 0 \leq \zeta_{0} < 1 \) are two hyperparameters; the function \( S_{\bm{y}_{i}, \tilde{\bm{y}}_{j}} \) is defined as:
\begin{align}
    S_{\bm{y}_{i}, \tilde{\bm{y}}_{j}} = \frac{\bm{y}_{n}^{T} \tilde{\bm{y}}_{i}}{\|\bm{y}_{n}\| \cdot \|\tilde{\bm{y}}_{i}\|},
\end{align}
which computes the cosine similarity between the sample \( \bm{y}_{n} \) and the augmented sample \( \tilde{\bm{y}}_{i} \), taking values between -1 and 1.

The final loss function for the \( m \)-th batch can be expressed as:
\begin{align}
    \mathcal{J}_{3,5}^{(m)} &= \sum_{n \in \mathcal{I}_{m}} \mathcal{J}_{3,5}(\bm{y}_{n}) \\
    &= \sum_{n \in \mathcal{I}_{m}} -\ln \frac{\expd{\alpha S_{\bm{y}_{n}, \tilde{\bm{y}}_{n}}}}{\expd{\alpha S_{\bm{y}_{n}, \tilde{\bm{y}}_{n}}} + \expd{\alpha \zeta_{0}} \sum_{i \neq n} \expd{\alpha S_{\bm{y}_{n}, \tilde{\bm{y}}_{i}}}}.
\end{align}
In practice, given a batch of data samples, we first perform data augmentation according to certain rules, then calculate the loss function for each sample based on the original and augmented samples, using the augmented samples to compute the gradients for backpropagation and completing the training of the network. As the network iterates and optimizes, it can achieve signal transformations resembling clustering effects.

\subsubsection{Codec Reconstruction Transformation}

During the signal transformation process, we generally aim for the transformation to be information-preserving, meaning that the transformed target domain signal \( \bm{y}_{n} \) should allow for the recovery of the original signal \( \bm{x}_{n} \). For such requirements, it is common to use a codec neural network structure combined with a ``consistency loss" to achieve signal transformation, where the encoder in the codec network serves as the neural network for implementing the signal transformation. Given the encoder network \( \mathcal{E}(\cdot) \) and the decoder network \( \mathcal{D}(\cdot) \), the consistency loss function can be described as:
\begin{align}
    \mathcal{J}_{3,6}(\bm{x}_{n}) = \left\|\bm{x}_{n} - \mathcal{D} \circ \mathcal{E}(\bm{x}_{n})\right\|_{2}^{2},
\end{align}
where the encoder \( \mathcal{E}(\cdot) \) is the target signal transformation network \( f(\cdot) \). Once the codec network has learned, given any sample \( \bm{x}_{n} \), the signal transformation can be achieved as follows:
\begin{align}
    \bm{y}_{n} = f(\bm{x}_{n}) = \mathcal{E}(\bm{x}_{n}).
\end{align}
In the training of such networks, to ensure that the signal transformed by the encoder has a specific meaning, additional loss functions are often added to assist in training the codec network. These auxiliary loss functions ensure that the transformed signal conforms to specific prior information (e.g., possessing a particular style).

The most accessible prior information relates to the samples in the target domain, i.e., \( \{\ddot{\bm{y}}_{n}\} \), allowing the transformed \( \bm{x}_{n} \) to inherit the characteristics encoded by the dataset \( \{\ddot{\bm{y}}_{n}\} \). Since \( \ddot{\bm{y}}_{n} \) and \( \bm{x}_{n} \) often do not have a one-to-one correspondence, the objective function can only aim to make the density function of \( \bm{y}_{n} = f(\bm{x}_{n}) \) approximate the density function of \( \{\ddot{\bm{y}}_{n}\} \) as closely as possible. Regarding the density function approximation problem, the most commonly used approach is the Generative Adversarial Training (GAN) strategy, which will be detailed in the upcoming section.

\subsubsection{Density Alignment}

In some practical applications, we may not be able to finely characterize the signal features in the target domain; instead, we can only inform the network of what characteristics the samples in the target domain should ``align" with. From a statistical perspective, this method aims to make the density function of the generated data in the target domain as close as possible to the sample set in the target domain. For convenience, we denote the sample set as \( \{\bm{x}_{n}\} \), the signal transformation network as \( f(\cdot) \), and the transformed signal as \( \bm{y}_{n} = f(\bm{x}_{n}) \), while the set of samples in the target domain is denoted as \( \{\ddot{\bm{y}}_{n}\} \). With the assistance of a discriminative network \( d(\cdot) \), the optimization problem for the function \( f(\cdot) \) can be described as\cite{chen2020simplehin,he2020momentum,xia2021self}:
\begin{align}
    \min_{\theta_{f}} \max_{\theta_{d}} \mathbb{E}_{\ddot{\bm{y}}} \ln d(\ddot{\bm{y}}) + \mathbb{E}_{\bm{x}} \ln \left[1 - d \circ f(\bm{x})\right].
\end{align}
This represents the optimization problem of GAN under binary cross-entropy loss. However, this problem is often difficult to converge and is unstable during training. Current mainstream approaches involve minimizing the Wasserstein distance; under this criterion, the corresponding optimization problem for GAN can be described as \cite{arjovsky2017wasserstein, gulrajani2017improved}:
\begin{align}
    \mathcal{J}_{3,7} &= \mathbb{E}_{\ddot{\bm{y}}} d(\ddot{\bm{y}}) - \mathbb{E}_{\bm{x}} d \circ f(\bm{x}) + \lambda \cdot \mathbb{E}_{\bm{z}} \left[\left\|\nabla d(\bm{z})\right\|_{2} - 1\right]^{2},
\end{align}
where \( \lambda > 0 \) is the gradient penalty coefficient; \( \bm{z} \) is the result of mixing signals from the source and target domains, defined as:
\begin{align}
    \bm{z} = \epsilon \cdot \ddot{\bm{y}} + (1 - \epsilon) \cdot f(\bm{x}) = \epsilon \cdot \ddot{\bm{y}} + (1 - \epsilon) \cdot \bm{y},
\end{align}
where the weighting coefficient \( \epsilon > 0 \) is uniformly sampled between 0 and 1. In the above equations, it is important to note that \( \ddot{\bm{y}} \) and \( \bm{y} \) do not have a one-to-one relationship; they are not inherently connected.

In practice, given a batch of data, the cost function and optimization problem can be described as:
\begin{align}
    \min_{\theta_{f}} \max_{\theta_{d}} \sum_{n \in \mathcal{I}_{m}} d(\ddot{\bm{y}}_{n}) - d \circ f(\bm{x}_{n}) + \lambda \cdot \left[\left\|\nabla d(\bm{z}_{n})\right\|_{2} - 1\right]^{2},
\end{align}
where \( \bm{z}_{n} = \epsilon_{n} \cdot \ddot{\bm{y}}_{n} + (1 - \epsilon_{n}) \cdot f(\bm{x}_{n}) \), and \( \mathcal{I}_{m} \) is the index set corresponding to the \( m \)-th batch of samples.

\subsubsection{Challenges in Signal Transformation and Feature Learning}

For tasks related to signal transformation and feature learning, there is often no one-to-one label correspondence, necessitating the automatic construction of objective functions for the network from the data. Thus, the challenges in these methods can be summarized in three main aspects:
\begin{itemize}
    \item The task objective itself is characterized by data; how to construct data that accurately represents the task objective is the first issue encountered in practice.
    
    \item The generalization problem: a network trained on one dataset may not yield effective features applicable to another dataset.
    
    \item The class imbalance problem: due to the lack of labels, various types of data often exhibit extreme imbalance, leading to poor representation of features for underrepresented classes. 
\end{itemize}
 
\section{Approaches to Source Localization}
 
Source localization has numerous applications, allowing the inference of the location or direction of a sound source based on observations from multiple microphone sensors. In applications related to microphone array signal processing, source localization often refers to the estimation of the direction of arrival (DOA) rather than true source localization. Source localization itself is a signal detection problem, focusing on detecting whether a source appears in a given direction. 
This section includes the following perspectives. \emph{1) Signal model: How do observations relate to the sample \( \bm{x}_{n} \), and how is the source angle encoded in the sample?
  2) Network training goals: How is the loss function constructed?
  3) What are the typical networks used}?

\subsection{Signal Model for Source Localization}

In previous sections, \( \bm{x} \) represents the input to the neural network, usually processed observation signals, while \( \bm{y} \) denotes the network output. This section aims to elucidate the relationship between \( \bm{x} \) and the source angle, establishing a basic signal model for source localization.

For convenience, let the signal captured by the \( m \)-th microphone sensor vary with time be denoted as \( p_{m}(t) \). Ignoring background noise, it can be expressed as:
\begin{align}
    p_{m}(t) &= h_{m}(t) \ast s(t) \\
    &= \sum_{i=0}^{L_{h}-1} h_{m}(i) s(t-i),
\end{align}
where \( h_{m}(t) \) is the impulse response from the source to the \( m \)-th microphone, and \( L_{h} \) is the length of the impulse response. In the impulse response sequence, there exists a direct sound path that describes the transmission of sound waves from the source to the microphone. Considering the direct sound path, the signal model can be simplified to:
\begin{align}
    p_{m}(t) &= h_{m}(\tau_{m}) s(t - \tau_{m}) + \sum_{i=0, i \neq \tau_{m}}^{L_{h}-1} h_{m}(i) s(t - i) \\
    &= h_{0} s(t - \tau_{m}) + \sum_{i=0, i \neq \tau_{m}}^{L_{h}-1} h_{m}(i) s(t - i),
\end{align}
where \( \tau_{m} \) is the propagation time of the direct sound path, and \( h_{m}(\tau_{m}) \) is its attenuation. The second equation assumes uniform attenuation for the direct sound path reaching all microphones, resulting in \( h_{1}(\tau_{m}) = h_{2}(\tau_{m}) = \cdots = h_{M}(\tau_{m}) = h_{0} \). The spatial angle information of the source is encoded in the arrival times \( \tau_{m} \), specifically in their differences.

Taking the propagation times of the first channel and the \( m \)-th channel as an example, if the source comes from the direction \( \theta \) \footnote{The illustration of the angle $\theta$ can be found in \cite{pan2013performance}. }, the direct sound signal for the \( m \)-th channel will be delayed relative to the first channel. The delay time is \( \delta_{m,1} \cos \theta / c \), where \( \delta_{m,1} \) is the distance between the two microphone sensors, and \( c \) is the speed of sound. In this case, we have \( \tau_{m} - \tau_{1} = \delta_{m,1} \cos \theta / c \). In other words, if we know the time differences of the source arrival in the microphone observation signals, we can determine the angle of the source relative to the axis of this pair of microphones. Different microphones can estimate the direction of the source and form a focal point in space, allowing us to determine the source's location. This is the fundamental principle behind how microphone arrays can estimate source angle.

Transforming time slices into the frequency domain, utilizing the property that ``time delays correspond to phase shifts in the frequency domain," we can further express:
\begin{align}
    P_{m}(\omega, t) &= \expd{-\jmath \omega \tau_{m}} h_{0} S(\omega, t) + V_{m}(\omega, t)  \\
    &= \expd{-\jmath \omega (\tau_{m} - \tau_{1} + \tau_{1})} h_{0} S(\omega, t) + V_{m}(\omega, t) \nonumber\\
    &= \expd{-\jmath \omega (\tau_{m} - \tau_{1})} \left[\expd{-\jmath \omega \tau_{1}} h_{0} S(\omega, t)\right] + V_{m}(\omega, t) \nonumber\\
    &= \expd{-\jmath \omega (\tau_{m} - \tau_{1})} S^{\prime}(\omega, t) + V_{m}(\omega, t) \nonumber\\
    &= \expd{-\jmath \omega \delta_{m,1} \cos \theta / c} S^{\prime}(\omega, t) + V_{m}(\omega, t),
\end{align}
where \( S^{\prime}(\omega, t) = \expd{-\jmath \omega \tau_{1}} h_{0} S(\omega, t) \) represents the direct sound signal observed by the first channel, and \( V_{m}(\omega, t) \) is the signal observed by the \( m \)-th channel, excluding the direct sound path.

To extend the two-channel case to multiple channels, we consider a more general scenario. First, we assume that the spatial three-dimensional coordinates of the \( m \)-th microphone are given by \( \bm{r}_{m} = [x_{m}~y_{m}~z_{m}]^{T} \). The direction of the sound source is defined as follows:
\begin{align}
    \bm{\varphi} = \left[\begin{array}{c}
    \sin\theta \cos\phi\\
    \sin\theta \sin \phi\\
    \cos \theta
    \end{array}\right],
\end{align}
where \( \theta \in [0, \pi] \) represents the elevation angle, and \( \phi \in [0, 2\pi] \) denotes the azimuth angle.
We can rewrite \( P_{m}(\omega, t) \) as:
\begin{align}
    P_{m}(\omega, t) = \expd{-\jmath \omega (\bm{r}_{1} - \bm{r}_{m})^{T} \bm{\varphi}} S^{\prime}(\omega, t) + V_{m}(\omega, t).
\end{align}
Next, we can arrange all \( M \) signals \( P_{m}(\omega, t) \) at the same time and frequency into a vector, yielding:
\begin{align}
    \bm{p}(\omega, t) &= \left[ \begin{array}{cccc}
    P_{1}(\omega, t) & P_{2}(\omega, t) & \cdots & P_{M}(\omega, t)
    \end{array} \right]^{T} \\
    &= \bm{d}(\omega, \bm{\varphi}) S^{\prime}(\omega, t) + \bm{v}(\omega, t),
\end{align}
where \( \bm{d}(\omega, \bm{\varphi}) \) is commonly referred to as the array manifold vector, with its \( m \)-th element defined as \( \expd{-\jmath \omega (\bm{r}_{1} - \bm{r}_{m})^{T} \bm{\varphi}} \);  The vector \( \bm{v}(\omega, t) \) represents noise and interference.

\subsubsection{Raw Data Features}

In the task of source localization, it is necessary to jointly analyze multiple frequencies to estimate the source's position. The network input is typically a two-dimensional matrix consisting of the number of channels and frequencies. Without any preprocessing of the observations, the network input slice \( \bm{x}(t) \) can be represented as:
\begin{align}
    \bm{x}(t) = \left[\begin{array}{cccc}
        \Re\left[\bm{p}(\omega_{1}, t)\right] & \Re\left[\bm{p}(\omega_{2}, t)\right] & \cdots & \Re\left[\bm{p}(\omega_{K}, t)\right] \\
        \Im\left[\bm{p}(\omega_{1}, t)\right] & \Im\left[\bm{p}(\omega_{2}, t)\right] & \cdots & \Im\left[\bm{p}(\omega_{K}, t)\right]
    \end{array}\right],
\end{align}
where \( \Re(\cdot) \) and \( \Im(\cdot) \) denote the extraction of the real and imaginary parts, respectively. Clearly, for each slice of the sample \( \bm{x}(t) \), it is a \( 2M \times K \) real-valued matrix.

For the task of source localization, aligning the phases of multi-channel observation signals can potentially enhance algorithm performance. Inspired by the relative transfer function vectors in sound propagation, one method of signal alignment is given by:
\begin{align}
    P_{m}(\omega, t) \leftarrow P_{m}(\omega, t) \frac{P_{\mathrm{ref}}^{\ast}(\omega, t)}{\left|P_{\mathrm{ref}}(\omega, t)\right|},
\end{align}
where the superscript \( (\cdot)^{\ast} \) denotes conjugation, and \( P_{\mathrm{ref}}(\omega, t) \) represents the observation of the reference channel. Since critical information is contained in the phase, the source localization task may sometimes exclude the signal from the aligned reference channel, changing the slice \( \bm{x}(t) \) to a dimension of \( 2(M-1) \times T \) instead of \( 2M \times T \).

\subsubsection{Spatial Spectral Features}

Given \( J \) discrete search spaces, with each space centered around the direction \( \bm{\varphi}_{j}, j = 1, 2, \ldots, J \). Assuming that a beamformer \( \bm{h}_{j}(\omega) \) is designed for each frequency band and direction \( \bm{\varphi}_{j} \) to extract the energy of signals from that direction, the spatial spectral features can be calculated using the following scanning method:
\begin{align}
    \varepsilon(\bm{\varphi}_{j}, t) = \sum_{k=1}^{K} \bm{h}_{j}^{H}(\omega_{k}) \bm{\Phi}(\omega_{k}, t) \bm{h}_{j}(\omega_{k}),
\end{align}
where
\begin{align}
    \bm{\Phi}(\omega, t) = \psi(\omega, t) \mathbb{E}[\bm{p}(\omega, t) \bm{p}^{H}(\omega, t)]
\end{align}
is proportional to the covariance matrix of the observed signals, and \( \psi(\omega, t) \) is some normalization coefficient; \( K \) is the total number of fused frequency bands.

Aside from this search method, for some high-quality time-frequency points, we can directly solve based on \( \bm{\Phi}(\omega, t) \). Suppose that the eigenvector corresponding to the largest eigenvalue of \( \bm{\Phi}(\omega, t) \) is \( \bm{u} \). Utilizing \( \bm{u} = \bm{d}(\omega, \bm{\varphi}) \), we can establish the following equation:
\begin{align}
    \left[\begin{array}{c}
        \bm{r}_{1}^{T} - \bm{r}_{1}^{T} \\
        \bm{r}_{1}^{T} - \bm{r}_{2}^{T} \\
        \vdots \\
        \bm{r}_{1}^{T} - \bm{r}_{M}^{T}
    \end{array}\right] \bm{\varphi} = \frac{c}{\omega} \angle(\bm{u}),
\end{align}
where \( \angle(\bm{u}) \) indicates the extraction of the phase; by solving this equation, we can derive the optimal \( \hat{\bm{\varphi}} \) and subsequently estimate the source direction \( (\hat{\theta}, \hat{\phi}) \). In this manner, each ``high-quality" time-frequency point can yield an estimate of a source's angle. By statistically analyzing past observations over multiple frames, we can compile the frequency of the source's occurrence in various spatial directions, thereby obtaining the spatial spectrum \( \varepsilon(\bm{\varphi}_{j}, t) \).

Whether through the search method or direct calculation, we can obtain the spatial spectra for each time slice from the observed signals. Therefore, the network input slice can be represented as:
\begin{align}
    \bm{x}(t) = \left[\begin{array}{cccc}
        \varepsilon(\bm{\varphi}_{1}, t) & \varepsilon(\bm{\varphi}_{2}, t) & \cdots & \varepsilon(\bm{\varphi}_{J}, t)
    \end{array}\right]^{T}. \label{x-input-doa-spatial}
\end{align}
This is a \( J \times 1 \) vector. Alternatively, if we do not pre-fuse the frequency \( \omega_{k} \), the slice \( \bm{x}(t) \) will be a \( J \times K \) two-dimensional matrix.

\subsubsection{Features Based on Correlation Coefficients}

Given the correlation coefficient \( \gamma_{n,m}(\tau, t) \) between the \( n \)-th and \( m \)-th channels at time \( t \), the calculation method for features based on correlation coefficients is as follows:
\begin{align}
    \varepsilon(\bm{\varphi}, t) = 2\pi \sum_{n=1}^{M} \sum_{m=1}^{M} \gamma_{n,m}[\tau_{n,m}(\bm{\varphi}, t)].
\end{align}
This can be expressed as:
\begin{align}
    \varepsilon(\bm{\varphi}, t) 
    &= \sum_{n=1}^{M} \sum_{m=1}^{M} \int_{-\infty}^{\infty} \psi_{n,m}(\omega) P_{n}(\omega, t) P_{m}^{\ast}(\omega, t) \expd{-\jmath \omega \tau_{n,m}(\bm{\varphi})} d\omega \nonumber  \\
    &= \int_{-\infty}^{\infty} \bm{d}(\omega, \bm{\varphi})^{H} \bm{\Phi}(\omega, t) \bm{d}(\omega, \bm{\varphi}) d \omega,
\end{align}
where \( \psi_{n,m}(\omega) \) is the weighting coefficient (under the principle of phase transformation, \( \psi_{n,m}(\omega) = \left|P_{n}(\omega, t) P_{m}^{\ast}(\omega, t)\right| \)); \( \bm{\Phi}(\omega, t) \) is an \( M \times M \) matrix, with its \( (n, m) \)-th element given by:
\begin{align}
    \left[\bm{\Phi}(\omega, t)\right]_{n,m} = \psi_{n,m}(\omega) P_{n}(\omega, t) P_{m}^{\ast}(\omega, t).
\end{align}
Considering that \( \bm{d}(\omega, \bm{\varphi})^{H} \bm{\Phi}(\omega, t) \bm{d}(\omega, \bm{\varphi}) \) essentially calculates the spatial spectrum at frequency \( \omega \), these features are effectively a fused spatial spectrum.

Assuming there are \( J \) discrete search spaces, with each space centered around the direction \( \bm{\varphi}_{j}, j=1,2,\ldots,J \). Under the condition that features based on correlation coefficients are used as input, the data slice \( \bm{x}(t) \) can be represented as (\ref{x-input-doa-spatial}), 
which is a \( J \times 1 \) vector.

\subsubsection{Modal Domain Features}

Transforming signals into the modal domain allows characteristics of signals from different frequency bands to have the same physical meaning and properties, theoretically enabling more efficient feature fusion and subsequent inference. Common modal domains include spherical harmonic domain and circular harmonic domain. Given \( J \) harmonic basis functions \( b_{j}(\bm{\varphi}) \) (\( j=1,2,\ldots,J \)), we can linearly transform \( \bm{p}(\omega,t) \) into:
\begin{align}
    \tilde{\bm{p}}(\omega, t) &= \bm{T}(\omega) \bm{p}(\omega, t) \\
    &= \bm{T}(\omega) \bm{d}(\omega, \bm{\varphi}) S^{\prime}(\omega, t) + \bm{T}(\omega) \bm{v}(\omega, t) \\
    &= \bm{b}(\bm{\varphi}) S^{\prime}(\omega, t) + \bm{T}(\omega) \bm{v}(\omega, t),
\end{align}
where \( \bm{\varphi} \) is the direction of the source and \( S^{\prime}(\omega, t) \) is the source signal. The matrix \( \bm{T}(\omega) \) is a transformation matrix constructed from the array structure and harmonic basis functions, with its \( j \)-th row designed to extract the \( j \)-th harmonic component of the observed signal; \( \bm{b}(\bm{\varphi}) = [b_{1}(\bm{\varphi})~b_{2}(\bm{\varphi})~\cdots~b_{J}(\bm{\varphi})]^{T} \) is a \( J \times 1 \) vector composed of harmonic components of various orders, and \( \bm{T}(\omega) \bm{v}(\omega, t) \) is the residual noise term in the harmonic decomposition.

If we ignore the influence of residual noise, it is evident that:
\begin{align}
    \tilde{\bm{p}}(\omega, t) \propto \bm{b}(\bm{\varphi}).
\end{align}
This means that modal domain features \( \tilde{\bm{p}}(\omega, t) \) across different frequency bands share the same characteristics. Merging features from \( K \) frequency bands leads to a \( J \times K \) feature matrix. Since \( b_{j}(\bm{\varphi}) \) is typically a complex signal, extracting the real and imaginary parts results in the final feature representation:
\begin{align}
    \bm{x}(t) = \left[\begin{array}{cccc}
        \Re\left[\tilde{\bm{p}}(\omega_{1}, t)\right] & \Re\left[\tilde{\bm{p}}(\omega_{2}, t)\right] & \cdots & \Re\left[\tilde{\bm{p}}(\omega_{K}, t)\right] \\
        \Im\left[\tilde{\bm{p}}(\omega_{1}, t)\right] & \Im\left[\tilde{\bm{p}}(\omega_{2}, t)\right] & \cdots & \Im\left[\tilde{\bm{p}}(\omega_{K}, t)\right]
    \end{array}\right],
\end{align}
resulting in a \( 2J \times K \) matrix. Alternatively, the magnitude and phase of the modal domain features \( \tilde{\bm{p}}(\omega, t) \) can be concatenated to form the final feature matrix.

\subsection{Loss Functions}

Assuming there exists a network \( g(\cdot) \) that transforms the original features \( \bm{x}(t) \) into higher-dimensional features \( \bm{z}(t) \), this high-dimensional feature is often referred to as an embedding, which can be understood as the high-dimensional characteristics encoded in the original features. To determine the optimal parameters \( \theta_{g} \) of the network \( g(\cdot) \), it is necessary to construct a loss function and train it using backpropagation methods based on the data. Common loss functions include norm distance, cross-entropy, and binary cross-entropy, each corresponding to different problems.

\subsubsection{Norm Distance}

Under the condition of a single sound source, the angle of the source \((\theta, \phi)\) is encoded in the vector \( \bm{\varphi} \); the goal of network training can be to directly estimate \( \theta \) and \( \phi \) or to estimate the vector \( \bm{\varphi} \). In this case, the subsequent layers are generally composed of fully connected networks, and the overall network can be represented as:
\begin{align}
    \hat{\bm{y}} &= f(\bm{x}) \\
    &= \mathcal{F} \circ g(\bm{x}).
\end{align}
For frame-by-frame prediction of the source angle, we have:
\begin{align}
    \hat{\bm{y}}(t) &= f[\bm{x}(t)] \\
    &= \mathcal{F} \circ g[\bm{x}(t)].
\end{align}
Given a batch of data \( \{(\bm{x}_{n}, \bm{y}_{n})\} \), where each data item has \( T \) slices, \( \bm{x}_{n}(0), \bm{x}_{n}(1), \ldots, \bm{x}_{n}(T-1) \), and each corresponding label is \( \bm{y}_{n}(t) = \bm{\varphi}_{n}(t), \forall t=0,1,2,\ldots,T-1 \), the network loss function can be expressed as:
\begin{align}
    \mathcal{J}_{1} &= \sum_{n} \sum_{t} \left\|\bm{y}_{n}(t) - \hat{\bm{y}}_{n}(t) \right\|_{2}^{2} \\
    &= \sum_{n} \sum_{t} \left\|\bm{\varphi}_{n}(t) - \hat{\bm{y}}_{n}(t) \right\|_{2}^{2} \\
    &= \sum_{n} \sum_{t} \left\|\bm{\varphi}_{n}(t) - f[\bm{x}_{n}(t)] \right\|_{2}^{2} \\
    &= \sum_{n} \sum_{t} \left\|\bm{\varphi}_{n}(t) - \mathcal{F} \circ g[\bm{x}_{n}(t)] \right\|_{2}^{2}.
\end{align}
By minimizing the loss function \( \mathcal{J}_{1} \), we can determine \( \theta_{\mathcal{F}} \) and \( \theta_{g} \), completing the learning of the functions \( \mathcal{F}(\cdot) \) and \( g(\cdot) \). After training the network, constructing \( f(\cdot) = \mathcal{F} \circ g(\cdot) \) allows us to achieve source localization.

If the source is known to have \( L \) categories, we can also construct a joint cost function to simultaneously address source detection and localization tasks. This joint cost function is termed Activity-coupled Cartesian Direction of Arrival (ACCDOA) \cite{shimada2021accdoa}. In the context of ACCDOA, the sample label \( \bm{y}(t) \) is a \( 3 \times L \) matrix, where the magnitude of the \( \ell \)-th column represents the probability of the \( \ell \)-th source's occurrence, and the normalized vector of this column indicates the direction of the source, specifically:
\begin{align}
    \bm{\varphi}_{n}(t) &\leftarrow \frac{1}{\left\|\left[\bm{y}(t)\right]_{:,\ell}\right\|}\left[\bm{y}(t)\right]_{:,\ell}, \\
    P[C_{\ell} | \bm{x}_{n}(t)] &\leftarrow \left\|\left[\bm{y}(t)\right]_{:,\ell}\right\|.
\end{align}
Here, \( P[C_{\ell} | \bm{x}_{n}(t)] \) represents the probability of the \( \ell \)-th class source's occurrence given \( \bm{x}_{n}(t) \). For a complete sample \( \bm{x}_{n} \), its label \( \bm{y}_{n} \) will be a \( 3 \times L \times T \) matrix, where \( T \) is the number of slices in sample \( \bm{x}_{n} \). The network loss function can also be expressed as:
\begin{align}
    \mathcal{J}_{2} &= \sum_{n} \left\|\bm{y}_{n} - f(\bm{x}_{n}) \right\|_{2}^{2} \\
    &= \sum_{n} \left\|\bm{y}_{n} - \mathcal{F} \circ g(\bm{x}_{n}) \right\|_{2}^{2} \\
    &= \sum_{n} \sum_{t} \left\|\bm{y}_{n}(t) - \mathcal{F} \circ g[\bm{x}_{n}(t)] \right\|_{2}^{2}.
\end{align}
In this case, \( \bm{y}_{n} \) encodes both the probability of a source's occurrence and the spatial position of the source. This approach allows for simultaneous training to accomplish both source localization and sound event detection tasks.

\subsubsection{Cross-Entropy for Single Source Localization}

The previous discussion framed source localization as a function fitting problem. Considering the practical scenario where three-dimensional space is divided into \( L \) finite angles, and assuming that there is only one source per moment and frequency, the source localization task can be viewed as a classification problem. In this case, the label \( \bm{y}(t) \) is a \( L \times 1 \) vector: if there is a signal in the \( \ell \)-th direction, the corresponding value at position \( \bm{y}(t) \) is 1; otherwise, it is 0. Under the condition of a single source, \( \bm{y}(t) \) is evidently a one-hot vector. For approximating one-hot vectors, the softmax function combined with cross-entropy is the most commonly used loss function construction method. The source localization network can be represented as:
\begin{align}
    \hat{\bm{y}} &= f(\bm{x}) \\
    &= \mathcal{S}_{x} \circ g(\bm{x}),
\end{align}
where \( \mathcal{S}_{x}(\cdot) \) is the softmax function. If the source angle needs to be output for each frame, we have:
\begin{align}
    \hat{\bm{y}}(t) &= f[\bm{x}(t)] \\
    &= \mathcal{S}_{x} \circ g[\bm{x}(t)].
\end{align}
Given a batch of training samples, the target loss function under cross-entropy can be characterized as:
\begin{align}
    \mathcal{J}_{3} &= -\sum_{n} \sum_{t} \bm{y}_{n}^{T} \ln \hat{\bm{y}}_{n}(t) \\
    &= -\sum_{n} \sum_{t} \bm{y}_{n}^{T} \ln f[\bm{x}_{n}(t)] \\
    &= -\sum_{n} \sum_{t} \bm{y}_{n}^{T} \ln \left\{ \mathcal{S}_{x} \circ g[\bm{x}_{n}(t)]\right\}.
\end{align}
In practice, after the network is trained, the source angle can be determined by finding the maximum value in \( \hat{\bm{y}}_{n}(t) \).

Beyond constructing loss functions for source angle, cross-entropy can also be utilized to construct loss functions for estimating the number of sources. In typical applications, the number of sources is often limited, and we can represent each source count using a network output. If the maximum number of sources in the sound field does not exceed 10, the softmax function \( \mathcal{S}_{x}(\cdot) \) will have 10 outputs. However, for the multi-source localization task, using cross-entropy to construct the loss function for source angle is no longer applicable; instead, we need to use the sigmoid function combined with binary cross-entropy to construct the loss.

\subsubsection{Binary Cross-Entropy for Multi-Source Localization}

The cross-entropy loss function is suitable for cases where labels are one-hot vectors, assuming that each slice of observation signals contains only one source signal, thereby enabling localization through classification. For multi-source localization, each slice can contain multiple sources, making this model clearly unsuitable.

The multi-source localization problem can be viewed as a detection problem: it divides three-dimensional space into \( L \) regions, each with a central direction \( \bm{\varphi}_{\ell} \); a neural network is then designed to determine whether a source appears in each region. The label \( \bm{y}(t) \) is now an \( L \times 1 \) vector, where the \( \ell \)-th element represents the probability of a source appearing in the \( \ell \)-th region at time \( t \). Since each value indicates whether a specific region contains a source, the task effectively becomes a binary classification problem. For this binary classification, using the sigmoid function as the network output is appropriate; when simultaneously detecting \( L \) regions, the network output can be expressed as \( \mathcal{S}_{d,\mathrm{multi}}(\cdot) \). The overall network can be represented as:
\begin{align}
    \hat{\bm{y}}(t) &= f[\bm{x}(t)] \\
    &= \mathcal{S}_{d,\mathrm{multi}} \circ g[\bm{x}(t)].
\end{align}
Given a batch of training data \( \{\bm{x}_{n}\} \), the corresponding loss function can be expressed as:
\begin{align}
    \mathcal{J}_{4} &= \sum_{n} - \bm{y}_{n}^{T}(t) \ln \hat{\bm{y}}_{n}(t) - \left[1 - \bm{y}_{n}^{T}(t)\right] \ln \left[1 - \hat{\bm{y}}_{n}(t)\right].
\end{align}
Since \( \bm{y}_{n}(t) \) usually contains only a few non-zero values, a substantial number of values in \( \left[1 - \bm{y}_{n}(t)\right] \) are non-zero. In this case, the contribution of the first term \( \bm{y}_{n}^{T}(t) \ln \hat{\bm{y}}_{n}(t) \) to the loss function is minimal. To balance the contributions from both parts of the loss function, it is often rewritten as:
\begin{align}
    \mathcal{J}_{4} &= \sum_{n} - \alpha \bm{y}_{n}^{T}(t) \ln \hat{\bm{y}}_{n}(t) - \left[1 - \bm{y}_{n}^{T}(t)\right] \ln \left[1 - \hat{\bm{y}}_{n}(t)\right].
\end{align}
Here, \( \alpha > 0 \) is a weighting coefficient, such as \( \alpha = 100 \).

Theoretically, the labels \( \bm{y}(t) \) can also be constructed using methods analogous to those in classification problems: when a source appears in a particular direction, the corresponding position in \( \bm{y}(t) \) is set to 1, with all other positions set to 0. However, due to the presence of noise, adjacent angles often do not directly map to the expected source direction, leading to certain errors. A better approach is to smooth the labels \( \bm{y}(t) \). One smoothing method is:
\begin{align}
    \left[\bm{y}(t)\right]_{\ell} = \max_{q=1,2,\ldots,Q} \expd{-\frac{1}{\sigma^{2}} \left\|\bm{\varphi}_{q} - \bm{\varphi}_{\ell}\right\|_{2}^{2}},
\end{align}
where \( \sigma > 0 \) controls the smoothing width, and \( Q \) is the total number of sources present at that moment, with \( \bm{\varphi}_{q} \) representing the source angle. Clearly, a larger \( \sigma \) results in more significant smoothing, while a smaller \( \sigma \) results in lighter smoothing.

\subsection{Typical Network Structures }

From the previous discussions, the network output layer can be a fully connected function \( \mathcal{F}(\cdot) \), a softmax function \( \mathcal{S}_{x}(\cdot) \), a sigmoid function \( \mathcal{S}_{d}(\cdot) \), or a multi-head sigmoid function \( \mathcal{S}_{d,\mathrm{multi}}(\cdot) \). However, we have yet to address the methods for the feature extraction network \( g(\cdot) \). This section will discuss several network structure design methods based on the input features \( \bm{x}(t) \) and target functions \( \mathcal{J}(\cdot) \) in conjunction with published literature. It is important to note that the networks discussed here are merely results reported in recent years and do not imply that they outperform other methods; in the author's opinion, there remains considerable room for improvement.

\subsubsection{Convolutional Network + Fully Connected}

For features such as spatial spectra, the signal input slice \( \bm{x}(t) \) can be a two-dimensional matrix, corresponding to frequency and spatial angles. Since spatial spectra possess structural information at local frequencies and angles, several layers of convolutional neural networks can be used to extract features from spatial spectra, followed by a fully connected neural network to achieve the transition from local features to global embeddings. Literature \cite{varanasi2020deep} provides a method for constructing functions based on spatial spectral features and convolutional networks, represented as:
\begin{align}
    g(\bm{x}) &= \mathcal{F}_{2} \circ \mathcal{F}_{1} \circ \mathrm{Flatten} \circ \mathcal{C}_{3} \circ \mathcal{C}_{2} \circ \mathcal{C}_{1}(\bm{x}),
\end{align}
where \( \mathcal{F}_{2}(\cdot) \) and \( \mathcal{F}_{1}(\cdot) \) are fully connected layers; \( \mathrm{Flatten}(\cdot) \) compresses and reorganizes the dimensions of the features; \( \mathcal{C}_{3} \), \( \mathcal{C}_{2} \), and \( \mathcal{C}_{1} \) are three convolutional layers. The convolutional kernels in \( \mathcal{C}_{i}(\cdot) \) have a width of \( 2 \times 2 \), convolving over angle and frequency dimensions. During convolution, the angle dimension remains unchanged while the frequency dimension is downsampled by half in each layer, completing feature compression. After three layers of convolution, the frequency and channel data are aligned (i.e., the flatten operation), and the resulting features are ultimately fed into the fully connected network \( \mathcal{F}_{1}(\cdot) \) to extract global embeddings.

\subsubsection{Multi-Objective Networks}

When training a network with multiple objectives, it is often beneficial to share part of the network. This section briefly discusses an example provided in literature \cite{nguyen2020robust}: this method divides the space into 72 segments, outputting an estimate of the source angle every 2 seconds (i.e., every 173 frames). The dimension of the sample slice \( \bm{x}(t) \) is \( 72 \times 1 \), with the sample \( \bm{x} \) having dimensions \( 173 \times 72 \); all 173 slices are fed into the network simultaneously to estimate the source angle. During the training process, this network needs to estimate two objectives concurrently: the number of sources and the source angle. The source count estimation employs a softmax function combined with cross-entropy, while the source angle estimation uses a multi-head sigmoid function combined with binary cross-entropy. The two network segments can be expressed as:
\begin{align}
    f_{1}(\bm{x}) &= \mathcal{S}_{x} \circ \mathcal{F}_{1,2} \circ \mathcal{F}_{1,1} \circ g(\bm{x}), \\
    f_{2}(\bm{x}) &= \mathcal{S}_{d,\mathrm{multi}} \circ \mathcal{F}_{2,2} \circ \mathcal{F}_{2,1} \circ g(\bm{x}),
\end{align}
where \( \mathcal{F}_{1,2}(\cdot) \) and \( \mathcal{F}_{1,1}(\cdot) \) are two fully connected networks designed to extract the final embeddings for the source count; \( \mathcal{F}_{2,2}(\cdot) \) and \( \mathcal{F}_{2,1}(\cdot) \) are also two fully connected networks for extracting the final embeddings for source angle. The source count estimation network \( f_{1}(\bm{x}) \) and the source angle estimation network \( f_{2}(\bm{x}) \) share part of the feature network \( g(\bm{x}) \), which consists of 10 layers of convolutional networks. Specifically,
\begin{align}
    g(\bm{x}) &= \mathrm{Flatten} \circ \mathcal{C}_{5} \circ \mathcal{C}_{4} \circ \mathcal{C}_{3} \circ \mathcal{C}_{2} \circ \mathcal{C}_{1}(\bm{x}),
\end{align}
where each \( \mathcal{C}_{i}(\cdot) \) is a two-layer convolutional network. Each convolutional network uses a kernel size of \( 3 \times 3 \), and the compression rate in both time and frequency dimensions is twofold. The output channels for \( \mathcal{C}_{1}(\cdot) \), \( \mathcal{C}_{2}(\cdot) \), \( \mathcal{C}_{3}(\cdot) \), \( \mathcal{C}_{4}(\cdot) \), and \( \mathcal{C}_{5}(\cdot) \) increase by a factor of two with each layer, specifically 16, 32, 64, 128, and 256. After the flatten operation, the network ultimately outputs a 512-dimensional feature.

\subsubsection{U-Net Architecture}

The U-Net type neural network takes as input an image constructed by concatenating multiple slices, outputting a tensor of the same two-dimensional image dimensions but different depths. This type of network can be viewed as an extension of the encoder-decoder network, differing in that it directly passes features of different scales from the encoder network to the decoder network. Literature \cite{chazan2019multi} presents a U-Net neural network constructed with 5 layers of convolutional networks for source localization and enhancement. Given a sample \( \bm{x} \), its ``encoding" part can be expressed as:
\begin{align}
    \bm{x}^{(5)} = \mathcal{C}_{5} \circ \mathcal{C}_{4} \circ \mathcal{C}_{3} \circ \mathcal{C}_{2} \circ \mathcal{C}_{1}(\bm{x}),
\end{align}
The relationships between the outputs of different layers can be described as:
\begin{align}
    \bm{x}^{(q)} = \mathcal{C}_{q}\left[\bm{x}^{(q-1)}\right], \quad q=1,2,3,4,5.
\end{align}
Through the encoder network, the features of the sample are gradually compressed, extracting structural features in the time-frequency domain step by step. Given that convolutional networks increase the receptive field with the number of layers, multiple scales of features \( \bm{x}^{(1)}, \bm{x}^{(2)}, \ldots, \bm{x}^{(5)} \) can ultimately be obtained.

The corresponding ``decoding" module consists of four parts:
\begin{align}
    \bm{c}^{(q)} = \mathcal{C}_{q}^{\prime} \left[\bm{x}^{(q)} \ddagger \bm{c}^{(q+1)}\right], \quad q=4,3,2,1.
\end{align}
Here, \( \ddagger \) denotes feature concatenation, and \( \mathcal{C}_{q}^{\prime}(\cdot) \) represents a transposed convolution module that expands the feature dimensions. The output of the final decoding module is the overall output of the network, i.e., \( \bm{y} = \bm{c}^{(1)} \).

Assuming the input \( \bm{x} \) has dimensions \( T \times K \times 2M \), where \( M \) is the number of observation channels, \( T \) is the total number of frames corresponding to a single data item, and \( K \) is the total number of frequency bands. The convolutional kernels in the U-Net network have a width of \( 3 \times 3 \), and the convolution occurs in the ``time \(\times\) frequency" dimensions, with the number of channels doubling at each layer while compressing features by half in both time and frequency dimensions. By analogy, the U-Net output depends on the task. Our task requires detecting the probability of a source appearing at each time-frequency point and direction, leading to an output dimension of \( T \times K \times L \), where \( L \) is the number of discrete spatial angles.

Since each element in the network output represents the probability of a source appearing at the corresponding time, frequency, and angle, once the network converges, we can calculate the source angle by summing the probabilities across all time-frequency points for each angle. This method also has the advantage of enabling signal extraction at specific directions. In practical applications, one can extract the source occurrence probabilities at various time-frequency points for a given direction, yielding a time-frequency mask. Applying this mask to the observed signals allows for the extraction of signals in that specific direction.

\section{Approaches to Sound Event Detection}
 
Sound event detection is a typical signal detection problem that identifies whether a specific class or multiple classes of events occur at a given time point. In real-world applications, sound event detection tasks often require the detection of dozens or even hundreds of sound events, along with the precise timing of their occurrences. This section focuses on constructing functions and designing learning strategies. The specific topics include:
 1) \emph{What are the basic features, and how can the sound event detection problem be described?
 2)  What is the fundamental framework? Why is aggregation necessary?
 3)  How is the loss function constructed? How is the data imbalance problem addressed?
 4) What are the typical networks used}?

\subsection{Sound Event Detection Problem}

\subsubsection{Signal Model and Detection Problem}

As in previous sections, the input to the network is represented as \( \bm{x} \), the output as \( \hat{\bm{y}} \), and the data labels as \( \bm{y} \). For a dataset and a batch of data, we can denote it simply as \( \{(\bm{x}_{n}, \bm{y}_{n}), \forall n\} \). In sound event detection, the network input is typically the Mel spectrogram, which is computed as follows: 1) segmenting the signal into frames; 2) performing a Fourier transform on each frame; 3) merging the frequency spectra of each band according to Mel frequencies (weighted summation with triangular windows); 4) if necessary, applying a logarithmic scale to convert the features. Each slice of the sample corresponds to a vector representing the Mel spectrogram for a specific time slice; for example, with a 40 ms frame length and 20 ms frame shift, each frame corresponds to a 64-point Mel spectrogram.

For convenience, we will omit the sample index \( n \). For a data sample composed of \( T \) slices, its sample, label, and network output can be represented as:
\begin{align}
    \bm{x} &= \left[\begin{array}{cccc}
    \bm{x}_{0} & \bm{x}_{1} & \cdots & \bm{x}_{T-1}
    \end{array}\right], \\
    \bm{y} &= \left[\begin{array}{cccc}
    y(0) & y(1) & \cdots & y(L-1)
    \end{array}\right]^{T}, \\
    \hat{\bm{y}} &= \left[\begin{array}{cccc}
    \hat{y}(0) & \hat{y}(1) & \cdots & \hat{y}(L-1)
    \end{array}\right]^{T}, \\
    \hat{\bm{y}}_{t} &= \left[\begin{array}{cccc}
    \hat{y}_{t}(0) & \hat{y}_{t}(1) & \cdots & \hat{y}_{t}(L-1)
    \end{array}\right]^{T}
\end{align}
where \( L \) is the total number of sound event categories. The label \( y(\ell) \) indicates whether the \( \ell \)-th type of event occurs in the sample, with \( y(\ell) = 1 \) if it occurs, and \( y(\ell) = 0 \) otherwise. The value \( \hat{y}(\ell) \) is the network's estimate, providing the probability of the \( \ell \)-th event occurring in the data sample, where values closer to 1 indicate a higher probability of occurrence. The value \( \hat{y}_{t}(\ell) \) indicates the probability of the \( \ell \)-th event occurring at the \( t \)-th moment in the data sample.

The fundamental problem of sound event detection is to determine, given the sample \( \bm{x} \), whether various types of events occur in the sample, i.e., to provide \( \hat{\bm{y}} \) and the timing of each event, represented by \( \hat{\bm{y}}_{t} \). Since \( \hat{\bm{y}}_{t} \) represents the output of a set of heads in a layer of the network, concatenating all heads from that layer gives us a matrix:
\begin{align}
    \hat{\bm{Y}} = \left[\begin{array}{cccc}
    \hat{\bm{y}}_{0} & \hat{\bm{y}}_{1} & \cdots & \hat{\bm{y}}_{T-1}
    \end{array}\right].
\end{align}
This matrix has dimensions \( L \times T \). The task of sound event detection is to construct a function \( f(\bm{x}) \) using a neural network such that:
\begin{align}
    \hat{\bm{Y}} = f(\bm{x}).
\end{align}
Here, the \( \ell \)-th row and \( t \)-th column of \( \hat{\bm{Y}} \) represent the probability of the \( \ell \)-th event occurring at time \( t \). The training goal of the network is to ensure that this probability reflects the true situation as accurately as possible.

\subsubsection{Detection Decision}

Each row of the matrix \( \hat{\bm{Y}} \) represents the probability of a specific event occurring at different times, which can also be characterized by \( \hat{y}_{t}(\ell), \forall t \). If we treat \( \hat{y}_{t}(\ell), \forall t \) as a function of time \( t \), we find that the function curve continuously varies between 0 and 1. However, from a practical perspective, we need to convert the continuous function \( \hat{y}_{t}(\ell), \forall t \) into a binary function: \( \hat{y}_{t}(\ell) = 1 \) indicates the event of type \( \ell \) has occurred, while \( \hat{y}_{t}(\ell) = 0 \) indicates it has not.

One simple approach is to set a threshold \( \epsilon_{0} \); when \( \hat{y}_{t}(\ell) \geq \epsilon_{0} \), we conclude that the event has occurred; when \( \hat{y}_{t}(\ell) < \epsilon_{0} \), we conclude it has not. However, this method does not account for the persistence of sound events, and the threshold setting always compromises between false alarm and miss rates. A better approach is to use multiple decision thresholds \cite{dinkel2021towards}. For convenience, we temporarily omit the category variable \( \ell \) and analyze how to convert \( \hat{y}_{t} \) into a binary function \( p_{t} \). The function \( p_{t} \) consists of three parts:
\begin{align}
    p_{t} &= p_{t}^{c} \cdot p_{t}^{g} + p_{t}^{l} \cdot p_{t}^{g} \\
    &= \left( p_{t}^{c} + p_{t}^{l}\right) \times p_{t}^{g},
\end{align}
where
\begin{align}
    p_{t}^{g} &= \left\{\begin{array}{ll}
    1, & \hat{y} \geq \epsilon_{g}; \\
    0, & \text{otherwise}
    \end{array}\right., \\
    p_{t}^{l} &= \left\{\begin{array}{ll}
    1, & \hat{y}_{t} \geq \epsilon_{\mathrm{high}}; \\
    0, & \text{otherwise}
    \end{array}\right., \\
    p_{t}^{c} &= \left\{\begin{array}{ll}
    1, & \exists (t_{1}, t_{2}), t \in [t_{1}, t_{2}], t_{2} - t_{1} \geq \zeta_{0}, \\
    & \text{ and } \forall t' \in [t_{1}, t_{2}], \epsilon_{\mathrm{low}} \leq \hat{y}_{t} \leq \epsilon_{\mathrm{high}}; \\
    0, & \text{otherwise}
    \end{array}\right.
\end{align} 
Here, \( \epsilon_{g}, \epsilon_{\mathrm{low}}, \text{ and } \epsilon_{\mathrm{high}} \) are three threshold values, such as \( \epsilon_{g} = 0.5, \epsilon_{\mathrm{low}} = 0.2, \epsilon_{\mathrm{high}} = 0.75 \).
\begin{itemize}
    \item \( \epsilon_{g} \) represents the threshold for the occurrence probability of events after time slice aggregation, serving as a global threshold; if \( \hat{y} \) exceeds this threshold, we consider the corresponding event triggered.
    \item \( \epsilon_{\mathrm{high}} \) is a local threshold used to determine whether a particular type of event occurs at a specific time slice; this value is generally set high to reduce the false alarm rate.
    \item For some time slices, the detection probability \( \hat{y}_{t} \) may not be high, but it may persist for a duration of \( \zeta_{0} \); if the detection probability remains above a certain threshold over a sufficiently long event segment, all slices corresponding to that segment are considered triggered. This introduces the third threshold, \( \epsilon_{\mathrm{low}} \), which is typically set lower than \( \epsilon_{\mathrm{high}} \).
\end{itemize}
It can be verified that \( p_{t}^{c} \) is used to detect the missed portions of \( p_{t}^{l} \), based on the principle that if an event can persist for a duration, the detection threshold can be lowered. Since \( p_{t}^{c} \) and \( p_{t}^{l} \) are mutually exclusive, they can be fused using \( p_{t} = (p_{t}^{c} + p_{t}^{l}) \times p_{t}^{g} \).
It is important to note that detection decisions only arise during the usage phase and do not involve training the network.

\subsubsection{Weak Label Problem}

In the task of sound event detection, the goal of network training is to ensure that \( \hat{\bm{Y}} = f(\bm{x}) \) approaches the true value as closely as possible. From the perspective of traditional neural network training strategies, this requires knowledge of the probabilities of all events occurring across all time slices for each sample, i.e., the true values corresponding to the detection probabilities \( \hat{y}_{t}(\ell), \forall t,\ell \). However, in practical data, it is often unrealistic to label every event slice; typically, only the types of events present in the data can be labeled without specifying the timing or exact occurrences. In other words, for each data sample \( \bm{x} \), we can only know \( \bm{y} \) but cannot ascertain the true values of \( \hat{\bm{Y}} \). This issue is known as the weak label problem in machine learning. Learning to achieve the effect of ``strong labeling" from weak labels and achieving detection probability super-resolution is a key focus of research in sound event detection.

\subsection{Detection Aggregation Methods}

The network output \( \hat{\bm{Y}} = f(\bm{x}) \) is an \( L \times T \) matrix, while the data label \( \bm{y} \) is an \( L \times 1 \) vector. To enable function learning, we first need to aggregate \( \hat{\bm{Y}} \) into \( \hat{\bm{y}} \) and then make \( \hat{\bm{y}} \) approximate \( \bm{y} \). Since multiple events may occur simultaneously within the same data sample, minimizing binary cross-entropy can achieve this approximation:
\begin{align}
    \min_{\theta_{f}} -\bm{y}^{T} \ln \hat{\bm{y}} - (1 - \bm{y}^{T}) \ln (1 - \hat{\bm{y}}).
\end{align}
The remaining issue is how to transform \( \hat{\bm{Y}} \) into \( \hat{\bm{y}} \). The basic principle is to assign a higher weight to slices with high detection probabilities and a lower weight to slices with low detection probabilities, aiming to statistically determine the average of the larger detection probabilities.

\subsubsection{Max Aggregation}

Max aggregation involves selecting the maximum value from different time slices for a given event as the result of the aggregation. The basic principle is that if an event occurs in any of the time slices of the data, it is considered to have occurred overall. Mathematically, this aggregation can be expressed as:
\begin{align}
    \hat{y}(\ell) = \max_{t} \hat{y}_{t}(\ell).
\end{align}
While max aggregation is intuitive, it has two issues. From a practical perspective, due to noise in actual data, the detection probabilities often exhibit fluctuations and randomness, leading to potential false alarms when estimating the occurrence probability based on a single slice. From the perspective of network training, max aggregation results in non-zero gradients only at the neuron outputting the maximum value, while gradients are zero elsewhere, significantly reducing the learning efficiency of the network and hindering its updates.

\subsubsection{Mean Aggregation}

Compared to max aggregation, mean aggregation computes the average detection probabilities across different time slices to derive the aggregated detection probability. This method can be represented as:
\begin{align}
    \hat{y}(\ell) = \frac{1}{T} \sum_{t=0}^{T-1} \hat{y}_{t}(\ell).
\end{align}
This approach can reduce detection disturbances for events with long durations. However, for short-duration events, this method tends to lower the overall \( \hat{y}(\ell) \) value, making it more challenging to determine whether the event is present in the sample from a global perspective.

Since the essence of network learning is function approximation, if we want the aggregated value for short-duration samples to approach the label value of 1, the overall \( \hat{y}_{t}(\ell) \) values must be elevated, which is not conducive to network training.

\subsubsection{Weighted Sorting Aggregation}

Given the issues with max and mean aggregation, a better approach is to select a subset of larger detection values for each event and average these values. This operation first requires sorting the detection probabilities from different time slices for the sample, yielding \( \hat{y}_{t}^{\prime}(\ell) \):
\begin{align}
    \hat{y}_{T-1}^{\prime}(\ell) \leq \hat{y}_{T-2}^{\prime}(\ell) \leq \cdots \leq \hat{y}_{1}^{\prime}(\ell) \leq \hat{y}_{0}^{\prime}(\ell),
\end{align}
Then, selecting the top \( N \) maximum values and averaging them gives:
\begin{align}
    \hat{y}(\ell) = \frac{1}{N} \sum_{t=0}^{N-1} \hat{y}_{t}^{\prime}(\ell),
\end{align}
where \( N \leq T \) is a predefined hyperparameter. When \( N=1 \), this method reduces to max aggregation; when \( N=T \), it reduces to mean aggregation.

In addition to averaging a subset of values, the sorted detection values can also be combined with non-uniform weights to achieve similar aggregation effects:
\begin{align}
    \hat{y}(\ell) = \frac{1 - \lambda}{1 - \lambda^{T}} \sum_{t=0}^{T-1} \lambda^{t-1} \hat{y}_{t}^{\prime}(\ell),
\end{align}
where \( 0 < \lambda < 1 \); \( \frac{1 - \lambda}{1 - \lambda^{T}} \) is a normalization factor to mitigate the impact of the weighting coefficient \( \lambda^{t-1} \) and ensures that the final aggregated value remains between 0 and 1. Similarly, as \( \lambda \) approaches 0, this method converges to max aggregation; as \( \lambda \) approaches 1, it converges to mean aggregation.

In practice, \( \tau \) can be set as a hyperparameter beforehand or designated as an optimizable parameter during network construction, allowing it to be learned from the data. Clearly, for events with longer durations (e.g., a fan), \( \tau \) can be set smaller to lean towards mean aggregation; for very short events (e.g., impacts), \( \tau \) can be set larger to elevate the function values of the higher detection probabilities.

\subsubsection{Softmax Weighted Aggregation}

The weighted sorting aggregation method requires sorting the values of each event slice to determine the necessary weighting coefficients for each time slice. This sorting introduces additional computational overhead, which can hinder the network's learning and inference processes. To address this issue, we can construct weighting coefficients based on the magnitudes of the detection values themselves for aggregation. The Softmax weighted aggregation method constructs a Softmax function based on the detection probabilities, as follows:

\begin{align}
    \hat{y}(\ell) &= \sum_{t=0}^{T-1} \frac{\expd{\tau \cdot \hat{y}_{t}(\ell)}}{\sum_{i=0}^{T-1} \expd{\tau \cdot \hat{y}_{i}(\ell)}} \hat{y}_{t}(\ell) \\
    &= \sum_{t=0}^{T-1} w_{t}(\ell) \hat{y}_{t}(\ell),
\end{align}
where \( \tau \geq 0 \) controls the behavior of the non-uniform weighting, and 
 \(
    w_{t}(\ell) = {\expd{\tau \cdot \hat{y}_{t}(\ell)}}/{\sum_{i=0}^{T-1} \expd{\tau \cdot \hat{y}_{i}(\ell)}}
 \)
represents the non-uniform weighting coefficient. As \( \tau \) approaches infinity, this method converges to maximum value aggregation; as \( \tau \) approaches zero, it converges to mean aggregation.

In practical applications, \( \tau \) can be set as a hyperparameter in advance, or it can be defined as an optimizable parameter within the network, learned from the data. Clearly, for events with longer durations (e.g., a fan), \( \tau \) can be set smaller to favor mean aggregation. Conversely, for events with short durations (e.g., a sporadic sound), \( \tau \) can be set larger to allow the function to emphasize the time slices with higher detection values.

\subsubsection{Linear Softmax Weighted Aggregation}

The core idea behind non-uniform weighted aggregation is to assign larger weights to larger detection probabilities while assigning smaller weights to smaller detection probabilities. Based on this principle, a simpler method for constructing weight functions is given by:
\begin{align}
    w_{t}(\ell) = \frac{\hat{y}_{t}(\ell)}{\sum_{i=0}^{T-1} \hat{y}_{i}(\ell)}.
\end{align}
Evidently, this weight function is determined by the relative values of the detection probabilities. Based on this weight function, the aggregated result can be expressed as:
\begin{align}
    \hat{y}(\ell) &= \sum_{t=0}^{T-1} w_{t}(\ell) \hat{y}_{t}(\ell) \\
    &= \sum_{t=0}^{T-1} \frac{\hat{y}_{t}(\ell)}{\sum_{i=0}^{T-1} \hat{y}_{i}(\ell)} \hat{y}_{t}(\ell) \\
    &= \sum_{t=0}^{T-1} \frac{\hat{y}_{t}^{2}(\ell)}{\sum_{i=0}^{T-1} \hat{y}_{i}(\ell)} \\
    &= \frac{\sum_{t=0}^{T-1}\hat{y}_{t}^{2}(\ell)}{\sum_{t=0}^{T-1} \hat{y}_{t}(\ell)}.
\end{align}
From a mathematical perspective, this aggregation method involves summing the squares of all detection probabilities and dividing by their direct sum. In addition to providing a non-uniform weighting effect, when combined with binary cross-entropy, it encourages \( \hat{y}_{t}(\ell) \) to converge toward 0 and 1, making larger values even larger and smaller values even smaller. This approach is particularly suitable for weakly supervised learning and detection probability super-resolution, and it has been widely applied in related tasks.

\subsection{Loss Function Construction}

The aggregation function maps the sound event detections \( \hat{\bm{Y}} \) from multiple time slices into the overall detection probability \( \hat{\bm{y}} \) for the entire data sample. By making each training sample's \( \hat{\bm{y}}_{n} \) as close to the sample's label \( \bm{y}_{n} \) as possible, we can complete the training of the network \( f(\cdot) \). In prior discussions about the network aggregation operation, we omitted the sample index \( _{n} \). Since the loss function is typically defined based on a batch of data, this section will address the construction of the loss function while reintroducing the sample index \( _{n} \). Here, \( \bm{x}_{n} \) denotes the \( n \)-th sample in the dataset, \( \bm{y}_{n} \) represents the corresponding label for that sample, and \( \hat{\bm{Y}}_{n} \) indicates the detection probabilities of each slice for the \( n \)-th sample. The term \( \hat{\bm{y}}_{n} \) is the result after aggregating \( \hat{\bm{Y}}_{n} \). Since the number of slices in each data sample may vary, we denote \( T_{n} \) as the number of slices in the \( n \)-th sample. Additionally, we introduce \( \mathcal{I}_{m} \) and \( N_{m} \) to represent the set of sample indices in the \( m \)-th batch of data within the entire dataset, with \( N_{m} \) being the total number of samples in the \( m \)-th batch.

\subsubsection{Binary Cross-Entropy}

If each sample originally contains only one sound event, the loss function can be constructed using the cross-entropy from multi-class problems, expressed as \( -\bm{y}_{n}^{T}\ln \hat{\bm{y}}_{n} \), where each element of \( \hat{\bm{y}}_{n} \) is greater than zero, and all elements sum to one, typically representing the output of a softmax function. However, in sound event detection, multiple sounds often overlap in the same audio segment, making it inappropriate to treat this as a multi-class problem. Instead, each sound event should be treated as a binary classification problem, essentially addressing the ``detection of presence or absence" for each event type at each time slice. Thus, we need to construct binary cross-entropy to formulate the loss function.

For the given \( m \)-th batch of samples \( \{(\bm{x}_{n}, \bm{y}_{n}), \forall n \in \mathcal{I}_{m}\} \), the corresponding loss function can be described as:
\begin{align}
    \mathcal{J}_{1} &= -\sum_{n \in \mathcal{I}_{m}} \left[\bm{y}_{n}^{T}\ln \hat{\bm{y}}_{n} + \left(1-\bm{y}_{n}^{T}\right)\ln \left(1- \hat{\bm{y}}_{n}\right)\right] \label{sed-cost-bce-1} \\
    &= -\sum_{n \in \mathcal{I}_{m}} \sum_{\ell=0}^{L-1} \left\{ y_{n}(\ell) \ln \hat{y}_{n}(\ell) + \left[1-y_{n}(\ell)\right]\ln \left[1- \hat{y}_{n}(\ell)\right]\right\}, \label{sed-cost-bce-2}
\end{align}
where \( y_{n}(\ell) \ln \hat{y}_{n}(\ell) + \left[1-y_{n}(\ell)\right]\ln \left[1- \hat{y}_{n}(\ell)\right] \) is the binary cross-entropy for the detection probability of the \( \ell \)-th event in the \( n \)-th sample. This is derived from aggregating the \( \ell \)-th row of \( \hat{\bm{Y}}_{n} \).

\subsubsection{Weighted Binary Cross-Entropy}

In some sound event detection problems, the number of event categories can be in the hundreds. However, within the same sample, only a few event types (2-3) typically occur simultaneously. This implies that only a few elements in \( \bm{y}_{n} \) are 1, while most are zero. Therefore, in equation (\ref{sed-cost-bce-1}), the actual number of contributing terms in the first term \( \bm{y}_{n}^{T}\ln \hat{\bm{y}}_{n} \) is much smaller than the second term \( \left(1-\bm{y}_{n}^{T}\right)\ln \left(1- \hat{\bm{y}}_{n}\right) \). As per equation (\ref{sed-cost-bce-2}), the summation terms concerning \( \ln \hat{y}_{n}(\ell) \) are significantly fewer than those for \( \ln \left[1- \hat{y}_{n}(\ell)\right] \). To achieve a better compromise in the loss function, it is common to rewrite it as:
\begin{align}
    \mathcal{J}_{2} &= -\sum_{n \in \mathcal{I}_{m}} \left[\beta \cdot \bm{y}_{n}^{T}\ln \hat{\bm{y}}_{n} + \left(1-\bm{y}_{n}^{T}\right)\ln \left(1- \hat{\bm{y}}_{n}\right)\right] \label{sed-cost-bce-1-weight} \\
    &= -\sum_{n \in \mathcal{I}_{m}} \sum_{\ell=0}^{L-1} \left\{ \beta \cdot y_{n}(\ell) \ln \hat{y}_{n}(\ell) + \left[1-y_{n}(\ell)\right]\ln \left[1- \hat{y}_{n}(\ell)\right]\right\}, \label{sed-cost-bce-2-weight}
\end{align}
where \( \beta \geq 1 \) is a predefined hyperparameter, such as \( \beta = 100 \). The parameter \( \beta \) increases the contribution of the summation term \( \ln \hat{y}_{n}(\ell) \) in the loss function\footnote{This portion corresponds to the moments when the sound events occur.}, enhancing its role in the backpropagation error, thus facilitating network convergence.

\subsubsection{Inverse Frequency Loss}

In sound event detection tasks, there is often a ``class imbalance" issue. In a given dataset, some classes may be very abundant while others are scarce. During random batch processing, the class imbalance problem can become exacerbated. Some classes may appear frequently in the same batch, while others may hardly appear at all, or not appear across several consecutive batches. This situation means that learning for the less frequent classes is very slow, and due to extreme data scarcity, the network may struggle to capture contributions from those samples in the backpropagation error. To address this problem, different weighting coefficients are often assigned to the loss functions of samples from different classes. Based on this principle, the inverse frequency loss (IFL) for binary cross-entropy can be described as:
\begin{align}
    \mathcal{J}_{3} = -\sum_{n \in \mathcal{I}_{m}} \sum_{\ell=0}^{L-1} & \left\{ \left( \frac{c_{0}}{K_{\ell} + c_{0}} \right)^{\eta} y_{n}(\ell) \ln \hat{y}_{n}(\ell) + \right.
    \nonumber\\ 
     & \left. \left[1-y_{n}(\ell)\right]\ln \left[1- \hat{y}_{n}(\ell)\right]\right\}, \label{sed-cost-bce-2-weight}
\end{align}
where \( K_{\ell} \) is the occurrence count of the \( \ell \)-th class event in the current batch of data, and \( c_{0} \geq 0 \) and \( \eta \geq 0 \) are two predefined hyperparameters. As can be seen from the equation, when a class event occurs frequently, the weight in front of \( \ln \hat{y}_{n}(\ell) \) will be relatively small; conversely, when a class event occurs infrequently, the weight will be larger. This means that the higher the occurrence frequency, the smaller the weight, which is the origin of this method's name.

\subsubsection{Asymmetric Focal Loss}

As the network training progresses, some samples may be easily distinguishable while others remain difficult. The contributions of easily distinguishable samples to the network's loss function are relatively small, so it is beneficial to consider excluding these samples, allowing the network to focus on improving the learning of difficult samples, thereby enhancing training efficiency. Based on this principle, the asymmetric focal loss can be formulated as:
\begin{align}
    \mathcal{J}_{4} = -\sum_{n \in \mathcal{I}_{m}} \sum_{\ell=0}^{L-1}& \left\{ \left[1-\hat{y}_{n}(\ell)\right]^{\eta} \cdot y_{n}(\ell) \ln \hat{y}_{n}(\ell) +\right.\nonumber\\
    &\left. \left[\hat{y}_{n}(\ell)\right]^{\eta} \cdot \left[1-y_{n}(\ell)\right] \ln \left[1- \hat{y}_{n}(\ell)\right]\right\}.
\end{align}
Here, \( \eta \geq 0 \) is a predefined hyperparameter, and \( \left[1-\hat{y}_{n}(\ell)\right]^{\eta} \) and \( \left[\hat{y}_{n}(\ell)\right]^{\eta} \) are two weighting functions.

When a class event occurs in the sample, \( y_{n}(\ell) = 1 \), and the loss function becomes \( -\ln \hat{y}_{n}(\ell) \); as training progresses and \( \hat{y}_{n}(\ell) \) approaches 1, the weight \( \left[1-\hat{y}_{n}(\ell)\right]^{\eta} \) approaches 0, effectively excluding this sample from the training set. Conversely, when a class event does not occur, \( y_{n}(\ell) = 0 \), and the loss function is \( -\ln \left[1-\hat{y}_{n}(\ell)\right] \); as training progresses and \( \hat{y}_{n}(\ell) \) approaches 0, the weight \( \left[\hat{y}_{n}(\ell)\right]^{\eta} \) decreases, thus reducing the contribution of this sample in the loss function.

\subsubsection{Dice Loss}

To address the data imbalance issue, in addition to modifying binary cross-entropy, alternative loss functions can be employed. One such substitute is the Dice loss function. For a given batch of samples, the Dice loss function can be described as:
\begin{align}
    \mathcal{J}_{5} &= 1 - \frac{2 \sum_{n \in \mathcal{I}_{m}} \sum_{\ell=0}^{L-1} y_{n}(\ell) \hat{y}_{n}(\ell)}{\sum_{n \in \mathcal{I}_{m}} \sum_{\ell=0}^{L-1} y_{n}^{2}(\ell) + \sum_{n \in \mathcal{I}_{m}} \sum_{\ell=0}^{L-1} \hat{y}_{n}^{2}(\ell)} \label{sed-cost-dice-1}
\end{align}
Clearly, the Dice loss is no longer constructed based on binary cross-entropy. Building on the aforementioned construction methods for binary cross-entropy, incorporating non-uniform weighting and asymmetric focal strategies, the Dice loss function can also be modified as \cite{imoto2021impact}:
\begin{align}
    &\mathcal{J}_{6} = 1 - \nonumber\\
    &\frac{\kappa_{0} + \sum\limits_{n \in \mathcal{I}_{m}} \sum\limits_{\ell=0}^{L-1} \varpi_{n}(\ell) \cdot y_{n}(\ell) \hat{y}_{n}(\ell)}{\kappa_{0} + \sum\limits_{n \in \mathcal{I}_{m}} \sum\limits_{\ell=0}^{L-1} (1-\alpha) \cdot y_{n}^{2}(\ell) + \sum\limits_{n \in \mathcal{I}_{m}} \sum\limits_{\ell=0}^{L-1} \alpha \cdot \varpi_{n}(\ell) \cdot \hat{y}_{n}^{2}(\ell)} \label{sed-cost-dice-2}
\end{align} 
where 
\begin{align}
    \varpi_{n}(\ell) = \left[1 - \hat{y}_{n}(\ell)\right]^{\eta}
\end{align}
is the weighting function from the asymmetric focal loss, \( \kappa_{0} \geq 0 \) is a smoothing parameter, \( \alpha \in (0,1) \) is a weighting coefficient, and \( \eta \geq 0 \) is a predefined hyperparameter. By setting \( \eta = 0, \alpha = 0.5, \kappa_{0} = 0 \), the loss function in equation (\ref{sed-cost-dice-2}) degrades to that in equation (\ref{sed-cost-dice-1}). Like the asymmetric focal loss, the weighting function \( \varpi_{n}(\ell) \) is also used to diminish the contributions of ``easy" samples in the loss function, allowing the network to focus more on learning from difficult samples.

\subsection{Typical Network Structures}

Sound event detection tasks typically do not demand real-time processing; given a signal containing multiple time slices, the task is to return a set of detection results \( \hat{\bm{Y}}_{n} \) and \( \hat{\bm{y}}_{n} \). If we need to determine whether a sound event occurs at a specific time slice, the aforementioned detection decision methods must be employed for further processing of \( \hat{\bm{Y}}_{n} \).

Networks for sound event detection usually incorporate aggregation layers to convert \( \hat{\bm{Y}}_{n} \) into \( \hat{\bm{y}}_{n} \). For convenience, we will denote the aggregation layer as \( \mathcal{A}(\cdot) \):
\begin{align}
    \hat{\bm{y}}_{n} = \mathcal{A}(\hat{\bm{Y}}_{n}).
\end{align}
Given a signal composed of \( T \) time slices, the network input is typically a \( F \times T \) ``image," where \( F \) is the number of Mel frequency bands. Assuming there are \( L \) types of sound events to detect, both \( \bm{y}_{n} \) and \( \hat{\bm{y}}_{n} \) are \( L \times 1 \) vectors, and \( \hat{\bm{Y}}_{n} \) is an \( L \times T \) matrix\footnote{Sometimes, \( \hat{\bm{Y}}_{n} \) may be a \( L \times T \times F \) tensor, representing the detection probabilities of various sound events at each time-frequency point.}

\subsubsection{Fully Convolutional Neural Network}

Literature \cite{kong2019sound} presents a sound event detection function constructed from multiple convolutional neural networks.
The network output consists of 5 convolutional blocks followed by an aggregation network. The neural network can be described as:
\begin{align}
    \hat{\bm{y}} &= \mathcal{A}(\hat{\bm{Y}}), \\
    \hat{\bm{Y}} &= f(\bm{x}) \\
    &= \mathcal{C}_{5} \circ \mathcal{C}_{4} \circ \mathcal{C}_{3} \circ \mathcal{C}_{2} \circ \mathcal{C}_{1}(\bm{x}).
\end{align}
Here, the convolutional blocks \( \mathcal{C}_{i}(\bm{x}), i=1,2,3,4 \) consist of two layers of identical convolutional networks, all with a kernel size of \( 3 \times 3 \), operating in both time and frequency dimensions. The number of output channels increases progressively from the 1st to the 4th block, specifically 32, 64, 128, and 128. In total, these 4 convolutional blocks consist of 8 layers of convolutional networks, with each layer followed by batch normalization and ReLU activation functions. The final convolutional block \( \mathcal{C}_{5}(\cdot) \) employs a \( 1 \times 1 \) convolutional kernel to reduce the number of feature channels from 128 to 41, followed by a Sigmoid activation function to match the probabilities of detecting 41 types of sound events.

In this case, \( \hat{\bm{Y}} \) is a \( L \times T \times F \) tensor. First, the frequency dimension of \( \hat{\bm{Y}} \) is averaged to yield an \( L \times T \) matrix, which is then processed through the aggregation operation to obtain the final \( \hat{\bm{y}} \). By comparing \( \hat{\bm{y}} \) and \( \bm{y}\), the network learning can be completed.

\subsubsection{Convolutional + Recurrent Neural Network}

Literature \cite{dinkel2021towards} presents a construction method that combines multi-layer convolutional neural networks with recurrent neural networks.
For each sound signal, the input remains an \( F \times T \) Mel frequency matrix. The network output is described as:
\begin{align}
    \hat{\bm{y}} &= \mathcal{A}(\hat{\bm{Y}}), \\
    \hat{\bm{Y}} &= f(\bm{x}) \\
    &= \mathcal{S}_{d,\mathrm{multi}} \circ \mathcal{G} \circ \mathcal{C}_{3} \circ \mathcal{C}_{2} \circ \mathcal{C}_{1}(\bm{x}).
\end{align}
Here, \( \mathcal{C}_{1}(\cdot), \mathcal{C}_{2}(\cdot) \), and \( \mathcal{C}_{3}(\cdot) \) are three convolutional blocks, all with a convolutional kernel size of \( 3 \times 3 \). The number of output channels for these blocks is 32, 128, and 128, respectively. The \( \mathcal{C}_{1}(\cdot) \) block consists of a single convolutional layer, followed by a 4-fold downsampling in the frequency dimension. The \( \mathcal{C}_{2}(\cdot) \) and \( \mathcal{C}_{3}(\cdot) \) blocks consist of two layers of identical convolutional networks, with each block performing a 4-fold downsampling in the frequency dimension. After processing through three convolutional blocks, the frequency dimension is downsampled by 64 times, leaving each event slice with only one point in the frequency dimension. Since the third convolutional block outputs 128 channels with a frequency dimension of 1, the features at each time slice are 128-dimensional vectors. These 128-dimensional features are then fed into the recurrent neural network \( \mathcal{G}(\cdot) \) for processing and transformation, followed by the multi-target detection layer \( \mathcal{S}_{d,\mathrm{multi}}(\cdot) \) to obtain \( \hat{\bm{Y}} \), which is an \( L \times T \) matrix.

\subsubsection{Convolutional + Attention Mechanism Network}

In scenarios where real-time computation is not required, transformer networks based on self-attention mechanisms can capture long-term and short-term dependencies better than recurrent neural networks. Thus, post-processing the features extracted by convolutional networks can yield improved results. Based on this observation, literature \cite{miyazaki2020weakly} presents a function construction method that combines convolutional networks with attention mechanisms.
\begin{align}
    \hat{\bm{y}} &= \mathcal{A}(\hat{\bm{Y}}), \\
    \hat{\bm{Y}} &= f(\bm{x}) \\
    &= \mathcal{S}_{d,\mathrm{multi}} \circ \mathcal{T} \circ \mathcal{C}(\bm{x}),
\end{align}
where \( \mathcal{C}(\cdot) \) is a feature extraction block composed of multiple layers of convolutional networks. For each Mel spectrogram at a time slice, it produces a feature vector. This feature vector is then input into a transformer network \( \mathcal{T}(\cdot) \) for post-processing, yielding the final feature vector used for detection. The last feature vector is subsequently fed into the multi-target detection function \( \mathcal{S}_{d,\mathrm{multi}}(\cdot) \) to obtain the probabilities of various events occurring at each time slice, represented by \( \hat{\bm{Y}} \).

\subsubsection{Multi-Objective Learning Approaches}

During the network learning process, utilizing multiple training objectives often leads to better feature learning. Based on this principle, literature \cite{imoto2020sound} proposes a function learning method that jointly addresses sound event detection and sound scene classification tasks.

For convenience, let the labels and detection probabilities for sound scene classification be denoted as \( \bm{z} \) and \( \hat{\bm{z}} \), respectively. The function consists of a multi-layer convolutional feature extraction network and two branches:
\begin{align}
    \hat{\bm{y}} &= \mathcal{A} \circ f(\bm{x}) \\
    &= \mathcal{A} \circ \mathcal{S}_{d,\mathrm{multi}} \circ \mathcal{F}_{1} \circ \mathcal{C}_{1} \circ \mathcal{G} \circ \mathcal{C}_{0} (\bm{x}), \\
    \hat{\bm{z}} &= \mathcal{S}_{x} \circ \mathcal{F}_{2} \circ \mathcal{C}_{2} \circ \mathcal{G} \circ \mathcal{C}_{0} (\bm{x}),
\end{align}
where \( \mathcal{C}_{0}(\bm{x}) \) is the shared feature network; \( \mathcal{G}(\cdot) \) is the recurrent neural network; \( \mathcal{C}_{1}(\cdot) \) and \( \mathcal{C}_{2}(\cdot) \) are convolutional networks for event detection and scene classification, respectively; \( \mathcal{F}_{1} \) and \( \mathcal{F}_{2} \) are two fully connected network blocks for feature mapping; \( \mathcal{S}_{d,\mathrm{multi}}(\cdot) \) is the multi-target detection network; and \( \mathcal{S}_{x}(\cdot) \) is the softmax layer for sound scene classification. For learning sound event detection, the binary cross-entropy can be employed to construct the loss function, while for sound scene classification, the cross-entropy \( -\bm{z}^{T} \ln \hat{\bm{z}} \) is utilized to complete the network training.

In situations where existing samples may not provide the scene labels, this presents challenges for backpropagation of errors. To address this issue, the loss function for the sound scene classification branch is characterized using pseudo-labels \( \tilde{\bm{z}} \), specifically \( -\tilde{\bm{z}}^{T} \ln \hat{\bm{z}} \). The required pseudo-labels are provided by a dedicated scene classification network \( g(\cdot) \), such that \( \tilde{\bm{z}} = g(\bm{x}) \). The learning of the function \( g(\cdot) \) can be performed using a specialized dataset for sound scene classification.

\section{Approaches to Voiceprint Extraction and Recognition}
 
The essence of voiceprint extraction is signal transformation, converting a segment of raw speech signal into a fixed-dimensional feature vector. Voiceprint recognition is based on the assumption that each individual's voice has unique characteristics. This area has a research history spanning several decades. This section focuses on deep learning methods, covering the following key topics: \emph{
1) Common basic features in voiceprint recognition.
    2) The basic process and applications of voiceprint recognition.
   3) Fundamental scientific issues in voiceprint recognition.
   4) Loss function construction and typical network structures.}

\subsection{Signal Model and Problem Description}

In applications related to speech communication and interaction, voiceprint extraction technology has three basic application points: speaker verification, speaker identification, and speaker diarization\cite{bai2021speaker}. Speaker verification helps users establish a private key for managing user logins, similar to single-target detection, where we determine if a given speech segment matches the registered voice. Speaker identification assists in creating smart access control systems managing multiple user keys, akin to multi-target detection, where we check if a given speech segment corresponds to a known user's key. Speaker diarization typically involves organizing meeting minutes by separating and organizing meeting speech based on differences in voiceprint features across different time slices, representing a classic clustering problem.

Even in speaker verification and identification, voiceprint recognition differs from traditional target detection problems. The voice to be recognized ultimately does not match the voice used during network training. For instance, if we train a method to convert sound signals into voiceprint features based on sounds produced by various auditory objects (e.g., cats, dogs), we obtain a function \( f(\cdot) \). During usage, we apply this function to convert the voice of a specific target (e.g., a human speaker) into a voiceprint. A significant reason for this issue is that users do not typically wish to upload their data for secondary development of voiceprint extraction technology. Therefore, despite voiceprint recognition being a signal detection problem, voiceprint extraction techniques often cannot rely on signal detection methods but require signal transformation techniques.

From the relationship between voiceprint extraction and recognition, it is evident that since the data during training and usage may not follow the same distribution, the features extracted for voiceprint recognition may not be particularly suited for recognizing a specific target's voiceprint, leading to domain mismatch issues. Domain mismatch is one of the key research focuses in voiceprint extraction and recognition.

\subsubsection{Data Features}

The commonly used feature in voiceprint recognition is the Mel Frequency Cepstrum Coefficient (MFCC). Given a \( 20 \) ms signal slice, an MFCC feature can be extracted, representing the signal slice as a vector, typically around 20 dimensions. If the original signal has a sampling rate of 8000 Hz, each \( 20 \) ms slice contains 160 points. After computation and transformation, this results in a feature vector of approximately 20 points. For convenience, we denote the MFCC coefficients at each time slice as \( \bm{x}(t), t=0,1,\ldots,T-1 \), where \( t \) is the time index; each \( \bm{x}(t) \) corresponds to a set of MFCC coefficients.

The term ``Mel frequency cepstrum coefficient" consists of two keywords: Mel frequency and cepstrum coefficient, each representing different concepts.
\begin{itemize}
    \item Mel frequency models the relationship between perceived frequency and actual frequency:
    \begin{align}
        f^{\mathrm{mel}} = 1125 \ln \left(1 + \frac{f^{\mathrm{Hz}}}{700}\right),
    \end{align}
    where \( f^{\mathrm{mel}} \) and \( f^{\mathrm{Hz}} \) represent Mel frequency and Hertz frequency, respectively. This is based on the fact that different positions in the human cochlea perceive signals at different frequency bands, but uniform sampling in the mid-to-high frequency range corresponds to logarithmic frequency sampling. This means that, in the mid-to-high frequency range, the perceived frequency differences are multiplicative rather than absolute. Based on this principle, the first step in extracting MFCC is to uniformly sample \( f^{\mathrm{mel}} \), and then to compute the power of the corresponding frequency band signals centered around the sampling points to obtain the Mel spectrum.
    
    \item In the Mel spectrum, useful information is typically found within the spectrum's envelope. Due to the slow transformation of the envelope, its inverse Fourier transform coefficients are usually concentrated in lower-order coefficients. Thus, by taking the logarithm of the Mel spectrum and then performing the inverse Fourier transform, followed by extracting the first \( L \) points\footnote{\( L \) corresponds to the number of MFCC coefficients. Taking the logarithm of the Mel spectrum converts ``multiplication" into ``addition," thus separating the envelope.}, we obtain the Mel Frequency Cepstrum Coefficients (MFCC).
\end{itemize}
Although MFCC coefficients reflect the spectral envelope on a logarithmic scale, they do not represent the spectrum in the traditional sense. Since all spectral envelope information is encoded in the MFCC coefficients, they are typically used directly as input features for the network.

\subsubsection{Basic Process of Voiceprint Recognition}

Voiceprint recognition generally comprises three basic stages: model construction and training, speaker registration, and speaker identification. Model construction and training involve creating a function \( g(\cdot) \) that maps speech signals to voiceprint features, typically achieved using a given dataset. Once the model is trained, the voiceprint extraction function \( g(\cdot) \) usually remains unchanged. During usage, users need to provide a segment of speech; the voiceprint extraction function \( g(\cdot) \) generates a voiceprint feature based on the provided speech and stores it, completing the speaker registration. For speaker identification, the machine similarly utilizes \( g(\cdot) \) to convert the currently obtained speech into a voiceprint feature, which is then compared against the existing voiceprint feature library to complete the final identification. Given any speech signal \( \bm{x}_{n} = [\bm{x}_{n}(0)~\bm{x}_{n}(1)~\cdots~\bm{x}_{n}(T-1)] \), its voiceprint feature can be represented as:
\begin{align}
    \bm{z}_{n} = g\left(\bm{x}_{n}\right).
\end{align}
Here, the index \( n \) indicates the index of the speech signal sample in the dataset.

In practical applications, the machine can first generate a voiceprint library \( \{\ddot{\bm{z}}_{i}, \forall i\} \) based on registered users. Each time identification is needed, the acquired speech \( \bm{x}_{n} \) is transformed to obtain the voiceprint feature \( \bm{z} = g\left(\bm{x}\right) \). The similarity between \( \bm{z} \) and \( \ddot{\bm{z}}_{i} \) is computed to determine whether the speaker is registered. A direct method for measuring similarity is:
\begin{align}
    S_{i} = \frac{\bm{z}^{T} \ddot{\bm{z}}_{i}}{\|\bm{z}\| \cdot \|\ddot{\bm{z}}_{i}\|}.
\end{align}
Define \( S_{\max} = \max_{i} S_{i} \). If \( S_{\max} \) exceeds a certain threshold, it is concluded that the speaker is registered; the corresponding \( i^{\star} = \arg \max_{i} S_{i} \) is the index of the registered speaker.

\subsection{Network Learning Strategies}

For voiceprint extraction, previous research has focused on classification and clustering methods. Classification methods require the construction of a labeled dataset, selecting representative speech samples from various speakers, and recording multiple samples for each speech to build the voiceprint extractor. In contrast, clustering methods do not require multiple recordings from each speaker; they assume that each speech in the dataset corresponds to a different speaker, treating each data sample as an independent speaker. By minimizing the distance between intra-class samples and maximizing the distance between inter-class samples, feature extraction can be accomplished. This section will briefly introduce both methods.

\subsubsection{Specific Dataset + Cross-Entropy Classification Method}

Voiceprint extraction technology based on classification methods requires constructing a special dataset \( \{(\bm{x}_{n}, \bm{y}_{n})\} \), consisting of \( L \) different speakers, each providing multiple speech samples \( \bm{x}_{n} \). The dataset label \( \bm{y}_{n} \) is an \( L \times 1 \) vector, where \( \bm{y}_{n} \) has only one element equal to 1, while the others are 0, forming a typical one-hot vector.

By adding a fully connected network \( \mathcal{F}(\cdot) \) and a softmax classification network \( \mathcal{S}_{x}(\cdot) \) after the voiceprint extraction network, the network function during training can be represented as:
\begin{align}
    \hat{\bm{y}} 
    &= f\left(\bm{x}\right) \\
    &= \mathcal{S}_{x} \circ \mathcal{F} \circ g\left(\bm{x}\right).
\end{align} 
Here, \( \mathcal{F}(\cdot) \) is typically a one or two-layer fully connected network.

During training, the cross-entropy loss function can be used to complete the training of the network \( f(\cdot) \). For a given batch of data \( \{(\bm{x}_{n}, \bm{y}_{n}), \forall n \in \mathcal{I}_{m}\} \), the network's loss function can be expressed as:
\begin{align}
    \mathcal{J}_{1} &= - \sum_{n \in \mathcal{I}_{m}}  \bm{y}_{n}^{T} \ln \hat{\bm{y}}_{n}  
    = - \sum_{n \in \mathcal{I}_{m}}   \bm{y}_{n}^{T} \ln f\left(\bm{x}_{n}\right)  .
\end{align} 
Here, \( \mathcal{I}_{m} \) is the index set for the \( m \)-th batch of data.

For methods based on classification, a larger dataset with diverse speaker features typically yields better voiceprint feature extraction results. If the dataset is too small, the training stage and usage stage may not align in distribution, leading to domain mismatch issues; the resulting feature extraction function \( g(\cdot) \) may perform poorly when used for speaker registration and identification.

\subsubsection{ Contrastive Voiceprint Extraction}

The preceding section outlines a method for voiceprint feature extraction based on classification principles, suggesting that if features can effectively distinguish various categories in the existing dataset, they should also be able to differentiate similar data during usage. Besides the classification principle, clustering principles can also be leveraged to design voiceprint feature extractors.

The essence of clustering is to minimize the distance between intra-class samples while maximizing the distance between inter-class samples. In other words, after transforming the data \( \bm{x}_{n} \) into \( \bm{z}_{n} \) via the feature extraction function \( g(\cdot) \), if two features \( \bm{z}_{n} \) and \( \bm{z}_{i} \) originate from the same speaker, their distance should be minimized. Conversely, if they come from different speakers, the distance should be maximized\cite{bertinetto2016fully}. Therefore:
\begin{itemize}
    \item For two features \( \bm{z}_{n} \) and \( \bm{z}_{i} \) from the same speaker, the loss function under the definition of Euclidean distance is given by:

    \begin{align}
        \mathcal{J}_{+} \left(\bm{z}_{n}, \bm{z}_{i}\right) &= \left\|\bm{z}_{n} - \bm{z}_{i}\right\|_{2}^{2}.
    \end{align}

    By optimizing \( \min_{\theta_{g}} \mathcal{J}_{+}\left(\bm{z}_{n}, \bm{z}_{i}\right) \), we can gradually reduce the distance between \( \bm{z}_{n} \) and \( \bm{z}_{i} \).

    \item For two features \( \bm{z}_{n} \) and \( \bm{z}_{i} \) from different speakers, the goal is to minimize the loss function to maximize their distance:

    \begin{align}
        \mathcal{J}_{-}\left(\bm{z}_{n}, \bm{z}_{i}\right) &= -\left\|\bm{z}_{n} - \bm{z}_{i}\right\|_{2}^{2}.
    \end{align}

    It follows that minimizing \( \mathcal{J}_{-} \) will maximize the distance between \( \bm{z}_{n} \) and \( \bm{z}_{i} \).

    In practical applications, once the distance reaches a certain threshold, it becomes unnecessary to further increase the distance, which would excessively consume the function's degrees of freedom. Therefore, a more flexible approach is:

    \begin{align}
        \mathcal{J}_{-}\left(\bm{z}_{n}, \bm{z}_{i}\right) &= \max\left(0, \zeta_{0} - \left\|\bm{z}_{n} - \bm{z}_{i}\right\|_{2}^{2}\right).
    \end{align}

    Here, \( \zeta_{0} > 0 \) is a hyperparameter. When the distance \( \left\|\bm{z}_{n} - \bm{z}_{i}\right\|_{2} \) exceeds \( \sqrt{\zeta_{0}} \), we no longer include them in the loss function, allowing the network to focus on training the more challenging samples.
\end{itemize}
Based on these two points, the network's loss function can be expressed as:
\begin{align}
    \mathcal{J}_{2}\left(\bm{z}_{n}, \bm{z}_{i}\right) &= p_{n,i} \cdot \mathcal{J}_{+} \left(\bm{z}_{n}, \bm{z}_{i}\right) + (1-p_{n,i})\cdot \mathcal{J}_{-}\left(\bm{z}_{n}, \bm{z}_{i}\right) \\
    &= p_{n,i} \cdot \left\|\bm{z}_{n} - \bm{z}_{i}\right\|_{2}^{2} +\nonumber\\
    &~~~~~~ (1-p_{n,i})\cdot \max\left(0, \zeta_{0} - \left\|\bm{z}_{n} - \bm{z}_{i}\right\|_{2}^{2}\right).
\end{align}
Here, \( p_{n,i} \) indicates whether samples \( \bm{x}_{n} \) and \( \bm{x}_{i} \) come from the same speaker. If they are from the same speaker, \( p_{n,i} = 1 \); otherwise, \( p_{n,i} = 0 \).

In practice, when given a dataset \( \{(\bm{x}_{n}, \bm{y}_{n}), \forall n\} \), we can pair the samples to form a new, larger dataset \( \{(\bm{x}_{n}, \bm{x}_{i}, p_{n,i}), \forall n, i, n \neq i\} \). The labels in this dataset are no longer one-hot vectors \( \bm{y}_{n} \) but rather reflect whether the sample pair comes from the same speaker, represented by \( p_{n,i} \). This approach offers a direct benefit of expanding the dataset; if the original dataset contains \( N \) training samples, the pairing process results in \( N(N-1)/2 \) pairs, effectively increasing the dataset size by \( (N-1)/2 \).

However, this pairing can also introduce severe data imbalance, as pairs where \( p_{n,i} = 0 \) will significantly outnumber those where \( p_{n,i} = 1 \), which is a contradiction stemming from combinatorial pairing. To address this issue, the final loss function is often rewritten as:
\begin{align}
    \mathcal{J}_{2} = \sum_{(n,i) \in \mathcal{I}_{m}}& \beta \cdot p_{n,i} \cdot \left\|\bm{z}_{n} - \bm{z}_{i}\right\|_{2}^{2} + \nonumber\\
    &(1-p_{n,i})\cdot \max\left(0, \zeta_{0} - \left\|\bm{z}_{n} - \bm{z}_{i}\right\|_{2}^{2}\right).
\end{align}
Here, \( \beta \geq 1 \) is used to mitigate the data imbalance introduced by sample pairing.

\subsubsection{Triplet Contrastive Voiceprint Extraction}

A common method in network learning involves assigning a unique label to each sample and training the network to match these labels. The construction of the binary contrastive loss suggests that the loss function for network training can, in fact, be constructed using other samples.

Based on the findings of binary contrastive loss, we can assign a positive sample \( \bm{x}_{n+} \) and a negative sample \( \bm{x}_{n-} \) for each sample \( \bm{x}_{n} \) to construct its loss function. Positive samples refer to ``samples of the same type," while negative samples refer to ``samples of different types." For voiceprint extraction tasks, positive samples are those ``from the same speaker," while negative samples are ``from different speakers." It is worth noting that although positive samples come from the same speaker, they are not the same speech segment. In practice, if there is only one speech segment available, we can artificially split it into two segments.

After voiceprint extraction, each sample \( \bm{x}_{n} \) is transformed into voiceprint features \( \bm{z}_{n} = g(\bm{x}_{n}) \). With positive and negative samples defined, the principle of ``minimizing the distance between intra-class samples and maximizing the distance between inter-class samples
" can be equivalently stated as ``minimizing the distance between the sample feature \( \bm{z}_{n} \) and the positive sample feature \( \bm{z}_{n+} \), while maximizing the distance between the sample feature \( \bm{z}_{n} \) and the negative sample feature \( \bm{z}_{n-} \). Given the positive and negative samples, we can directly apply the binary loss function to obtain\cite{schroff2015facenet}:
\begin{align}
    \mathcal{J}_{2}(\bm{z}_{n},\bm{z}_{n+},\bm{z}_{n-}) &= \left\|\bm{z}_{n} - \bm{z}_{n+}\right\|_{2}^{2} + \nonumber\\
    &~~~~~\max\left(0, \zeta_{0} - \left\|\bm{z}_{n} - \bm{z}_{n-}\right\|_{2}^{2}\right).
\end{align}
Since the loss function for positive and negative samples shares a common sample \( \bm{x}_{n} \), there is a more compact representation of their loss function:
\begin{align}
    &\mathcal{J}_{3}(\bm{z}_{n},\bm{z}_{n+},\bm{z}_{n-})\nonumber \\ &= \max \left(0, \zeta_{0} - \left\|\bm{z}_{n} - \bm{z}_{n-}\right\|_{2}^{2} + \left\|\bm{z}_{n} - \bm{z}_{n+}\right\|_{2}^{2}\right) \\
    &= \max \left[0, \zeta_{0} - \left(D_{n-} - D_{n+}\right)\right],
\end{align}
where
\begin{align}
    D_{n+} &= \left\|\bm{z}_{n} - \bm{z}_{n+}\right\|_{2}^{2}, \\
    D_{n-} &= \left\|\bm{z}_{n} - \bm{z}_{n-}\right\|_{2}^{2}.
\end{align}
Here, \( D_{n+} \) and \( D_{n-} \) represent the distances between the sample feature \( \bm{z}_{n} \) and the positive sample feature \( \bm{z}_{n+} \), and the distance between the sample feature \( \bm{z}_{n} \) and the negative sample feature \( \bm{z}_{n-} \), respectively. By minimizing the triplet loss, we can ``increase the inter-class boundary and decrease the distance between intra-class samples", as explained in Section \ref{subsect-triple-cluster}.

In constructing the triplet contrastive loss, in addition to assigning positive and negative samples for each sample \( \bm{x}_{n} \), we can also allocate a set of positive samples and a set of negative samples for each sample in the original dataset. During training, a positive sample \( \bm{x}_{n+} \) can be sampled from the positive sample set; similarly, a negative sample \( \bm{x}_{n-} \) can be sampled from the negative sample set. Although this approach may seem cumbersome, it enhances the diversity among sample pairs, facilitating better optimization and learning by the network.

\subsubsection{Label-Free Auto-Clustering}

When designing voiceprint feature extraction, we ideally want a large labeled dataset to provide a feature extractor with better generalization performance. However, large datasets imply high costs, making them challenging to construct in practice. Considering that many public datasets exist across various fields, the only limitation is the lack of labels for specific tasks. A crucial question is whether we can effectively utilize existing datasets for developing voiceprint feature extraction without requiring labels.

Given an unlabeled dataset \( \{\bm{x}_{n}, \forall n\} \), we assume that each data sample corresponds to a unique user, meaning that each user's voice appears only once in the dataset. Under the contrastive loss training strategy, we need to construct positive and negative samples. Since each user appears only once, negative samples can be directly sampled from the dataset. Positive samples require splitting each data sample into two, resulting in two datasets \( \{\bm{x}_{n}, \forall n\} \) and \( \{\tilde{\bm{x}}_{n}, \forall n\} \). For each sample's loss function, we can characterize it using its positive sample and all negative samples in the same batch.

Of course, when given positive and negative sample pairs, we can directly employ the binary contrastive loss and triplet contrastive loss to construct the loss function. In this specific case, since each user appears only once within each batch of data, we can also utilize classification methods to construct the loss function. Given a batch of data \( \{\bm{x}_{n}, \forall n \in \mathcal{I}_{m}\} \) and \( \{\tilde{\bm{x}}_{n}, \forall n \in \mathcal{I}_{m}\} \), the loss function for sample \( \bm{x}_{n} \) can be expressed as:
\begin{align}
    J_{4}(\bm{z}_{n}) &= -\ln \frac{\expd{\bm{z}_{n}^{T} \tilde{\bm{z}}_{n} / \tau}}{\sum_{i \in \mathcal{I}_{m}} \expd{\bm{z}_{n}^{T} \tilde{\bm{z}}_{i} / \tau}}.
\end{align}
Here, the variable \( \tau \) can be set directly or learned from the data. If we treat \( \tilde{\bm{z}}_{i}, i=1,2,\ldots \) as the weight coefficients in the softmax function, the above expression can be represented as\footnote{At this point, \( \bm{y}_{n} \) is a one-hot vector, and \( \mathcal{S}_{x}(\cdot) \) is the softmax function.} \( -\bm{y}_{n}^{T}\ln \mathcal{S}_{x}(\bm{z}_{n})  \), which corresponds to the typical construction method of softmax combined with cross-entropy for loss functions.
For this batch of data, the overall loss function can be expressed as:
\begin{align}
    J_{4} &= -\sum_{n \in \mathcal{I}_{m}} \ln \frac{\expd{\bm{z}_{n}^{T} \tilde{\bm{z}}_{n} / \tau}}{\sum_{i \in \mathcal{I}_{m}} \expd{ \bm{z}_{n}^{T} \tilde{\bm{z}}_{i} / \tau}}.
\end{align}
Under the constraints \( \|\bm{z}_{n}\| = 1 \) and \( \|\tilde{\bm{z}}_{n}\| = 1 \), the final loss function for the network can be expressed as:
\begin{align}
    J_{4} &= -\sum_{n \in \mathcal{I}_{m}} \ln \frac{\expd{S_{\bm{z}_{n}, \tilde{\bm{z}}_{n}} / \tau}}{\sum_{i \in \mathcal{I}_{m}} \expd{ S_{\bm{z}_{n}, \tilde{\bm{z}}_{i}} / \tau}} \label{loss-SR-simCLR-batch},
\end{align}
where
\begin{align}
    S_{\bm{z}_{n}, \tilde{\bm{z}}_{i}} &= \frac{\bm{z}_{n}^{T} \tilde{\bm{z}}_{i}}{\|\bm{z}_{n}\| \cdot \|\tilde{\bm{z}}_{i}\|}.
\end{align}
This measures the cosine similarity between the sample features, i.e., the cosine of the angle between vectors \( \bm{z}_{n} \) and \( \bm{z}_{i} \); it assesses the similarity between sample features. Feature normalization has an additional potential benefit: during the voiceprint recognition process, we can directly compare the similarity between voiceprint features without needing to build an additional discriminator.

It is worth noting that in the original literature\cite{chen2020simplehin}, the loss function differs slightly; through simplification, it can be expressed as:
\begin{align}
    J_{5}(\bm{z}_{n}) &= -\ln \frac{\expd{S_{\bm{z}_{n}, \tilde{\bm{z}}_{n}} / \tau}}{\sum_{i \in \mathcal{I}_{m}} \expd{ S_{\bm{z}_{n}, \tilde{\bm{z}}_{i}} / \tau} + \sum_{j \in \mathcal{I}_{m}, j \neq n} \expd{ S_{\bm{z}_{n}, {\bm{z}}_{i}} / \tau}}.
\end{align}
This form corresponds to the SimCLR representation learning method, where SimCLR stands for ``A simple framework for contrastive learning of visual representations."
In subsequent literature\cite{he2020momentum,xia2021self}, the function form in (\ref{loss-SR-simCLR-batch}) gradually emerged.

However, in terms of the loss function representation, the SimCLR method constructs a loss function for a sample using a single positive sample pair and numerous negative sample pairs, with the number of negative sample pairs depending on the number of samples in each batch of data. From this perspective, it is entirely feasible to break the limitation that negative samples must come from the same batch of data, artificially constructing a dictionary to assist in designing the loss function corresponding to negative sample pairs. The MoCo\footnote{MoCo is an abbreviation for Momentum Contrast\cite{he2020momentum}.} representation learning method effectively utilizes this principle. The corresponding loss function can be expressed as:
\begin{align}
    \mathcal{J}_{6}(\bm{z}_{n}) &= -\sum_{n \in \mathcal{I}_{m}} \ln \frac{\expd{S_{\bm{z}_{n}, \tilde{\bm{z}}_{n+}} / \tau}}{\sum_{\tilde{\bm{z}}_{i} \in \mathbb{D}} \expd{ S_{\bm{z}_{n}, \tilde{\bm{z}}_{i}} / \tau}}.
\end{align}
Here, \( \mathbb{D} \) is a dictionary containing multiple negative samples, which is updated gradually during training; \( \tilde{\bm{z}}_{n+} \) is the feature corresponding to the positive sample of \( \bm{x}_{n} \). During the training process of the MoCo method, the extraction methods for voiceprint features \( \bm{z}_{n} \) and \( \tilde{\bm{z}}_{i} \) are not identical; they are obtained through two different networks:
\begin{align}
    \bm{z}_{n} &= g(\bm{x}_{n}), \\
    \tilde{\bm{z}}_{n} &= g^{\prime}(\tilde{\bm{x}}_{n}),
\end{align}
where the structure of network \( g^{\prime}(\cdot) \) is the same as that of network \( g(\cdot) \), but the parameters differ. Specifically, all parameters of network \( g^{\prime}(\cdot) \) are derived from a moving average of the parameters of network \( g(\cdot) \):
\begin{align}
    \theta_{g^{\prime}} \leftarrow \alpha \theta_{g^{\prime}} + (1-\alpha) \theta_{g},
\end{align}
where \( \alpha \in (0,1) \) is a smoothing coefficient, referred to in the original text as the momentum coefficient; this is also the source of the abbreviation ``Mo" in the MoCo method. Typically, the smoothing coefficient \( \alpha \) is close to 1, e.g., \( \alpha = 0.999 \). In practice, as the network trains, we only update the parameters of network \( g(\cdot) \) using the backpropagation error, i.e., \( \theta_{g} \); the parameters of network \( g^{\prime}(\cdot) \) are not updated using backpropagation.

\subsubsection{Robust Voiceprint Feature Extraction Methods}

During the voiceprint transformation process, the goal is to convert the same speaker's voice in different scenarios into the same voiceprint. Different scenarios can include at least three aspects: 1) varying lengths of speech signals; 2) different signal-to-noise ratios and noise backgrounds; 3) different room transfer functions. In other words, voiceprint feature extraction methods should be able to handle signals of various lengths while effectively removing the influences of different noise backgrounds and room transfer functions. This means that voiceprint extraction methods must exhibit a certain level of robustness against various disturbances and changes.

In traditional voiceprint extraction methods (e.g., i-vector\cite{dehak2010front}), robust methods, especially those addressing variations in room transfer functions, are often difficult to model and handle. However, under a data-driven approach, utilizing data augmentation techniques can make such issues quite intuitive. The overall principle is that if we can map the original signals and the augmented signals to the same voiceprint, then the voiceprint extraction method is robust against the disturbances introduced by data augmentation. For convenience, we denote the data augmentation operation as \( \hbar(\cdot) \), with the augmented sample expressed as:
\begin{align}
    \tilde{\bm{x}}_{n+} = \hbar(\bm{x}_{n}).
\end{align}
For example, in channel augmentation, we have \( \hbar(\bm{x}_{n}) = h(t) \ast \bm{x}_{n}(t) \), where \( \ast \) denotes convolution, and \( h(t) \) is the impulse response from the sound source to the microphone; if noise is added as an augmentation method, then \( \hbar(\bm{x}_{n}) = \bm{x}_{n} + \bm{v}_{n} \), where \( \bm{v}_{n} \) is the artificially added noise. Additionally, segments of events within the sample \( \bm{x}_{n} \) can be selectively removed for data augmentation.

After data augmentation, if we can minimize the distance between \( \bm{z}_{n} = g(\bm{x}_{n}) \) and \( \tilde{\bm{z}}_{n+} = g(\tilde{\bm{x}}_{n+}) \), the trained network will naturally be able to counteract the disturbances introduced during data augmentation. Data augmentation is one of the most common techniques used in robust voiceprint extraction methods.

\subsection{Typical Network Structures}

In the previous discussions regarding the construction of loss functions and network training strategies, we have not addressed the construction of the voiceprint feature extraction network \( g(\cdot) \). In the voiceprint feature extraction process, the input speech sample \( \bm{x}_{n} \) may vary in length, with each sample containing a different number of slices. However, after voiceprint extraction, the feature \( \bm{z}_{n} = g(\bm{x}_{n}) \) is expected to have a consistent length. Therefore, the network structure typically consists of three components: feature extraction, feature aggregation, and feature transformation. In this framework, the network \( g(\cdot) \) can typically be expressed as:
\begin{align}
    \bm{z} &= g(\bm{x}) \\
    &= \mathcal{F} \circ \mathcal{A} \circ \mathcal{C}(\bm{x}).
\end{align}
Here, \( \mathcal{F}(\cdot) \) is a fully connected network used for feature transformation; \( \mathcal{A}(\cdot) \) is an aggregation network for aggregating features corresponding to different time slices; and \( \mathcal{C}(\cdot) \) is similar to a convolutional network for extracting complex high-dimensional features.

\subsubsection{Time-Delay Neural Network Model}

Literature \cite{snyder2018x} presents a Time-Delay Neural Network (TDNN) model. The TDNN model's input is the MFCC coefficients, with each time slice \( \bm{x}_{n}(t) \) being a \( 24 \times 1 \) vector. Therefore, \( \bm{x}_{n} \) can be seen as a ``one-dimensional" dataset with 24 channels. If we extract the feature of any channel and plot its variation over time, we can obtain 24 curves.

In the TDNN model, the network \( g(\cdot) \) is defined as:
\begin{align}
    \bm{z}_{n} &= g(\bm{x}_{n}) \\
    &= \mathcal{F} \circ \mathcal{A} \circ \mathcal{C}_{5} \circ \mathcal{C}_{4} \circ \mathcal{C}_{3} \circ \mathcal{C}_{2} \circ \mathcal{C}_{1}(\bm{x}_{n}).
\end{align}
Here, \( \mathcal{C}_{i}(\cdot), i=1,2,3,4,5 \) are five layers of one-dimensional convolutional neural networks; \( \mathcal{A}(\cdot) \) is the aggregation network; and \( \mathcal{F}(\cdot) \) is the fully connected network.
\begin{itemize}
    \item The convolutional kernel width in \( \mathcal{C}_{1}(\cdot) \) is 5, and the output channel count is 512. Thus, when calculating the output at time \( t \), it requires inputs from five time slices: \( \{t-2, t-1, t, t+1, t+2\} \). Each time slice has a dimension of 24, so the equivalent dimension for the five time slices is \( 24 \times 5 = 120 \).
    \item The convolutional kernel width in \( \mathcal{C}_{2}(\cdot) \) is also 5, with an output channel count of 512. Unlike \( \mathcal{C}_{1}(\cdot) \), it does not use all five consecutive time slices; it only requires data from \( \{t-2, t, t+2\} \). The equivalent dimension for the three input time slices to \( \mathcal{C}_{2}(\cdot) \) is \( 512 \times 3 = 1536 \).
    \item The convolutional kernel width in \( \mathcal{C}_{3}(\cdot) \) is 7, with an output channel count remaining at 512. Although the width has increased, it only uses data from the first, last, and middle time slices: \( \{t-3, t, t+3\} \). Given that the output channel count from \( \mathcal{C}_{2}(\cdot) \) is 512, the equivalent dimension for the three input time slices to \( \mathcal{C}_{3}(\cdot) \) is \( 512 \times 3 = 1536 \).
    \item Both \( \mathcal{C}_{4}(\cdot) \) and \( \mathcal{C}_{5}(\cdot) \) have a convolutional kernel width of 1, focusing on merging channel information, with output channel counts of 512 and 1500, respectively.
    \item After passing through the five layers of convolutional networks, the original \( 24 \times T \) dimensional MFCC features are transformed into \( 1500 \times T \) dimensional features. The aggregation network \( \mathcal{A}(\cdot) \) then performs time-dimensional fusion by computing the mean and standard deviation, resulting in a \( 3000 \times 1 \) dimensional feature vector.
    \item Finally, the 3000-dimensional feature vector undergoes a nonlinear mapping through a fully connected network, yielding the final voiceprint feature \( \bm{z}_{n} \), which is a \( 512 \times 1 \) vector.
\end{itemize}
Thus, the process of transforming an input speech signal of variable length \( \bm{x}_{n} \) into a fixed-length feature \( \bm{z}_{n} \) is completed; the feature \( \bm{z}_{n} \) is known as the famous x-vector feature. The key to converting variable-length data into fixed-length features lies in the aggregation network \( \mathcal{A}(\cdot) \). Assuming the feature at time \( t \) after \( \mathcal{C}_{5}(\cdot) \) is \( \bm{x}_{n}^{(5)}(t) \), the mean and standard deviation are calculated as follows:
\begin{align}
    \bm{\mu} &= \frac{1}{T} \sum_{t=0}^{T-1} \bm{x}_{n}^{(5)}(t), \\
    \bm{\sigma} &= \sqrt{\frac{1}{T} \sum_{t=0}^{T-1} \left[\bm{x}_{n}^{(5)}(t) - \bm{\mu}\right] \odot \left[\bm{x}_{n}^{(5)}(t) - \bm{\mu}\right]}.
\end{align}
Here, \( \odot \) denotes element-wise multiplication, and the result remains a vector of the same dimension. Once the mean and standard deviation are obtained, the output of the aggregation network is:
\begin{align}
    \bm{x}_{n}^{(6)} = \mathcal{A} \left[\bm{x}_{n}^{(5)}(t), \forall t\right] = \left[\begin{array}{c}
    \bm{\mu} \\
    \bm{\sigma}
    \end{array}\right].
\end{align}
The TDNN model provides a concise aggregation approach. In practice, many aggregation methods are available, and researchers can utilize the statistical properties of \( \bm{x}_{n}^{(5)}(t) \) to fine-tune the data features.

\subsubsection{Residual Network Model}

Since each time slice \( \bm{x}_{n}(t) \) is a vector, the overall data \( \bm{x}_{n} \) can be viewed as an image. Literature \cite{wang2020data} treats the input data \( \bm{x}_{n} \) as an image and employs two-dimensional convolutions combined with a standard 34-layer residual network structure to construct a voiceprint feature extractor, referred to as r-vector\cite{he2016deep}. Within the framework of the residual network model, the voiceprint feature extractor can be described as:
\begin{align}
    \bm{z}_{n} &= g(\bm{x}_{n}) \\
    &= \mathcal{F} \circ \mathcal{A} \circ \mathcal{R}_{4} \circ \mathcal{R}_{3} \circ \mathcal{R}_{2} \circ \mathcal{R}_{1} \circ \mathcal{C}_{1}(\bm{x}_{n}),
\end{align}
where \( \mathcal{R}_{i}(\cdot), i=1,2,3,4 \) are residual blocks, \( \mathcal{C}_{1}(\cdot) \) is a convolutional neural network, \( \mathcal{A}(\cdot) \) is the aggregation network, and \( \mathcal{F}(\cdot) \) is the fully connected network. Each residual block consists of multiple sub-residual networks, specifically:
\begin{align}
    \mathcal{R}_{1}(\cdot) &= \mathcal{R}_{1,3} \circ \mathcal{R}_{1,2} \circ \mathcal{R}_{1,1}(\cdot), \\
    \mathcal{R}_{2}(\cdot) &= \mathcal{R}_{2,4} \circ \mathcal{R}_{2,3} \circ \mathcal{R}_{2,2} \circ \mathcal{R}_{2,1}(\cdot), \\
    \mathcal{R}_{3}(\cdot) &= \mathcal{R}_{3,6} \circ \mathcal{R}_{3,5} \circ \mathcal{R}_{3,4} \circ \mathcal{R}_{3,3} \circ \mathcal{R}_{3,2} \circ \mathcal{R}_{3,1}(\cdot), \\
    \mathcal{R}_{4}(\cdot) &= \mathcal{R}_{4,3} \circ \mathcal{R}_{4,2} \circ \mathcal{R}_{4,1}(\cdot).
\end{align}
Each sub-residual network consists of two layers of convolutional networks with kernels of size \( 3 \times 3 \). Assuming the input to \( \mathcal{R}_{i,j}(\bm{x}) \) is \( \bm{x}^{\prime} \), its output can be expressed as:
\begin{align}
    \mathcal{R}_{i,j}(\bm{x}^{\prime}) &= \bm{x}^{\prime} + \mathcal{C}^{(i,j)}_{2} \circ \mathcal{C}^{(i,j)}_{1}(\bm{x}^{\prime}).
\end{align}
Here, \( \mathcal{C}^{(i,j)}_{1} \) and \( \mathcal{C}^{(i,j)}_{2} \) are the stacked two-layer convolutional networks. The total number of layers in the feature extraction module is \( (3 + 6 + 4 + 3) \times 2 + 2 = 34 \) layers. Since the data dimensions remain unchanged within the same residual block, and downsampling is applied only after larger residual blocks \( \mathcal{R}_{i} \), we will provide a brief introduction to the inputs and outputs of each large residual block.
\begin{itemize}
    \item The input dimension of the network is \( 40 \times T \times 1 \). Unlike the TDNN model, where data is treated as a one-dimensional signal, here it is treated as an image; thus, the channel count is 1, and the data dimension is \( 40 \times T \).
    \item In \( \mathcal{C}_{1}(\cdot) \), the convolutional kernel width is \( 3 \times 3 \), with the convolution operating in both time and feature dimensions, and the output channel count is 32. After passing through \( \mathcal{C}_{1}(\cdot) \), the feature dimension becomes \( 40 \times T \times 32 \).
    \item The first residual block \( \mathcal{R}_{1}(\cdot) \) does not compress the data; its output channel count remains 32, so the output is still \( 40 \times T \times 32 \).
    \item The second residual block \( \mathcal{R}_{2}(\cdot) \) performs a 2-fold downsampling, doubling the output channel count to 64, resulting in an output dimension of \( 20 \times \frac{T}{2} \times 64 \).
    \item The third residual block \( \mathcal{R}_{3}(\cdot) \) also performs 2-fold downsampling, doubling the output channel count to 128, yielding an output dimension of \( 10 \times \frac{T}{4} \times 128 \).
    \item The fourth residual block \( \mathcal{R}_{4}(\cdot) \) performs another 2-fold downsampling, doubling the output channel count to 256, resulting in an output dimension of \( 5 \times \frac{T}{8} \times 256 \).
    \item The aggregation layer \( \mathcal{A}(\cdot) \) performs time-dimensional fusion, calculating the mean and standard deviation of the feature points, ultimately yielding a feature representation of \( 5 \times 2 \times 256 \). The final step involves flattening all feature points to obtain a \( 2560 \times 1 \) vector, which is the output of the aggregation layer.
    \item After aggregation, a fully connected neural network is used to finally transform the \( 2560 \times 1 \) feature vector into a \( 256 \times 1 \) feature vector, which represents the final voiceprint feature.
\end{itemize}
Thus, the process of converting a variable-length speech signal \( \bm{x}_{n} \) into a fixed-length feature \( \bm{z}_{n} \) is achieved; the feature \( \bm{z}_{n} \) is the final voiceprint feature.

\subsubsection{Direct Extraction of Voiceprint Features from Waveforms}

Literature \cite{xu2021target} introduces a method for directly extracting voiceprint features from speech waveforms. The overall network structure is as follows:
\begin{align}
    \bm{z}_{n} &= g(\bm{x}_{n}) \\
    &= \mathcal{A} \circ R_{3} \circ R_{2} \circ R_{1} \circ \mathcal{C}_{2} \circ \mathcal{C}_{1}(\bm{x}_{n}).
\end{align}
Here, \( \mathcal{C}_{1}(\cdot) \) consists of three groups of convolutional networks with different kernel sizes, each having 256 output channels. If we denote the three groups of convolutional networks as \( \mathcal{C}_{1,\mathrm{short}}, \mathcal{C}_{1,\mathrm{middle}} \), and \( \mathcal{C}_{1,\mathrm{long}} \), the output of \( \mathcal{C}_{1}(\cdot) \) can be expressed as:
\begin{align}
    \mathcal{C}_{1}(\bm{x}_{n}) &= \mathcal{C}_{1,\mathrm{short}}(\bm{x}_{n}) \ddag \mathcal{C}_{1,\mathrm{middle}}(\bm{x}_{n}) \ddag \mathcal{C}_{1,\mathrm{long}}(\bm{x}_{n}).
\end{align}
Here, \( \mathcal{C}_{1,\mathrm{short}}(\cdot), \mathcal{C}_{1,\mathrm{middle}}(\cdot) \), and \( \mathcal{C}_{1,\mathrm{long}}(\cdot) \) have convolutional kernel widths of 2.5 ms, 10 ms, and 20 ms, respectively, which correspond to 40, 160, and 320 points at a 16 kHz sampling rate. Each convolution effectively filters the data; after three groups of convolutions, the total number of channels becomes \( 256 \times 3 = 768 \). The next step involves downsampling the \( T \times 768 \) dimensional data by selecting one sample every 2.5 ms; under a 16 kHz sampling rate, this results in one sample being taken every 20 points, yielding data of dimension \( \frac{T}{2.5 \text{ ms}} \times 768 \). This data is then passed to \( \mathcal{C}_{2}(\cdot) \), which is a convolutional network with a kernel width of 1, effectively merging channels and transforming the input dimension from \( \frac{T}{2.5 \text{ ms}} \times 768 \) into \( \frac{T}{2.5 \text{ ms}} \times 256 \) features before feeding into the residual network \( \mathcal{R}_{1}(\cdot) \). 

The structures of the three residual network modules are identical; let the input to the residual block be \( \bm{x}^{\prime}_{n} \), introducing an intermediate variable \( \bm{x}^{\prime\prime}_{n} \), which can be expressed as:
\begin{align}
    \bm{x}_{n}^{\prime\prime} &= \bm{x}^{\prime} + \mathcal{N}_{\mathrm{batch}} \circ \mathcal{C}_{2}^{\prime} \circ \sigma \circ \mathcal{N}_{\mathrm{batch}} \circ \mathcal{C}_{1}^{\prime}(\bm{x}^{\prime}), \\
    \mathcal{R}(\bm{x}_{n}^{\prime}) &= \mathcal{C}_{3}^{\prime} \circ \sigma \left(\bm{x}_{n}^{\prime\prime}\right).
\end{align}
Here, \( \mathcal{C}_{2}^{\prime}(\cdot) \) and \( \mathcal{C}_{1}^{\prime}(\cdot) \) are both convolutional networks with a kernel width of 1, \( \mathcal{N}_{\mathrm{batch}}(\cdot) \) is the batch normalization layer, and \( \sigma(\cdot) \) is the nonlinear activation layer; the convolutional network \( \mathcal{C}_{3}^{\prime}(\cdot) \) has a kernel width of 3. It is important to note that the residual network does not apply the nonlinear activation before summation; instead, it applies nonlinear activation and downsampling after summation. Finally, the network aggregation layer \( \mathcal{A}(\cdot) \) directly applies a linear transformation to the input and computes the mean to obtain the voiceprint features for the given data.

\section{Approaches to Noise Reduction}
 
The goal of noise reduction is to remove background noise from the observed signal, preserving the clean source signal. The corresponding English term is Noise Reduction. While speech enhancement encompasses a broader scope, focusing not only on enhancing speech from observed signals but also on removing background noise, the term is often used interchangeably with Speech Enhancement in the literature. Essentially, noise reduction can be viewed as a signal estimation problem, typically accomplished in three steps: 1) transforming the time-domain observed signal to a special domain; 2) retaining components with a high signal-to-noise ratio (SNR) and removing those with a low SNR; 3) transforming the denoised signal back to the time domain. This section covers the following topics.\emph{ 
1) Signal analysis, filtering, and reconstruction methods;
2) Training strategies and loss functions;
3) Time-domain neural network structures;
4) Frequency-domain neural network structures.}

\subsection{Signal Analysis, Filtering, and Reconstruction}

Microphone sensors typically capture the variation of sound pressure at their location over time. In real acoustic environments, various noise sources, in addition to the desired source, affect the sound pressure changes at the microphone's position. Thus, for the time-domain observed signal \( x(t) \), it usually contains both the desired source signal \( s(t) \) and background noise \( v(t) \):
\begin{align}
    x(t) = s(t) + v(t).
\end{align}
Since speech signals exhibit short-term stationary characteristics, they can be considered stationary over short time intervals. Therefore, it is common to apply windowing and framing to the original time-domain signal. Assuming each frame length is \( L_{w} \) and the frame shift is \( L_{s} \), the \( i \)-th frame can be represented as:
\begin{align}
    x_{i}(t) = x(t) \psi(t - iL_{s}), \quad \forall t,
\end{align}
where \( \psi(t) \) for \( t=0,1,2,\ldots,L_{w}-1 \) is a window of length \( L_{w} \). To avoid information loss, it is necessary to reconstruct the original observed signal from \( x_{i}(t) \), meaning that \( \sum_{i} x_{i}(t) \) must equal \( x(t) \). Thus, the choice of the window function must satisfy:
\begin{align}
    x(t) = \sum_{i=-\infty}^{\infty} x_{i}(t), \quad \forall t \label{overlap-add}
\end{align}
This implies:
\begin{align}
    \sum_{i=-\infty}^{\infty} \psi(t - iL_{s}) = 1, \quad \forall t.
\end{align}
This equation represents a system with infinite equations. In fact, under the condition that \( L_{w} \) is an integer multiple of \( L_{s} \), denoted as \( L_{w} = Q L_{s} \), this infinite number of equations can be condensed into \( L_{w} \) equations. The original problem can be reduced to \( \bm{A}\bm{\psi} = \bm{1} \), where \( \bm{A} \) is an \( L_{s} \times L_{w} \) matrix, \( \bm{\psi} \) is a \( L_{w} \times 1 \) vector of window coefficients, and \( \bm{1} \) is a vector with all elements equal to 1.

In summary, each frame signal \( x_{i}(t) \) is a fixed-length signal, with a length equal to the window length \( L_{w} \); adjacent frames (i.e., \( x_{i}(t) \) and \( x_{i+1}(t) \)) often overlap. The more overlap there is, the smaller the frame shift \( L_{s} \), which leads to fewer constraints on the window function \( L_{w} \). If there is no overlap, i.e., \( L_{s} = L_{w} \), the matrix \( \bm{A} \) in the constraint \( \bm{A}\bm{\psi} = \bm{1} \) becomes an identity matrix, and the window function \( \bm{\psi} \) can only be a rectangular window; if the frame shift \( L_{s} = 1 \), then \( \bm{A} = \bm{1}^{T} \), and any window function that normalizes to satisfy \( \bm{1}^{T}\bm{\psi} = 1 \) can be used for framing.

For convenience, we temporarily represent the windowed frame signal as a vector:
\begin{align}
    \bm{x}_{i} = \left[ \begin{array}{c}
    x(iL_{s}) \psi(0) \\
    x(iL_{s}+1) \psi(1) \\
    \vdots \\
    x(iL_{s}+L_{w}-1) \psi(L_{w}-1)
    \end{array} \right].
\end{align}
Clearly, \( \bm{x}_{i} \) is a \( L_{w} \times 1 \) vector. The index \( i \) denotes the frame number, which corresponds to the time slice concept discussed in previous sections. Therefore, in subsequent uses of the index, we will denote the time slice index as \( t \), and represent \( \bm{x}_{i} \) as \( \bm{x}(t) \). 

\subsubsection{Short-Time Fourier Transform  }

For each time slice, the short-time Fourier transform method utilizes the processes of signal analysis, filtering, and reconstruction to achieve denoising. For convenience, we first define an \( L_{w} \times L_{w} \) Fourier transform matrix \( \bm{W} \), where the \( (i,j) \)-th element is given by:
\begin{align}
    \left[\bm{W}\right]_{i,j} = \expd{-\jmath \frac{ij}{L_{w}} 2\pi}, \quad i,j=0,1,2,\ldots,L_{w}-1.
\end{align}
Correspondingly, the inverse Fourier transform matrix can be expressed as \( \frac{1}{L_{w}} \bm{W}^{\ast} \), where the superscript \( \ast \) denotes the complex conjugate. It can be verified that \( \bm{W}^{T} = \bm{W} \) and \( \frac{1}{L_{w}} \bm{W}^{\ast} \bm{W} = \bm{I} \).

Given a data sample \( \bm{x} = \left[\bm{x}(0)~\bm{x}(1)~\cdots~\bm{x}(T-1)\right] \), the \( t \)-th time slice \( \bm{x}(t) = \bm{p}_{t} \). Utilizing the Fourier transform, the time-domain signal is transformed into the frequency domain:
\begin{align}
    \overleftarrow{\bm{x}}(t) &= \left[\begin{array}{cccc}
    X(\omega_{0},t) & X(\omega_{1},t) & \cdots & X(\omega_{L_{w}-1},t)
    \end{array}\right]^{T} \\
    &= \bm{W}^{T} \bm{x}(t).
\end{align}
Here, \( X(\omega_{k},t) \) represents the value of the signal at the \( k \)-th frequency band for the \( t \)-th time slice, while \( \overleftarrow{\bm{x}}(t) \) represents the frequency domain slice signal. Within the framework of the short-time Fourier transform, the basic principle of filtering involves selectively retaining and removing components in the frequency domain, keeping frequency bands with high SNR while discarding those with low SNR. Thus, the frequency domain denoising can be expressed as:
\begin{align}
    \overleftarrow{\hat{\bm{s}}}(t) = \bm{h}(t) \odot \overleftarrow{\bm{x}}(t),
\end{align}
where 
\begin{align}
    \bm{h}(t) = \left[\begin{array}{cccc}
    H(\omega_{0},t) & H(\omega_{1},t) & \cdots & H(\omega_{L_{w}-1},t)
    \end{array}\right]^{T}
\end{align}
is the frequency domain filter. In a data-driven denoising framework, the coefficients of the filter \( H(\omega, t) \) for each frequency band are estimated by the corresponding neural network. Since the SNR of speech signals varies across time slices and frequency bands, the filter coefficients \( H(\omega, t) \) naturally differ across time slices and frequency bands, making them functions of time and frequency. As we have not introduced concepts such as time-domain filtering frameworks and filter coefficients, there is no need to denote \( \bm{h}(t) \) with the \( \overleftarrow{(\cdot)} \) notation to indicate that it is a frequency domain filter coefficient.

Since \( \overleftarrow{\hat{\bm{s}}}(t) \) is an estimate of the desired source signal in frequency domain form, we can ultimately obtain the time-domain slice estimate of the source signal \( \hat{\bm{s}}(t) \) using the inverse Fourier transform:
\begin{align}
    \hat{\bm{s}}(t) &= \frac{1}{L_{w}} \bm{W}^{\ast} \overleftarrow{\hat{\bm{s}}}(t) \\
    &= \frac{1}{L_{w}} \bm{W}^{\ast} \left[\bm{h}(t) \odot \overleftarrow{\bm{x}}(t)\right] \\
    &= \frac{1}{L_{w}} \bm{W}^{\ast} \mathrm{diag}\left[\bm{h}(t)\right] \overleftarrow{\bm{x}}(t) \\
    &= \frac{1}{L_{w}} \bm{W}^{\ast} \mathrm{diag}\left[\bm{h}(t)\right] \bm{W}^{T} \bm{x}(t).
\end{align}
The operation here indicates that the average output of the signal over the time slices is influenced by the filter applied to the observed signal. When ignoring the filtering for denoising, i.e., \( H(\omega, t) = 1, \forall \omega, t \), it can be verified that \( \hat{\bm{s}}(t) = \bm{x}(t) \). If we denote the \( k \)-th column of the Fourier transform matrix \( \bm{W} \) as \( \bm{w}_{k} \), we can also represent the processes of signal analysis, filtering, and reconstruction as:
\begin{align}
    \hat{\bm{s}}(t) &= \sum_{k=0}^{L_{w}-1} \bm{w}_{k}^{\ast} \cdot \left[ H(\omega_{k},t) \bm{w}_{k}^{T} \bm{x}(t)\right] \\
    &= \sum_{k=0}^{L_{w}-1} \bm{w}_{k}^{\ast} \cdot \left[ H(\omega_{k},t) X(\omega_{k},t)\right].
\end{align}
In this representation, \( X(\omega_{k},t) = \bm{w}_{k}^{T} \bm{x}(t) \) denotes the process of signal analysis, \( H(\omega_{k},t) X(\omega_{k},t) \) represents the process of signal filtering, and the final summation \( \sum_{k} \bm{w}_{k}^{\ast} \cdot [\cdot] \) signifies the reconstruction process according to the basis functions \( \bm{w}_{k}^{\ast} \).

Ultimately, each time slice only restores the estimated source signal for that specific time slice. To obtain a complete time series signal, the signals from different slices must be combined using the equation (\ref{overlap-add}). The overlapping addition operation is analogous to the overlapping framing operation and will not be discussed further here.

\subsubsection{Filtering Principle and Wiener Filter}

The purpose of filtering is to remove background noise from the observed signal. Within the framework of signal analysis, filtering, and reconstruction, the basic filtering mode involves multiplying the observed signal \( X(\omega, t) \) by the coefficient \( H(\omega,t) \):
\begin{align}
    \hat{S}(\omega,t) 
    &= H(\omega, t) X(\omega, t) \\
    &= H(\omega, t) S(\omega, t) + H(\omega, t) V(\omega, t).
\end{align}
In this expression, the first term \( H(\omega, t) S(\omega, t) \) represents the alteration of the desired signal due to filtering, while the second term \( H(\omega, t) V(\omega, t) \) indicates the alteration of noise by the filter. This shows that while the filter removes noise, it will inevitably attenuate the corresponding desired signal at the same time-frequency point. Therefore, designing an effective filter \( H(\omega,t) \) becomes crucial.

To design the optimal filter \( H(\omega, t) \), we need to define the error signal \( \varepsilon(\omega,t) = S(\omega, t) - H(\omega, t) X(\omega, t) \). From a statistical perspective, we can define the mean squared error (MSE) as follows:
\begin{align}
    \mathcal{J}_{\mathrm{MSE}} &= \mathbb{E}\left[\left|\varepsilon(\omega,t)\right|^2\right] \\
    &= \mathbb{E}\left[\left| S(\omega, t) - H(\omega, t) X(\omega, t)\right|^2\right] \\
    &= \left| H(\omega, t)\right|^2 \phi_{XX}(\omega,t) - \nonumber\\
    &~~~ H(\omega, t) \phi_{XS}(\omega, t) - H^{\ast}(\omega, t)\phi_{XS}^{\ast}(\omega, t) + \phi_{SS}(\omega,t)\nonumber \\
    &= \left| H(\omega, t)\right|^2 \phi_{XX}(\omega,t) - \nonumber\\
    &~~~ H(\omega, t) \phi_{SS}(\omega, t) - H^{\ast}(\omega, t)\phi_{SS}^{\ast}(\omega, t) + \phi_{SS}(\omega,t),
\end{align}
where
\begin{align}
    \phi_{XX}(\omega,t) &= \mathbb{E}\left[\left|X(\omega,t)\right|^2\right], \\
    \phi_{SS}(\omega,t) &= \mathbb{E}\left[\left|S(\omega,t)\right|^2\right], \\
    \phi_{XS}(\omega, t) &= \mathbb{E}\left[X(\omega,t)S^{\ast}(\omega,t)\right] \\
    &= \mathbb{E}\left\{\left[S(\omega,t) + V(\omega,t)\right]S^{\ast}(\omega,t)\right\} \\
    &= \phi_{SS}(\omega,t).
\end{align}
In the derivation of \( \phi_{XS}(\omega, t) \), we assume that \( \mathbb{E}\left[V(\omega,t)S^{\ast}(\omega,t)\right]=0 \), indicating that the desired signal and noise are uncorrelated. Under this assumption, the variance among the observed signal, desired signal, and noise satisfies \( \phi_{XX}(\omega,t) = \phi_{SS}(\omega, t) + \phi_{VV}(\omega,t) \). By taking the derivative of \( \mathcal{J}_{\mathrm{MSE}} \) with respect to \( H^{\ast}(\omega, t) \) and setting the result to zero, we can derive:
\begin{align}
    H(\omega, t) &= \frac{\phi_{SS}(\omega,t)}{\Phi_{XX}(\omega,t)} \\
    &= \frac{\phi_{SS}(\omega,t)}{\phi_{SS}(\omega,t) + \phi_{VV}(\omega,t)} \\
    &= \frac{\mathrm{iSNR}(\omega,t)}{1 + \mathrm{iSNR}(\omega,t)},
\end{align}
where \( \mathrm{iSNR}(\omega,t) = \phi_{SS}(\omega, t) / \phi_{VV}(\omega, t) \) represents the SNR at the current time-frequency point. Clearly, under the minimum mean square error criterion, the optimal filter coefficients are real numbers between 0 and 1. The term \( H(\omega, t) \) represents the Wiener gain in classical time-frequency filtering methods.

When \( \mathrm{iSNR}(\omega,t) \gg 1 \), \( H(\omega,t) \) approaches 1; when \( \mathrm{iSNR}(\omega,t) \ll 1 \), \( H(\omega,t) \) approaches the input SNR, tending toward zero. Thus, the optimal filter \( H(\omega, t) \) effectively acts as a gate function: it selectively retains time-frequency points with high SNR while removing those with low SNR to achieve noise reduction. When visualizing the values of \( H(\omega, t) \) over time and frequency, it resembles a ``mask" overlaying a time-frequency spectrogram to obscure points with lower SNR. Consequently, in deep learning-related papers, it is more common to refer to this as a mask rather than Wiener gain.

During network training, the datasets related to denoising are typically constructed artificially. In the construction process, the original signals \( S(\omega, t) \) and noise \( V(\omega, t) \) are known; single-frame samples can replace expected values, meaning  \( \phi_{SS}(\omega,t) = \left|S(\omega, t)\right|^2 \) and \( \phi_{VV}(\omega,t) = \left|V(\omega, t)\right|^2 \). The corresponding optimal gain can be expressed as:
\begin{align}
    H(\omega, t) = \frac{\left|S(\omega, t)\right|^2}{\left|S(\omega, t)\right|^2 + \left|V(\omega, t)\right|^2} \label{H-wiener-label}.
\end{align}
If the optimal gain \( H(\omega, t) \) is the target output of the network, then the labels \( \bm{y}(t) \) are constructed using the equation above, resulting in the following label vector:
\begin{align}
    \bm{y}(t) &= \bm{h}(t)\\
    & = \left[\begin{array}{cccc}
    H(\omega_{0},t) & H(\omega_{1},t) & \cdots & H(\omega_{L_{w}-1},t)
    \end{array}\right]^{T}.
\end{align}
Here, \( H(\omega_{k},t) \) is computed using equation (\ref{H-wiener-label}) based on the instantaneous variances/powers of the expected and noise signals. Once the filter labels are calculated for all time slices, they can be combined to yield the overall label for each speech signal:
\begin{align}
    \bm{y} = \left[\bm{y}(0)~\bm{y}(1)~\cdots ~\bm{y}(T-1)\right],
\end{align}
which is now a matrix of size $L_{w}\times T$.

\subsubsection{Data-Driven Signal Analysis and Reconstruction}

In addition to using Fourier transforms and their inverses, neural networks can be trained to learn transformations and inversions directly from data. Assume there exist matrices \( \tilde{\bm{U}} \) and \( \bm{U} \), where the \( k \)-th columns are \( \tilde{\bm{u}}_{k} \) and \( \bm{u}_{k} \); \( \tilde{\bm{U}} \) is used for signal reconstruction, while \( \bm{U} \) is used for signal analysis, both being \( L_{w} \times K \) matrices. Analogous to the Fourier transform, signal analysis can be expressed as:
\begin{align}
    \overleftarrow{\bm{x}}(t) = \bm{U}^{T}\bm{x}(t).
\end{align}
After signal analysis, the original signal slice \( \bm{x}(t) \) is transformed into a \( K \times 1 \) vector. Unlike Fourier analysis, all elements of the matrix \( \bm{U} \) are real numbers, and the transformed signal \( \overleftarrow{\bm{x}}(t) \) is also a real signal. Analogous to short-time Fourier transform analysis and reconstruction, the estimates of source signals in the transform domain and time domain are given by:
\begin{align}
    \overleftarrow{\hat{\bm{s}}}(t) &= \bm{h}(t) \odot \overleftarrow{\bm{x}}(t), \\
    \hat{\bm{s}}(t) &= \tilde{\bm{U}}\overleftarrow{\hat{\bm{s}}}(t) \\&= \tilde{\bm{U}} \left[\bm{h}(t) \odot \overleftarrow{\bm{x}}(t)\right] \\
    &= \tilde{\bm{U}} \mathrm{diag}\left[\bm{h}(t)\right] \bm{U}^{T} \bm{x}(t) \\
    &= \sum_{k=0}^{K-1} \tilde{\bm{u}}_{k} \left[ H_{k}(t) \cdot \bm{u}_{k}^{T} \bm{x}(t)\right].
\end{align}
Here, \( H_{k}(t) \) is the \( k \)-th element of the vector \( \bm{h}(t) \). When all elements of \( \bm{h}(t) \) are equal to 1, if \( \tilde{\bm{U}} \bm{U}^{T} = \bm{I} \), the original signal can be perfectly restored through analysis and reconstruction. In practice, the signal analysis process sometimes introduces a nonlinear mapping \( \sigma(\cdot) \) to eliminate negative values after transformation\cite{luo2019conv}; thus, the framework for signal analysis becomes:
\begin{align}
    \overleftarrow{\bm{x}}(t) = \sigma\left[ \bm{U}^{T}\bm{x}(t)\right].
\end{align}
To prevent information loss during the analysis process, it is usually required that \( K > L_{w} \). From the filtering perspective, it is indeed unnecessary to introduce nonlinear mappings during the signal analysis process.

During actual training, both the reconstruction matrix \( \tilde{\bm{U}} \) and the analysis matrix \( \bm{U} \) are learned from the training data using the loss function, without needing to construct them based on expert's experience.

\subsection{Loss Function Construction}

\subsubsection{Distance Between Filters}

In the process of enhancing signals in the short-time Fourier transform domain, the optimal Wiener gain is a real number between 0 and 1. Assuming the neural network's output is \( \hat{\bm{h}}(t) \), we can construct the loss function for the neural network using binary cross-entropy:
\begin{align}
    \mathcal{J}_{1} &= -\sum_{t} \left\{ \bm{h}^{T}(t) \ln \hat{\bm{h}}(t) + \left[1 - \bm{h}^{T}(t)\right] \ln \left[1 - \hat{\bm{h}}(t)\right]\right\}.
\end{align}
In this case, the \( k \)-th element of the vector \( \hat{\bm{h}}(t) \) approximates the optimal Wiener gain for the \( k \)-th frequency band.

\subsubsection{Distance Between  Spectral  }

Since the optimal Wiener gain only adjusts the amplitude of the observed signal without altering its phase, it can be viewed as filtering to bring the amplitude of the observed signal closer to that of the desired source signal. Thus, from this perspective, the loss function for the neural network can be defined as:
\begin{align}
    \mathcal{J}_{2} &= \sum_{t} \sum_{k} \left[ \hat{H}(\omega_{k},t) |X(\omega_{k},t)| - |S(\omega_{k},t)|\right]^{2} \\
    &= \sum_{t} \left\|\hat{\bm{h}}(t) \odot |\bm{y}(t)| - |\bm{s}(t)|\right\|^{2},
\end{align}
where \( |\bm{y}| \) denotes the absolute value of each element in the vector \( \bm{y} \). Within the framework of short-time Fourier transform analysis, processing, and reconstruction, since the expected signal and noise are known at each time-frequency point during training, the network output mask is initially approximated to the ideal mask, and then the network is fine-tuned based on the differences in spectra.

\subsubsection{Distance Between Waveforms}

If we aim to approximate signals from the waveform perspective, the most straightforward approach is to minimize the distance between \( \bm{s}(t) \) and \( \hat{\bm{s}}(t) \), such as \( \left\|\bm{s}(t) - \hat{\bm{s}}(t)\right\|^2 \). However, during the processing of speech signals, if only amplitude changes occur in the waveform, it is generally not considered a deviation from the expected result. For example, for a non-zero constant \( c \), if \( \hat{\bm{s}}(t) = c \cdot \bm{s}(t) \), we consider that the distance between \( \bm{s}(t) \) and \( \hat{\bm{s}}(t) \) is zero. To eliminate the impact of scale, a degree of freedom \( \alpha \) is artificially introduced in the objective function, rewriting the expected signal \( \bm{s}(t) \) as \( \alpha \bm{s}(t) \). The distance between the network estimate and the expected signal then becomes \( \left\|\alpha \bm{s}(t) - \hat{\bm{s}}(t)\right\|^2 \). Treating this distance as the instantaneous power of noise, we can derive the loss function:
\begin{align}
    \mathcal{J}_{3} &= -\sum_{t} 10 \log \left(\frac{\left\|\alpha_{t} \bm{s}(t)\right\|^{2}}{\left\|\alpha_{t} \bm{s}(t) - \hat{\bm{s}}(t)\right\|^{2}}\right) \label{sI-SDR-org} \\
    &= -\sum_{t} \frac{\rho_{\bm{s}\hat{\bm{s}}}^{2}(t)}{1 - \rho_{\bm{s}\hat{\bm{s}}}^{2}(t)} \label{sI-SDR},
\end{align}
where,
\begin{align}
    \alpha_{t} &= \arg \min_{\alpha} \left\|\alpha \bm{s}(t) - \hat{\bm{s}}(t)\right\|^{2} \\
    &= \frac{\hat{\bm{s}}^{T}(t) \bm{s}(t)}{\left\|\bm{s}(t)\right\|^2} \label{sI-SDR-alpha}, \\
    \rho_{\bm{s}\hat{\bm{s}}}(t) &= \frac{\hat{\bm{s}}^{T}(t) \bm{s}(t)}{\left\|\bm{s}(t)\right\| \cdot \left\|\hat{\bm{s}}(t)\right\|} \label{sI-SDR-rho}.
\end{align}
The simplification of equation (\ref{sI-SDR}) relies on the optimal \( \alpha_{t} \) from equation (\ref{sI-SDR-alpha}) and the definition of the coefficient \( \rho_{\bm{s}\hat{\bm{s}}}(t) \) from equation (\ref{sI-SDR-rho}). From the perspective of traditional signal processing, \( \rho_{\bm{s}\hat{\bm{s}}}(t) \) can be seen as the correlation coefficient between zero-mean random signals \( \bm{s}(t) \) and \( \hat{\bm{s}}(t) \). Therefore, minimizing the original loss function is equivalent to maximizing the correlation coefficient between the two vectors \( \bm{s}(t) \) and \( \hat{\bm{s}}(t) \)\cite{cohen2009pearson}.

The summation term in equation (\ref{sI-SDR-org}) is commonly referred to in the literature as the Scale-Invariant Signal-to-Distortion Ratio (SI-SDR)\cite{le2019sdr}.

\subsection{Typical Network Framework}

\subsubsection{Fully Connected Network with Multi-Frame Concatenation}

For denoising tasks in the short-time Fourier transform domain, the simplest approach is to directly input a single frame of the spectrum into the network, allowing the network to learn the optimal Wiener filter coefficients for each frequency point. This method requires the network to infer which frequency points have high SNR and which have low SNR based on the distribution of the signals across different frequency bands within a single frame. Given the strong structural information present in the time-frequency spectrogram between adjacent frames/slices, a better approach is to concatenate multiple frames, thereby obtaining a larger input dataset for the network to automatically capture high-SNR time-frequency points from multiple frames\cite{han2015learning}. 

Specifically, for the time slice at time \( t \), we first perform a Fourier transform to obtain \( \overleftarrow{\bm{x}}(t) \), and then concatenate the results from the Fourier transform. The method provided in literature \cite{han2015learning} involves concatenating the logarithmic magnitude spectra as follows:
\begin{align}
    \tilde{\bm{x}}(t) = \left[\begin{array}{c}
    \log \left| \overleftarrow{\bm{x}}(t-Q)\right| \\
    \vdots \\
    \log \left| \overleftarrow{\bm{x}}(t)\right| \\
    \vdots \\
    \log \left| \overleftarrow{\bm{x}}(t+Q)\right|
    \end{array}\right].
\end{align}
This concatenation of the logarithmic magnitude spectra from \( 2Q \) adjacent frames results in a vector of length \( (2Q + 1)L_{w}/2 \). The vector is then fed into a neural network to predict the optimal Wiener gain \( \hat{\bm{h}}(t) \) for all frequency bands at the current time slice. The original paper used a sampling rate of 16 kHz, with a window length of 20 ms and a frame shift of 10 ms. Each window contains \( 16 \times 20 = 320 \) samples. After the Fourier transform, the negative frequency axis is discarded, yielding 161 valid frequency points. By taking five frames on either side for concatenation (i.e., \( Q = 5 \)), the length of \( \tilde{\bm{x}}(t) \) becomes \( (2 \times 5 + 1) \times 161 = 1771 \). Ultimately, each time slice corresponds to a vector of length 1771, which serves as the input to the neural network.

Given that the data dimensions are not large, a three-layer fully connected network suffices. The resulting network can be represented as:
\begin{align}
    \hat{\bm{h}}(t) &= f\left[\tilde{\bm{x}}(t)\right] \\
    &= \mathcal{S}_{d} \circ \mathcal{F}_{3} \circ \mathcal{F}_{2} \circ \mathcal{F}_{1} \left[\tilde{\bm{x}}(t)\right],
\end{align}
where each of the three fully connected networks \( \mathcal{F}_{i}(\cdot) \) has 1600 output units, and the nonlinear activation function used is the ReLU function. The final layer is a Sigmoid output layer \( \mathcal{S}_{d}(\cdot) \). Since this network needs to predict the optimal Wiener gain for a total of 161 frequency points at the current time slice, the output layer of the Sigmoid function has 161 output units. The training of the network is accomplished using binary cross-entropy or time-frequency spectrum loss.

During usage, this method requires introducing system delay. It waits for data from \( Q \) frames, then concatenates a total of \( 2Q + 1 \) frames to get the input \( \tilde{\bm{x}}(t) \) for the trained \( f(\cdot) \). This yields the optimal Wiener gain \( \hat{\bm{h}}(t) \) for that time slice, which is then applied to the observed signal to estimate the desired source signal in the time-frequency domain, i.e., \( \overleftarrow{\hat{\bm{s}}}(t) = \hat{\bm{h}}(t) \odot \overleftarrow{\bm{x}}(t) \). Finally, performing the inverse Fourier transform on \( \overleftarrow{\hat{\bm{s}}}(t) \) yields the estimated desired source signal \( \hat{\bm{s}}(t) \) for the current time slice.

\subsubsection{Fully Connected + Recurrent Neural Network}

To capture the long-term dependencies in the signals more effectively, a recurrent neural network can be integrated into the fully connected network. Even when using recurrent neural networks to capture correlations, literature \cite{tan2018convolutional} still employs the multi-frame concatenation approach to obtain better features \( \tilde{\bm{x}}(t) \). Specifically, the corresponding neural network can be expressed as:
\begin{align}
    \hat{\bm{h}}(t) &= f\left[\tilde{\bm{x}}(t)\right] \\
    &= \mathcal{S}_{d} \circ \mathcal{F}_{2} \circ \mathcal{R}_{3} \circ \mathcal{R}_{2} \circ \mathcal{R}_{1} \circ \mathcal{L} \circ \mathcal{F}_{1} \left[\tilde{\bm{x}}(t)\right].
\end{align}
Here, \( \tilde{\bm{x}}(t) \) is the input feature, and the original paper used 23 frames of 64-dimensional Mel spectrum features\footnote{It can also be considered to replace it with time-frequency domain logarithmic magnitude spectrum features.}. The output unit of the fully connected network \( \mathcal{F}_{1}(\cdot) \) is 1024; after a linear transformation via layer \( \mathcal{L}(\cdot) \), the output is sent to the recurrent neural network. The output unit of the linear layer is 512. The three layers of the recurrent neural network utilize LSTM structures, each with an output unit count of 512. Following the recurrent neural networks, a fully connected layer \( \mathcal{F}_{2}(\cdot) \) is used to normalize the data, which also has 512 output units. Finally, the network employs a Sigmoid function output layer \( \mathcal{S}_{d}(\cdot) \) to predict the optimal Wiener gain/time-frequency mask for each frequency band.

\subsubsection{Time-Frequency Stepwise Fusion Structure}

The aforementioned methods utilize fully connected networks to simultaneously capture temporal and frequency correlations from the observed signal's spectrum. The input to the network is a concatenation of multiple frames, and the output is the optimal time-frequency mask for the current time slice across various frequency bands. An alternative approach is to capture temporal and frequency information separately. Literature \cite{hao2021fullsubnet} presents a scheme for this.

To facilitate understanding, we introduce intermediate variables \( \bm{z}(t) \) and \( Z(\omega_{k},t) \), where \( \bm{z}(t) = \left[Z(\omega_{0},t) ~Z(\omega_{1},t) ~\cdots ~Z(\omega_{K-1},t)\right]^{T} \), and \(K \) is the number of valid frequency bands. In the original paper, the input slice has a window length of 512 samples (32 ms) with an overlap of 256 samples. The network construction uses the magnitude spectrum of the signal as input, i.e., \( \left|\overleftarrow{\bm{x}}(t)\right| \), which then passes through two layers of LSTM followed by a fully connected network, resulting in:
\begin{align}
    \bm{z}(t) &= \left[\begin{array}{cccc}
    Z(\omega_{0},t) & Z(\omega_{1},t) & \cdots & Z(\omega_{K-1},t)
    \end{array}\right]^{T} \\
    &= g_{1} \left[\left|\overleftarrow{\bm{x}}(t)\right|\right] \\
    &= \mathcal{F}_{1} \circ \mathcal{R}_{2} \circ \mathcal{R}_{1} \left[\left|\overleftarrow{\bm{x}}(t)\right|\right].
\end{align}
Here, \( \mathcal{R}_{1}(\cdot) \) and \( \mathcal{R}_{2} \) are two layers of LSTM recurrent neural networks, both having 512 output units; \( \mathcal{F}_{1}(\cdot) \) is a fully connected network with the number of output units equal to the number of valid frequency bands, which is 257. After a nonlinear transformation via \( g_{1}(\cdot) \), we obtain \( \bm{z}(t) \), where each element \( Z(\omega_{k},t) \) can be viewed as prior information about the corresponding frequency band's source signal. Using this prior information and considering the magnitude spectra of \( 2Q + 1 \) adjacent frequency bands, we can concatenate a new feature vector for each frequency band:
\begin{align}
\bm{c}(\omega_{k},t) = \left[\begin{array}{cc}
    Z(\omega_{k},t) & \bm{x}^{\prime}(\omega_{k},t)
    \end{array}\right]^{T}, 
\end{align}
where $\bm{x}^{\prime}(\omega_{k},t)  =  \left[ ~
      |X(\omega_{k-Q},t)|  ~ \cdots ~ |X(\omega_{k+Q},t)|
    ~\right]^{T}$. 
The length of the  vector $\bm{c}(\omega_{k},t)$ is \( 2Q + 2 \). This vector is then input into the network \( f(\cdot) \) to estimate the time-frequency mask \( \hat{H}(\omega_{k},t) \) for the current frequency band. Specifically, the structure of the network \( f(\cdot) \) can be represented as:
\begin{align}
    \hat{H}(\omega_{k},t) &= f\left[\bm{c}(\omega_{k},t)\right] \\
    &= \mathcal{S}_{d} \circ \mathcal{R}_{4} \circ \mathcal{R}_{3} \left[\bm{c}(\omega_{k},t)\right], \quad \forall k \label{SE-two-step}.
\end{align}
Here, \( \mathcal{R}_{3}(\cdot) \) and \( \mathcal{R}_{4} \) are also two layers of LSTM recurrent neural networks, each with 384 output units; the output layer \( \mathcal{S}_{d}(\cdot) \) uses a Sigmoid activation function, while the original paper used a linear layer without a nonlinear transformation.

It is important to note that in this approach, the concatenation of features captures adjacent frequency band information while leveraging the prior estimated information. During inference, each frequency band shares a common \( f(\cdot) \) and individually infers based on the concatenated input. Thus, for each frequency band, the first step is to combine \( \bm{c}(\omega_{k},t) \), followed by utilizing equation (\ref{SE-two-step}) to complete the inference. Regarding the construction of the network's objective function, either binary cross-entropy or magnitude spectrum approximation can theoretically be employed.

\subsubsection{Fully Convolutional Network}

The aforementioned methods are based on the framework of transformation domain analysis, filtering, and reconstruction. For deep learning methods, if the desired source signal within the observed signal is known, it is possible to construct an end-to-end network for signal estimation without needing to perform separate signal analysis and reconstruction steps. Literature \cite{pascual2017segan,pandey2019new} presents an end-to-end voice denoising network framework based on the U-Net architecture. Specifically, the encoder consists of 9 layers of convolutional neural networks. Let the \( q \)-th layer be represented as \( \mathcal{C}_{q}(\cdot) \), with its output expressed as:
\begin{align}
    \bm{x}^{(q)}(t) &= \mathcal{C}_{q} \left[\bm{x}^{(q-1)}(t)\right] \\
    &= \mathcal{C}_{q} \circ \mathcal{C}_{q-1} \circ \cdots \circ \mathcal{C}_{1}\left[\bm{x}(t)\right], \quad q=2,3,\ldots, 9.
\end{align}
The input is still a time slice, but it is relatively longer compared to the previous frequency-domain methods; instead of 20 ms, each slice has a length of 128 ms, resulting in 2048 samples at a 16 kHz sampling frequency. The network input \( \bm{x}(t) \) is thus a \( 2048 \times 1 \) vector. The first layer \( \mathcal{C}_{1}(\cdot) \) expands the channel count from one channel to 64 channels without downsampling, resulting in data dimensions of \( 2048 \times 64 \). Each subsequent convolutional layer undergoes 2-fold downsampling while increasing the channel count. The parameters (sample points \( \times \) output channels) from the first to the ninth layer are as follows:
\begin{align}
    & (2048 \times 64) \rightarrow (1024 \times 64) \rightarrow (512 \times 128) \rightarrow (256 \times 128) \rightarrow \nonumber \\
    & (128 \times 128) \rightarrow (64 \times 128) \rightarrow (32 \times 256) \rightarrow (16 \times 256) \nonumber\\
    & \rightarrow (8 \times 256).
\end{align}
Thus, after passing through the ninth convolutional neural network layer, the original signal of length 2048 is transformed into a feature \( \bm{x}^{(9)}(t) \) of dimension \( 8 \times 256 \), where 8 represents the time slice index and 256 represents the channel count.

After encoding, eight transposed convolution modules are needed to merge the features from the encoding stage. For convenience, we define the \( q \)-th transposed convolution module as \( \mathcal{C}^{\prime}_{q}(\circ) \), with its output denoted as \( \bm{z}^{(q)}(t) \). The first transposed convolution module near the output yields the estimate of the desired signal:
\begin{align}
    \hat{\bm{s}}(t) &= \bm{z}^{(1)}(t) \\
    &= \mathcal{C}^{\prime}_{1}\left[\bm{z}^{(2)}(t) \ddag \bm{x}^{(1)}(t) \right].
\end{align}
The input consists of two parts: the output \( \bm{z}^{(2)}(t) \) from the previous module in the decoding network and the corresponding layer output \( \bm{x}^{(1)}(t) \) from the encoding network. For the remaining seven decoding modules, their outputs can be expressed as:
\begin{align}
    \bm{z}^{(q)}(t) &= \mathcal{C}^{\prime}_{q}\left[\bm{z}^{(q+1)}(t) \ddag \bm{x}^{(q)}(t) \right], \quad q= 2,3,\ldots, 7.
\end{align}
The eighth decoding module, closest to the last layer of the encoding network, can be expressed as \( \bm{z}^{(8)}(t) = \mathcal{C}^{\prime}_{q}\left[ \bm{x}^{(9)}(t) \ddag \bm{x}^{(8)}(t) \right] \), where the dimensions of \( \bm{x}^{(8)}(t) \) and \( \bm{x}^{(9)}(t) \) are \( 16 \times 256 \) and \( 8 \times 256 \), respectively. Corresponding to the encoding phase, the decoding phase progressively expands the data dimensions. The input signal dimensions from the eighth to the first transposed convolution block are as follows:
\begin{align}
    & (16 \times 512) \rightarrow (32 \times 512) \rightarrow (64 \times 256) \rightarrow (128 \times 256) \rightarrow \nonumber \\
    & (256 \times 256) \rightarrow (512 \times 128) \rightarrow (1024 \times 128) \rightarrow (2048 \times 128).
\end{align}
The final transposed convolution network receives an input of \( 2048 \times 128 \) data and outputs a \( 2048 \times 1 \) signal, which is the estimated desired source signal within the observed signal.

In the construction of the above one-dimensional convolutional neural network, the convolution kernels all have the same width of 11 points, resulting in a relatively simple structure. In terms of loss function construction, the most intuitive metric in the time domain is the Scale-Invariant Signal Distortion Ratio; of course, one can also transform \( \hat{\bm{s}}(t) \) to the frequency domain and define the loss function based on magnitude spectrum approximation, then compute the backpropagation error to learn the coefficients of the convolutional kernels in the network\cite{pandey2019new}.

\subsubsection{Multimodal Network Architecture}

Multimodal fusion can leverage correlations between different dimensions of information to provide mutual prior information. In the context of denoising tasks, features such as the spatial orientation of the source, the voiceprint of the source, and visual information about the source can all be fused. The most common approach in multimodal fusion is feature-level integration, which we briefly introduce using the architecture of a ``specified source" signal extraction network as an example\cite{wang2018voicefilter}.

For the task of extracting specified source signals, a segment of desired source speech is typically provided, ranging from several seconds to tens of seconds. This speech signal is used to characterize the desired source, requiring extraction from the observed signal. To enhance fusion efficiency, the original desired source is first subjected to feature compression, generally using a voiceprint extraction network. Utilizing the voiceprint extraction network described in previous sections, any length of speech waveform can be converted into a fixed-length feature vector, denoted as \( \bm{c} \). Our task then becomes extracting the desired source signal \( \bm{s} \) from the observed signal \( \bm{x} \) given \( \bm{c} \).

Using the framework of short-time Fourier transform analysis, filtering, and reconstruction, the first step involves utilizing the neural network \( g(\cdot) \) to convert the observed signal slices \( \overleftarrow{\bm{x}}(t) \) into features \( \bm{z}(t) \). Subsequently, the voiceprint feature is concatenated with \( \bm{z}(t) \), and the resulting input is fed into the mask estimation network to obtain estimates of the optimal masks at each time-frequency point. Assuming the mask estimation network is represented by \( f(\cdot) \), the operation flow of the network can be expressed as:
\begin{align}
    \bm{z}(t) &= g\left[\overleftarrow{\bm{x}}(t)\right] \label{z-g-multimodel}, \\
    \hat{\bm{h}}(t) &= f\left[\bm{z}(t) \ddag \bm{c}\right].
\end{align}
Of course, in equation (\ref{z-g-multimodel}), a multi-frame concatenation approach can also be used as input to the network; this serves merely as a case example. The feature extraction network \( g(\cdot) \) can be any previously mentioned network types, including fully connected networks, recurrent neural networks, convolutional neural networks, or self-attention mechanisms. As for the mask estimation network, its input is already a high-dimensional feature, typically utilizing recurrent neural networks followed by two layers of fully connected networks.

\textbf{Understanding Feature Concatenation:} If the features \( \bm{z} \) and \( \bm{c} \) are concatenated and then subjected to a linear transformation via matrix \( \bm{A} = [\bm{A}_{1} ~ \bm{A}_{2}] \), we can obtain:
\begin{align}
     \bm{A}\left(\bm{z} \ddag\bm{c}\right)= \bm{A} \left[\begin{array}{c}
    \bm{z} \\
    \bm{c}
    \end{array}\right] = \bm{A}_{1}\bm{z} + \bm{A}_{2}\bm{c}.
\end{align}
Thus, this direct concatenation can be viewed, in a certain sense, as performing a linear transformation on the features before summing them directly.

Beyond this direct concatenation approach, another fusion method is often employed. Assuming the source feature vector \( \bm{c} \) has a dimension of \( K \); the \( k \)-th element of this vector is \( \alpha_{k} \), such that \( \bm{c} = \left[\alpha_{0}~\alpha_{1}~\cdots~\alpha_{K-1}\right]^{T} \). Before feature fusion, the signal's features \( \bm{z}(t) \) undergo different linear transformations, mapping them to different spaces, and then using the source feature coefficients for weighted summation to obtain the fused feature\cite{delcroix2018single}. In this case, the mask estimation network can be expressed as:
\begin{align} 
    \hat{\bm{h}}(t) &= f\left[\sum_{k=0}^{K-1} \alpha_{k} \bm{A}_{k} \bm{z}(t)\right].
\end{align}
The matrices \( \bm{A}_{k} \) are transformation matrices learned from the task, treated as part of the network parameters. In practice, it is also possible to apply a series of nonlinear mappings, such as \( \mathcal{F}_{k}(\cdot) \), to \( \bm{z}(t) \) first; finally, the results can be weighted by \( \alpha_{k} \) to obtain the fused features, i.e., \( \sum_{k} \alpha_{k} \mathcal{F}_{k}(\cdot) \).

In the training structure of the multimodal fusion network, the voiceprint extraction network can be pre-trained. During the construction of the denoising network training dataset, the voiceprint features \( \bm{c} \) of the desired source signal must also be processed and included in the training set. From the perspective of attention mechanisms, we can also treat the feature vector \( \bm{c} \) as a query vector, constructing a self-attention layer to achieve feature fusion; this will not be detailed further.

\section{Approaches to Source Separation}
 
Sound source separation has a research history spanning decades, evolving from classical independent component analysis methods to current deep learning approaches. The problem it addresses is how to isolate each speaker's voice when multiple sources are present in the microphone's observation. These sources can be multiple speakers, various musical instruments, or even different vibrating components of the same device. The applications of sound source separation are extensive, with some techniques overlapping with those discussed in previous sections. This section will focus on three main points: \emph{
1) The basic framework for sound source separation;
2) The permutation ambiguity problem;
3)Time-domain and frequency-domain methods.}

\subsection{Signal Model}

Consider a microphone observation signal \( \bm{x} \), which can be decomposed into \( T \) time slices, with the \( t \)-th time slice represented as \( \bm{x}(t) \). Assume there are \( J \) source signals in the observation, with the \( j \)-th source signal denoted as \( \bm{s}^{(j)} \); it shares the same dimension as \( \bm{x} \). The relationship between \( \bm{x} \) and \( \bm{s}^{(j)} \) is expressed as:
\begin{align}
    \bm{x} = \sum_{j=1}^{J} \bm{s}^{(j)}.
\end{align}
Similarly, for each time slice, the relationship holds:
\begin{align}
    \bm{x}(t) = \sum_{j=1}^{J} \bm{s}^{(j)}(t).
\end{align}
The goal of sound source separation is to estimate all or a specific source signal \( \bm{s}^{(j)} \) from the observed signal \( \bm{x} \). For convenience, we denote the estimated source signal from the network as \( \hat{\bm{s}} \). In sound source separation tasks, most part of the network is typically shared. For example, in an end-to-end output framework, the network's output can be represented as:
\begin{align}
    \hat{\bm{s}}^{(j)} = \mathcal{S}_{j} \circ g(\bm{x}),
\end{align}
where \( g(\cdot) \) represents the shared part of the network, and \( \mathcal{S}_{j}(\cdot) \) is the output layer specific to the \( j \)-th source.

\subsection{Permutation Ambiguity Problem}

Due to the uncertainty regarding which \( \hat{\bm{s}}^{(j)} \) corresponds to which source signal, the sound source separation framework introduces a permutation ambiguity problem in the construction of dataset labels. Forcing a label design to specify the network's training target can lead to convergence issues. For instance, in the case of separating two sources, for sample \( \bm{x}_{n} \), the source estimates \( \hat{\bm{s}}_{n}^{(1)} \) and \( \hat{\bm{s}}_{n}^{(2)} \) correspond to \( \bm{s}^{(1)} \) and \( \bm{s}^{(2)} \) in the sample set. If there happens to be another sample in the dataset, \( \bm{x}_{n^{\prime}} \), which is also composed of \( \bm{s}^{(1)} \) and \( \bm{s}^{(2)} \), but in this case, we treat \( \bm{s}^{(2)} \) as the first source and \( \bm{s}^{(1)} \) as the second source. For the first data sample, the network training aims for \( \bm{s}^{(1)} \leftarrow \mathcal{S}_{1} \circ g(\bm{x}) \); however, for the second data sample, the training target becomes \( \bm{s}^{(2)} \leftarrow \mathcal{S}_{1} \circ g(\bm{x}) \). Clearly, there exists a significant ``one-to-many" issue between the inputs and outputs during network training, making convergence difficult.

To address the permutation ambiguity problem, literature \cite{yu2017permutation} proposes a permutation-invariant training method. The main principle is to use optimal matching instead of forced matching. Before explaining this in detail, we define a subset of the source indices of the current sample, i.e.,   \( \mathcal{P} \). For example, with three sources, its elements are:
\begin{align}
    \bm{p} &= \left[\begin{array}{c}
    p_{1} \\ p_{2} \\ p_{3}
    \end{array}\right] \nonumber \\
    &\in \left\{\left[\begin{array}{c}
    1 \\ 2 \\ 3
    \end{array}\right], \left[\begin{array}{c}
    1 \\ 3 \\ 2
    \end{array}\right], \left[\begin{array}{c}
    2 \\ 1 \\ 3
    \end{array}\right], \left[\begin{array}{c}
    2 \\ 3 \\ 1
    \end{array}\right], \left[\begin{array}{c}
    3 \\ 1 \\ 2
    \end{array}\right], \left[\begin{array}{c}
    3 \\ 2 \\ 1
    \end{array}\right]\right\}.
\end{align}
Here, \( p_{i} \) denotes the \( i \)-th element of vector \( \bm{p} \); the index vector \( \bm{p} \) is an element of the index set \( \mathcal{P} \). Under the principle of permutation-invariant training, the network's loss function can be expressed as:
\begin{align}
    \mathcal{J} = \min_{\bm{p} \in \mathcal{P}} \sum_{j=1}^{J} d\left[\hat{\bm{s}}^{(j)}, {\bm{s}}^{(p_{j})}\right] \label{J-PIT-1}.
\end{align}
In this equation, \( \bm{p} \in \mathcal{P} \) indicates that all elements \( \bm{p} \) in the set \( \mathcal{P} \) are traversed. For the separation of two sources, the number of traversable samples in the dataset is 2; for three sources, it is 6; and for four sources, it is 24. When there are \( J \) sources, the number of traversable samples is \( J! \). Generally, we discuss sound source separation problems involving up to four sources.

\subsection{Transform Domain Sound Source Separation Framework}

The core of the transform domain analysis, separation, and reconstruction method is to estimate the masks for each source across all time and feature dimensions within the transform domain. Based on the mask information, the observed signal in the transform domain is selectively retained, allowing the original signal to be reconstructed from the resultant signal. Taking short-time Fourier analysis as an example, the analysis, separation, and reconstruction steps for each time slice \( \bm{x}(t) \) are as follows:
\begin{align}
    \overleftarrow{\bm{x}}(t) &= \bm{W}^{T} \bm{x}(t), \\
    \overleftarrow{\bm{s}}^{(j)}(t) &= \hat{\bm{h}}_{j}(t) \odot \overleftarrow{\bm{x}}(t), \quad j=1,2,\ldots,J, \\
    \hat{\bm{s}}^{(j)}(t) &= \frac{1}{L_{w}} \bm{W}^{\ast} \overleftarrow{\bm{s}}^{(j)}(t).
\end{align}
Here, \( \bm{W} \) is the Fourier transform matrix, \(L_{w}\) is the length of the windowed signal, \( \odot \) denotes element-wise multiplication, and \( \hat{\bm{h}}_{j}(t) \) is the mask value for the current time slice, which has the same dimension as \( \overleftarrow{\bm{x}}(t) \) and is estimated by the neural network \( f(\cdot) \). The network \( f(\cdot) \) processes single-frame or multi-frame features of \( \overleftarrow{\bm{x}}(t) \) (such as magnitude spectrum, logarithmic magnitude spectrum, real part, and imaginary part).

\subsection{Mask Estimation-Based Methods}

Using various deep denoising network architectures, we can construct the network \( g(\cdot) \) with appropriate output layers \( \mathcal{S}_{j} \) to represent the mask estimation network in the transform domain as:
\begin{align}
    \hat{\bm{h}}_{j}(t) = f_{j}\left[\tilde{\bm{x}}(t)\right] = \mathcal{S}_{j} \circ g\left[\tilde{\bm{x}}(t)\right].
\end{align}
Here, \( \tilde{\bm{x}}(t) \) is the feature vector/matrix derived from single-frame/multi-frame \( \overleftarrow{\bm{x}}(t) \).

During the network training phase, ideal masks can be constructed based on the relative power of the sources at each time-frequency point:
\begin{align}
    \bm{h}_{i}(t) = \frac{\left|{\bm{s}}^{(j)}(t)\right|^2}{\epsilon_{0} + \sum_{i=1}^{J} \left|{\bm{s}}^{(i)}(t)\right|^2}, \quad i=1,2,\ldots,J.
\end{align}
Here, \( \epsilon_{0} \) represents the power of the noise, \( |\cdot| \) denotes the absolute value of each element in the vector, and \( \frac{|\bm{a}|}{\bm{b}} \) indicates element-wise division between vectors \( \bm{a} \) and \( \bm{b} \). Under the permutation-invariant training strategy, the network output \( \hat{\bm{h}}_{j}(t) \) can be made to approximate \( \bm{h}_{i}(t) \) to complete the training.

Considering that the ultimate goal of the network is to approximate the signals' amplitudes, in conjunction with the permutation-invariant strategy, the loss function in the context of the \( L \)-norm distance can be described as:
\begin{align}
    \mathcal{J}_{1} &= \min_{\bm{p} \in \mathcal{P}} \sum_{j=1}^{J} \left\| \left|\overleftarrow{\bm{x}}(t)\right| \odot \hat{\bm{h}}_{j}(t) - \left|\overleftarrow{\bm{s}}^{(p_{j})}(t)\right| \right\|^{2} \\
    &= \min_{\bm{p} \in \mathcal{P}} \sum_{j=1}^{J} \left\| \left|\overleftarrow{\bm{x}}(t)\right| \odot \left\{f_{j}\left[\tilde{\bm{x}}(t)\right]\right\} - \left|\overleftarrow{\bm{s}}^{(p_{j})}(t)\right| \right\|^{2}.\nonumber
\end{align}
In this equation, \( |\cdot| \) still denotes the absolute value of each element in the vector. The difference between the two vectors represents the error between the estimated source magnitude spectra and the actual source magnitude spectra at each frequency band. By minimizing this loss function, we can optimize the network \( f_{j}(\cdot) \) and obtain the optimal network parameters \( \theta_{f_{j}} \). It is important to note that while there are \( J \) networks \( f_{j}(\cdot) \), they differ only in their output layers; the remaining parts are shared in the network \( g(\cdot) \).

\subsection{Prior-Based Methods}

In sound source separation tasks, to extract sources that are meaningful and relevant, prior information about the desired sources is often utilized. This prior information can be incorporated into the network framework as part of the conditional probability. Common prior information from an auditory perspective can be categorized into two types: the voiceprint features of the source and the spatial location of the source.

Methods for extracting voiceprint features and embedding these features into the network have been discussed in previous sections and will not be elaborated further here. For spatial location features, these are typically derived from scenarios with multiple microphone observations. Common features include: 1) the phase of the relative transfer functions between microphone observations; 2) the inner product of the relative transfer functions with the array steering vector corresponding to the source; 3) designing beamformers based on the spatial location of the source and the geometric structure of the array to preprocess the observed signals.

For convenience, we unify the information obtained from source priors into the variable \( \bm{c} \). The approach to utilizing such prior information is straightforward: simply concatenate \( \bm{c} \) with \( \tilde{\bm{x}} \) along the feature dimension and input the combined data into the network. The mask estimation function can be expressed as:
\begin{align}
    \hat{\bm{h}}_{j}(t) = f_{j}\left[\tilde{\bm{x}}(t) \ddag \bm{c}\right], \quad j=1,2,\ldots,J.
\end{align}
In literature \cite{chen2018multi}, the concatenated features are fed into a four-layer LSTM recurrent neural network, followed by a Sigmoid layer to complete the network modeling.

\subsection{Clustering-Assisted Separation Methods}

Clustering-assisted separation methods incorporate the clustering loss function into the network training process to assist the separation network in extracting more robust features. Suppose each time-frequency point can extract an \( L \)-dimensional feature vector \( \bm{v}(\omega,t) \), i.e., $L$ channels. If we view the time-frequency domain signal \( \overleftarrow{\bm{x}} \) as an image of size \( K \times T \), where \( K \) is the number of frequency bands and \( T \) is the number of time slices, then the feature can be considered a tensor with dimensions \( K \times T \times L \).

To facilitate the description of the loss function, we first reshape the three-dimensional feature tensor into a matrix, resulting in an \( L \times KT \) matrix, which we temporarily denote as \( \bm{V} \). Analogous to \( \bm{V} \), we define a matrix \( \bm{U} \), where each column is a one-hot vector representing the category of the source. The basic idea of clustering is to minimize the distance between samples within the same class while maximizing the distance between samples from different classes. Based on the definitions of \( \bm{U} \) and \( \bm{V} \), the clustering optimization problem can be described as follows \cite{hershey2016deep}:
\begin{align}
    \mathcal{J}_{2} = \left\|\bm{V}^{T}\bm{V} - \bm{U}^{T}\bm{U}\right\|^2.
\end{align}
In this equation, the \( (i,j) \)-th element of \( \bm{V}^{T}\bm{V} \) represents the inner product of feature vectors \( \bm{v} \) at the corresponding time-frequency points, while the elements of \( \bm{U}^{T}\bm{U} \) equal 1 only when the corresponding time-frequency points originate from the same source; all other positions are zero. Minimizing the loss function \( \mathcal{J}_{2} \) encourages the feature vectors between different sources to be as orthogonal as possible. Additionally, other loss functions can also be employed, such as the improved version in literature \cite{wang2018combining} given by:
\begin{align}
    \mathcal{J}_{2} = L - \mathrm{tr}\left[(\bm{V}\bm{V}^{T})^{-1} \bm{V} \bm{U}^{T} (\bm{U}\bm{U}^{T})^{-1} \bm{U}\bm{V}^{T}\right].
\end{align}
We can also use the previously mentioned contrastive loss to construct a loss function for clustering operations. The introduction of clustering networks facilitates alignment of features across different frequency bands, mitigating permutation issues between bands.

Combining the aforementioned mask estimation network construction methods, the clustering-assisted separation method can be expressed as:
\begin{align}
    \bm{V} &= \mathcal{F}_{1} \circ g(\tilde{\bm{x}}), \\
    \hat{\bm{h}}_{j} &= \mathcal{S}_{j} \circ \mathcal{F}_{2} \circ g(\tilde{\bm{x}}), \quad j=1,2,\ldots,J.
\end{align}
Here, \( \mathcal{F}_{1}(\cdot) \) and \( \mathcal{F}_{2}(\cdot) \) are two fully connected network modules used for nonlinear mapping, while \( g(\cdot) \) is the feature extraction network. Considering the loss function for mask estimation, the overall loss function for the network can be described as:
\begin{align}
    \mathcal{J}_{3} &= \beta \left\|\bm{V}^{T}\bm{V} - \bm{U}^{T}\bm{U}\right\|^2 \nonumber \\
    &~~~+ \min_{\bm{p} \in \mathcal{P}} \sum_{j=1}^{J} \left\| \left|\overleftarrow{\bm{x}}\right| \odot \left\{ \mathcal{S}_{j} \circ \mathcal{F}_{2} \circ g\left[\tilde{\bm{x}}\right]\right\} - \left|\overleftarrow{\bm{s}}^{(p_{j})}\right| \right\|^{2}.
\end{align}
Here, \( \beta \geq 0 \) is used to control the trade-off between the loss functions.

It is noteworthy that clustering operations can assist in constructing the loss function. Simultaneously, the output feature vector \( \bm{v}(\omega,t) \) possesses similarity within the same source while being distant between different sources. Based on this property, the clustering center for the same source can be viewed as the voiceprint of that source. This voiceprint information can be incorporated as a condition into the denoising neural network discussed in the previous sections to extract the specified desired source signal \cite{zeghidour2021wavesplit}.

In addition to the frequency-domain analysis, separation, and reconstruction framework, sound source separation also has time-domain end-to-end methods. However, time-domain methods often struggle with waveform distortion, and most approaches still utilize the analysis, separation, and reconstruction framework, where the basis functions for analysis and reconstruction could be either learned from data or using the Fourier transform and its inverse \cite{veluri2023real}. Aside from the differences in network frameworks, the underlying principles are similar and will not be discussed further here.

   \section{Other Data-Driven Signal Processing Methods}
 
The previous sections discussed the fundamental principles of deep learning, typical network architectures, methods for constructing loss functions, and tasks related to sound signal processing. This section explores some other classic, data-driven machine learning concepts, aiming to inspire readers. The section includes following  topics.
\emph{ 
1) Generative adversarial learning: how it utilizes data to characterize loss functions and its potential applications.
2) Optimal transport theory: what it entails and how it aligns different density functions. 
3) Optimization of the area under the ROC curve: how to compute this curve in practice and whether it can be used to construct loss functions.
4) Diffusion models: how a sample evolves from meaningful data to noise and vice versa.
5) Data visualization: how to display high-dimensional data on a two-dimensional plane, providing an intuitive sense of data distribution.}

\subsection{Generative Adversarial Learning: Characterizing Targets with Data}

This section introduces Generative Adversarial Networks (GANs) \cite{goodfellow2014generative} and several of their typical variants, aiming to understand the basic principles of GANs, feature learning methods, and applications. Key points include:   1) the architecture and basic principles of GANs, including mathematical notation, optimization objectives, and fundamental issues; 2) directional data generation methods, explaining the principles and objective function construction of Conditional GANs (CGAN) \cite{mirza2014conditional}; 3) style-specific data generation, discussing the principles of CycleGAN and introducing consistency constraints and data boundary constraints \cite{zhu2017unpaired}; 4) generative adversarial strategies that assist small-sample feature learning, detailing how GANs can effectively utilize unlabeled data to enhance feature learning performance in small-sample scenarios.

\subsubsection{Principles of Generative Adversarial Networks}

Given a dataset \( \{\bm{x}_{i}, \forall i\} \) and a random variable \( \bm{z} \) following a specific distribution, the challenge is to find a function \( g(\bm{z}) \) such that the output \( \hat{\bm{x}} = g(\bm{z}) \) follows the same distribution as the dataset \( \{\bm{x}_{i}\} \). This problem involves two fundamental questions:
\begin{itemize}
    \item How to construct the function \( g(\bm{z}) \) using basic neural network modules?
    \item How to define the target function for training the neural network based on the dataset \( \{\bm{x}_{i}\} \)?
\end{itemize}
Assume there exists a function \( d(\bm{x}) \) that can distinguish whether any input \( \bm{x} \) follows the same distribution as the dataset \( \{\bm{x}_{i}\} \): if it does, then \( d(\bm{x})=1 \); otherwise, \( d(\bm{x})=0 \). This function \( d(\bm{x}) \) is also constructed through a neural network; to restrict its output values within \( [0, 1] \), the final layer of \( d(\bm{x}) \) is typically a sigmoid activation function, represented as:
\begin{align}
    \sigma(\bm{v}) = \frac{1}{1 + \expd{\bm{w}^{T} \bm{v}}},
\end{align}
where \( \bm{v} \) is the feature from the layer before the last layer\footnote{The dimension of \( \bm{v} \) is actually one more than that of the features, with the additional dimension being 1, used to accommodate the bias term of the sigmoid function.}, and \( \bm{w} \) are the trainable parameters of the network.

Given \( d(\bm{x}) \), the objective for the function \( g(\bm{z}) \) becomes: maximize \( d\left(\hat{\bm{x}}\right) \), or \( d \circ g(\bm{z}), \forall \bm{z} \). Common objective functions for learning the parameters of the generative network include:
\begin{align}
    \max_{\theta_{g}} \mathbb{E}_{\bm{z}} d\circ g(\bm{z}), \label{GAN-opt-4-g-1} \\
    \max_{\theta_{g}} \mathbb{E}_{\bm{z}} \ln d\circ g(\bm{z}), \label{GAN-opt-4-g-2} \\
    \min_{\theta_{g}} \mathbb{E}_{\bm{z}} \ln\left[1-d\circ g(\bm{z})\right]. \label{GAN-opt-4-g}
\end{align}
Optimizing any of the three problems can be achieved by adjusting the parameters \( \theta_{g} \) to maximize \( d \circ g(\bm{z}) \), but the specific choice depends on the context.

In the GAN framework, the discriminator network \( d(\bm{x}) \) also needs to be learned from data. Its training objective is to distinguish between real data \( \bm{x} \) and generated data \( \hat{\bm{x}} = g(\bm{z}) \), which is a typical binary classification problem. Therefore, we can define the binary cross-entropy as its cost function:
\begin{align}
    \mathcal{J}_{\mathrm{BCE}}(\theta_{d}) &= -\mathbb{E}_{\bm{x}} \left\{ 1\cdot \ln d(\bm{x}) + 0\cdot \ln \left[1-d(\bm{x})\right]\right\} \nonumber \\
    &~~~ - \mathbb{E}_{\hat{\bm{x}}} \left\{ 0\cdot \ln d(\hat{\bm{x}}) + 1\cdot \ln \left[1-d(\hat{\bm{x}})\right]\right\} \\
    &= -\mathbb{E}_{\bm{x}} \ln d(\bm{x}) - \mathbb{E}_{\hat{\bm{x}}} \ln\left[1-d(\hat{\bm{x}})\right] \nonumber \\
    &= -\mathbb{E}_{\bm{x}} \ln d(\bm{x}) - \mathbb{E}_{\bm{z}} \ln\left[1-d\circ g(\bm{z})\right].
\end{align}
The parameters \( \theta_{d} \) of the discriminator network can be learned by minimizing \( \mathcal{J}_{\mathrm{BCE}}(\theta_{d}) \), i.e., \( \min_{\theta_{d}} \mathcal{J}_{\mathrm{BCE}}(\theta_{d}) \). For convenience in analysis, the optimization problem is often expressed as \( \max_{\theta_{d}} -\mathcal{J}_{\mathrm{BCE}}(\theta_{d}) \), leading to:
\begin{align}
    \max_{\theta_{d}} \mathbb{E}_{\bm{x}} \ln d(\bm{x}) + \mathbb{E}_{\bm{z}} \ln\left[1-d\circ g(\bm{z})\right].
\end{align}
Combining the optimization problems of the generative network from (\ref{GAN-opt-4-g}), the overall optimization problem for GANs can be summarized as:
\begin{align}
    \min_{\theta_{g}} \max_{\theta_{d}} \mathbb{E}_{\bm{x}} \ln d(\bm{x}) + \mathbb{E}_{\bm{z}} \ln\left[1-d\circ g(\bm{z})\right].
\end{align}
Clearly, for the same objective function, the optimization goals for the two sets of parameters are  opposed: adjusting the parameters \( \theta_{g} \) of the generative network aims to minimize the value of the objective function, while adjusting the parameters \( \theta_{d} \) of the discriminator network aims to maximize it. This contradictory optimization presents significant challenges for training generative adversarial networks. Additionally, GANs face the following challenges:
\begin{itemize}
    \item Convergence time is difficult to determine. Due to the contradictory nature of the training objective, it is challenging to establish a stopping criterion for training.
    \item Convergence is often not guaranteed, and the training process heavily relies on the engineer's experiential knowledge.
\end{itemize}
Fortunately, there is a wealth of existing work to reference; readers attempting to implement GANs can benefit from widely reading about training configurations in related literature to grasp basic training techniques.

\subsubsection{Conditional Data Generation}

In the previously discussed generative adversarial networks (GANs), data is generated from a random variable \( \bm{z} \) through a generator network \( g(\bm{z}) \). However, this approach does not allow control over the characteristics or categories of the generated samples. To address this issue, the Conditional Generative Adversarial Network (CGAN) was introduced \cite{mirza2014conditional}.

For convenience, we define the conditional variable in CGAN as \( \bm{y} \). If we aim to generate a specific type of data, \( \bm{y} \) can be a one-hot vector. For directed data generation applications, the dataset typically consists of three parts:

\begin{itemize}
    \item A labeled dataset \( \{(\bm{x}_{n}, \mathbf{y}_{n}), \forall n\} \), where \( \bm{y}_{n} \) describes the condition of the data and serves as the label.
    \item An unlabeled dataset \( \{\ddot{\bm{x}}_{n}, \forall n\} \); in practice, many data samples lack conditional annotations, resulting in a relatively large dataset.
    \item A dataset generated by the network \( \{(\bm{z}_{n}, \bm{y}_{n}, \hat{\bm{x}}_{n})\} \), where \( \bm{z}_{n} \) is a random variable following a certain distribution and \( \bm{y}_{n} \) is the condition expected for sample \( \hat{\bm{x}}_{n} \).
\end{itemize}
Within the framework of conditional GANs, the generator's input includes not only the random variable but also the given conditional variable. The corresponding output can be described as:

\begin{align}
    \hat{\bm{x}} = g(\bm{z}, \bm{y}).
\end{align}

For the discriminator network, since it expects different outputs for different data, the corresponding objective functions must be separated: 1) the basic feature learning module defined as \( d(\bm{x}) \); 2) the module for discriminating conditions defined as \( S_{c}(\cdot) \), where the subscript \( c \) indicates "Conditional"; 3) the module for distinguishing real from fake data defined as \( S_{d}(\cdot) \), where the subscript \( d \) indicates the Sigmoid function. Accordingly, the expected output of the discriminator can be expressed as:

\begin{align}
    \mathbf{y} &\leftarrow S_{c} \circ d(\bm{x}),\\
    1 &\leftarrow S_{d} \circ d(\ddot{\bm{x}}),\\
    0 &\leftarrow S_{d} \circ d(\hat{\bm{x}}) = S_{d} \circ d \circ g(\bm{z}, \bm{y}).
\end{align}
The objective function for conditional prediction is defined as \( \mathcal{J}_{1}[S_{c} \circ d(\bm{x}), \bm{y}] \). For a classification task, this can be expressed as \( \mathcal{J}_{1}[S_{c} \circ d(\bm{x}), \bm{y}] = \bm{y}^{T} \ln S_{c} \circ d(\bm{x}) \). The optimization problem for learning the parameters of the discriminator network can be formulated as:

\begin{align}
    \max_{\theta_{d};\theta_{S_{c}},\theta_{S_{d}}} & \mathbb{E}_{\bm{x}} \mathcal{J}_{1}[S_{c} \circ d(\bm{x}), \bm{y}] + \nonumber\\
    & \mathbb{E}_{\ddot{\bm{x}}} \ln S_{d} \circ d(\ddot{\bm{x}}) + \mathbb{E}_{\bm{z}} \ln\left[1 - S_{d} \circ d \circ g(\bm{z}, \bm{y})\right]. \label{sgan-cost-4-d}
\end{align}
The corresponding optimization problem for the generator network can be expressed as:

\begin{align}
    \max_{\theta_{g}} &~~~~ \mathbb{E}_{\hat{\bm{x}}} \mathcal{J}_{1}[S_{c} \circ d(\hat{\bm{x}}), \bm{y}],
\end{align}
which is equivalent to:
\begin{align}
    \max_{\theta_{g}} & ~~~~\mathbb{E}_{\bm{z}|\bm{y}} \mathcal{J}_{1}[S_{c} \circ d \circ g(\bm{z}, \bm{y}), \bm{y}].
\end{align}
For data generation tasks of specific categories, the generator's optimization problem can be described as:
\begin{align}
    \max_{\theta_{g}}~~~~ \mathbb{E}_{\bm{z}|\bm{y}} \bm{y}^{T}\ln S_{c} \circ d \circ g(\bm{z}, \bm{y}),
\end{align}
where the output layer of the discriminator network for this specific condition is suitably chosen as the Softmax function.

In learning the conditional adversarial generator network, a core issue is quantifying the conditions and defining an error function that measures the deviation from these conditions—specifically, the function \( \mathcal{J}_{1}[S_{c} \circ d(\bm{x}), \bm{y}] \) mentioned earlier. We use labeled data, unlabeled data, and generated pseudo-data to jointly train the feature extraction function \( d(\bm{x}) \) of the discriminator network. Subsequently, the discriminator network is used to progressively train the generator network, enabling the training of the directed data generation network.

\subsubsection{Style-Specified Data Generation}

The mainstream network for style transformation is the Cycle Consistent Generative Adversarial Network (CycleGAN) \cite{zhu2017unpaired}. This type of adversarial generative network typically requires at least two datasets:
\begin{itemize}
    \item The dataset to be transformed \( \{\bm{z}_{n}, \forall n\} \);
    \item The target dataset \( \{\bm{x}_{n}, \forall n\} \).
\end{itemize}
CycleGAN achieves style transformation using two generator networks and two discriminator networks, defined for convenience as \( g_{1}(\bm{z}) \), \( g_{2}(\bm{x}) \), \( d_{1}(\bm{x}) \), and \( d_{2}(\bm{z}) \). Here, \( g_{1}(\bm{z}) \) is the generator of primary interest, used to generate data samples of the specified style, \( \hat{x} = g_{1}(\bm{z}) \); \( g_{2}(\bm{x}) \) is used to restore the original data \( \bm{z} \), ensuring consistency during style transformation. The discriminator networks \( d_{1}(\bm{x}) \) and \( d_{2}(\bm{z}) \) guide the designs of \( g_{1}(\bm{z}) \) and \( g_{2}(\bm{x}) \), respectively.

Given the conditions set by the discriminator networks, the design objectives for the generator networks \( g_{1}(\bm{z}) \) and \( g_{2}(\bm{x}) \) can typically be described as follows:

\begin{align}
    1 &\leftarrow d_{1} \circ g_{1}(\bm{z}),\\
    1 &\leftarrow d_{2} \circ g_{2}(\bm{x}),\\
    \bm{z} &\leftarrow g_{1}(\bm{z}),\\
    \bm{x} &\leftarrow g_{2}(\bm{x}),\\
    \bm{x} &\leftarrow g_{1}(\hat{\bm{z}}) = g_{1} \circ g_{2}(\bm{x}),\\
    \bm{z} &\leftarrow g_{2}(\hat{\bm{x}}) = g_{2} \circ g_{1}(\bm{z}).
\end{align} 
These six design objectives can be categorized into three parts: 1) The first two equations aim for the generator to produce high-quality data through the discriminator; 2) The middle two equations ensure that the generator does not make unnecessary changes, altering only what is indeed required; 3) The final two equations impose cycle consistency constraints, allowing data to transform to the target domain and back to the original domain, thus ensuring that the generated data maintains a certain consistency. Consequently, the optimization problem for the generator networks can be formulated as:
\begin{align}
    \min_{\theta_{g_{1}}, \theta_{g_{2}}} &  -\mathbb{E}_{\bm{z}} \ln d_{1} \circ g_{1}(\bm{z}) - \mathbb{E}_{\bm{x}} \ln d_{2} \circ g_{2}(\bm{x}) \nonumber \\
    & + \mathbb{E}_{\bm{z}} \left\|\bm{z} - g_{1}(\bm{z})\right\|_{2}^{2} + \mathbb{E}_{\bm{x}} \left\|\bm{x} - g_{2}(\bm{x})\right\|_{2}^{2} \nonumber \\
    & + \mathbb{E}_{\bm{z}} \left\|\bm{z} - g_{2} \circ g_{1}(\bm{z})\right\|_{2}^{2} + \mathbb{E}_{\bm{x}} \left\|\bm{x} - g_{1} \circ g_{2}(\bm{x})\right\|_{2}^{2}.
\end{align}
This can be expressed as:
\begin{align}
   & \min_{\theta_{g_{1}}, \theta_{g_{2}}}  \mathbb{E}_{\bm{z}} \left[-\ln d_{1} \circ g_{1}(\bm{z}) + \left\|\bm{z} - g_{1}(\bm{z})\right\|_{2}^{2} + \left\|\bm{z} - g_{2} \circ g_{1}(\bm{z})\right\|_{2}^{2} \right] \nonumber \\
    & + \mathbb{E}_{\bm{x}} \left[-\ln d_{2} \circ g_{2}(\bm{x}) + \left\|\bm{x} - g_{2}(\bm{x})\right\|_{2}^{2} + \left\|\bm{x} - g_{1} \circ g_{2}(\bm{x})\right\|_{2}^{2}\right].\nonumber
\end{align}
To adjust the weights of different parts of the objective function, proportional coefficients can be introduced between the various objective functions, such as \( -\ln d_{1} \circ g_{1}(\bm{z}) + \alpha_{1} \left\|\bm{z} - g_{1}(\bm{z})\right\|_{2}^{2} + \alpha_{2} \left\|\bm{z} - g_{2} \circ g_{1}(\bm{z})\right\|_{2}^{2} \), where \( \alpha_{1} > 0 \) and \( \alpha_{2} > 0 \) are two constants.

In contrast, the training objectives for the two discriminator networks in CycleGAN are relatively straightforward. They align with the objectives of discriminator networks in classical GANs, described as follows:

\begin{align}
    1 &\leftarrow d_{1}(\bm{x}),\\
    0 &\leftarrow d_{1}(\hat{\bm{x}}) = d_{1} \circ g_{1}(\bm{z}),\\
    1 &\leftarrow d_{2}(\bm{z}),\\
    0 &\leftarrow d_{2}(\hat{\bm{z}}) = d_{2} \circ g_{2}(\bm{x}).
\end{align}
Using binary cross-entropy, the corresponding optimization problems can be expressed as:
\begin{align}
    \max_{\theta_{d_{1}}} & \mathbb{E}_{\bm{x}} \ln d_{1}(\bm{x}) + \mathbb{E}_{\bm{z}} \ln \left[1 - d_{1} \circ g_{1}(\bm{z})\right],\\
    \max_{\theta_{d_{2}}} & \mathbb{E}_{\bm{x}} \ln d_{2}(\bm{x}) + \mathbb{E}_{\bm{z}} \ln \left[1 - d_{2} \circ g_{2}(\bm{z})\right].
\end{align}
The CycleGAN framework constructs multiple objective functions for the generator \( g_{1}(\bm{z}) \), incorporating cycle consistency constraints and data consistency constraints, thereby constraining the offset between the generated signals and the original signals. Unlike conditional GANs, CycleGAN's training strategy provides boundary constraints for the generator based on data from the target domain, aligning more closely with data-driven deep learning approaches.

\subsubsection{Generative Adversarial Strategies for Few-Shot Classification and GAN-Assisted Feature Learning}

In many practical applications, the data samples themselves may not be scarce; instead, it is the reliable labeled samples that are limited. The challenge of building classification or detection networks with only a small number of labeled samples has been a hot topic in the relevant fields. One of the core issues is how to effectively utilize unlabeled samples during the training process.

Since GANs are capable of modeling data distributions, the discriminator within the adversarial network serves as a feature learning network. Therefore, if the training strategy of the GAN (specifically, the objective function related to the discriminator) can be integrated into the construction of the classification network, this would effectively leverage the information provided by unlabeled data. This is the fundamental idea behind the Semi-supervised Generative Adversarial Network (SGAN) \cite{odena2016semi}.

For convenience, we define the generator network as \( g(\bm{z}) \) and the discriminator network as \( S_{d} \circ d(\bm{x}) \). Unlike the previous definitions of the discriminator network, here we separate the last layer of the discriminator network and define it as \( S_{d}(\cdot) \), where the subscript \( d \) indicates the Sigmoid function. Analogously, we define the output layer constructed using the Softmax function as \( S_{x}(\cdot) \). Under this series of definitions, the classifier to be trained \( f(\bm{x}) \) can be represented as:
\begin{align}
    f(\bm{x}) = S_{x} \circ d(\bm{x}).
\end{align}
By completing the parameter learning for functions \( S_{x}(\cdot) \) and \( d(\bm{x}) \), namely \( \theta_{S_{x}} \) and \( \theta_{d} \), the construction of the classifier \( f(\bm{x}) \) can be finalized. The parameter \( \theta_{S_{x}} \) can only be learned using the labeled dataset, while \( \theta_{d} \) can utilize all available data.

In the framework of generative adversarial strategies for few-shot classification, the data typically consists of the following three parts:

\begin{itemize}
    \item A labeled dataset \( \{(\bm{x}_{n}, \mathbf{y}_{n}), \forall n\} \), where \( \bm{x}_{n} \) represents the \( n \)-th sample data and \( \bm{y}_{n} \) is its label, typically a one-hot vector; this part of the dataset is usually quite small.
    \item An unlabeled dataset \( \{\ddot{\bm{x}}_{n}, \forall n\} \), where the superscript \( \ddot{(\cdot)} \) is used to differentiate it from the labeled data \( \bm{x}_{n} \); this dataset is often very large.
    \item A dataset generated by the network \( \{(\bm{z}_{n}, \hat{\bm{x}}_{n})\} \), where \( \bm{z}_{n} \) is a random variable following a certain distribution; this dataset is output by the generator in the GAN.
\end{itemize}
All parameter learning requires an objective function, which in turn requires data labels. For the three categories of data mentioned, the labels are defined as follows:
\begin{align}
    \bm{y} &\leftarrow S_{x} \circ d(\bm{x}),\\
    1 &\leftarrow S_{d} \circ d(\ddot{\bm{x}}),\\
    0 &\leftarrow S_{d} \circ d(\hat{\bm{x}}) = S_{d} \circ d \circ g(\bm{z}).
\end{align}
Clearly, the objective function corresponding to the first equation is cross-entropy, while the objective functions for the latter two equations are binary cross-entropy. Thus, the training for the classification network can be summarized as:
\begin{align}
    \min_{\theta_{d};\theta_{S_{x}},\theta_{S_{d}}} & -\mathbb{E}_{\bm{x}} \bm{y}^{T} \ln S_{x} \circ d(\bm{x})- \nonumber \\
    &  \mathbb{E}_{\ddot{\bm{x}}} \ln S_{d} \circ d(\ddot{\bm{x}})   - \mathbb{E}_{\hat{\bm{x}}} \ln\left[1 - S_{d} \circ d(\hat{\bm{x}})\right],
\end{align} 
which can be expressed as:
\begin{align}
	\min_{\theta_{d};\theta_{S_{x}},\theta_{S_{d}}}&
	-\mathbb{E}_{\bm{x}} \bm{y}^{T}\ln  S_{x}\circ d(\bm{x})
	- \nonumber\\
	&\mathbb{E}_{\ddot{\bm{x}}}\ln S_{d}\circ d(\ddot{\bm{x}})
	- \mathbb{E}_{\bm{z}} \ln\left[1- S_{d}\circ d\circ g(\bm{z})\right],
\end{align}
where \( \theta_{d} \) and \( \theta_{S_{x}} \) are the parameters needed to construct the classifier \( f(\bm{x}) \), while \( \theta_{S_{d}} \) are the parameters required for auxiliary training, which are not needed after the network learning is complete.

To complete the learning of the function \( d(\bm{x}) \) in the classifier, we also need to construct the generator network \( g(\bm{z}) \). Since the objective of the generator network is to maximize \( S_{d} \circ d \circ g(\bm{z}) \), its objective function can be any of those in equations (\ref{GAN-opt-4-g}), (\ref{GAN-opt-4-g-1}), or (\ref{GAN-opt-4-g-2}). By selecting equation (\ref{GAN-opt-4-g-2}) as the objective function for the generator network, the training of the network can ultimately be organized as follows:

\begin{align}
    \max_{\theta_{g}} & \mathbb{E}_{\bm{z}} S_{d} \circ d \circ g(\bm{z}),\\
    \max_{\theta_{d};\theta_{S_{x}},\theta_{S_{d}}} & \mathbb{E}_{\bm{x}} \bm{y}^{T} \ln S_{x} \circ d(\bm{x}) + \nonumber\\
    & \mathbb{E}_{\ddot{\bm{x}}} \ln S_{d} \circ d(\ddot{\bm{x}}) + \mathbb{E}_{\bm{z}} \ln\left[1 - S_{d} \circ d \circ g(\bm{z})\right]. \label{sgan-cost-4-d}
\end{align}
By alternately updating the parameter sets \( \theta_{g} \) and \( \{\theta_{d};\theta_{S_{x}},\theta_{S_{d}}\} \), the network training can be completed. From these two equations, it is evident that both the core networks \( S_{x}(\cdot) \) and \( d(\bm{x}) \) learn from all three types of datasets. Under the condition of network convergence, this approach often yields good performance. Once the network is constructed, simply extracting the modules \( S_{x}(\cdot) \) and \( d(\bm{x}) \) allows the construction of \( f(\bm{x}) = S_{x} \circ d(\bm{x}) \), thereby completing the learning of the semi-supervised few-shot classifier.

In addition to few-shot classification tasks, similar functions such as interpolation and fitting can leverage generative adversarial strategies to learn more effective features. However, the Softmax function following \( d(\bm{x}) \) should be replaced with a task-related network function, such as a fully connected network \( F(\cdot) \). Simultaneously, the first objective function in equation (\ref{sgan-cost-4-d}) can be modified to a mean squared error objective, expressed as \( \mathbb{E}_{\bm{x}} \left\|\bm{y} - F_{2} \circ F_{1} \circ d(\bm{x})\right\|_{2}^{2} \).
    
\subsection{Optimal Transport: Aligning Data Densities without Labels}

Optimal transport theory has a wide range of applications and can define distances between different variable probability distributions, which can guide the training of network functions through error backpropagation. The optimal transport problem focuses not on the transportation issue itself but on the cost function under optimal transport, which can be used in neural network training to align data distributions.

Given two random variables \( \bm{x} \) and \( \breve{\bm{x}} \), and their probability distributions \( p(\bm{x}) \) and \( q(\breve{\bm{x}}) \), along with a distance \( c(\bm{x}, \breve{\bm{x}}) \) between the two variables, we typically have \( \bm{x} \) and \( \breve{\bm{x}} \) sharing the same physical meaning and having identical data dimensions. If we have the joint distribution \( \hbar(\bm{x}, \breve{\bm{x}}) \) of the two random variables, we can define the average distance as follows:
\begin{align}
    \varepsilon(p, q) = \int_{\bm{x}} \int_{\breve{\bm{x}}} \hbar(\bm{x}, \breve{\bm{x}}) c(\bm{x}, \breve{\bm{x}}).
\end{align}
In practice, we do not know the joint distribution \( \hbar(\bm{x}, \breve{\bm{x}}) \), only the marginal distributions \( p(\bm{x}) \) and \( q(\breve{\bm{x}}) \). For a given marginal distribution, multiple joint distributions may correspond to it. Among these, one optimal distribution minimizes the average distance \( \varepsilon(p, q) \), while satisfying the constraints:
\begin{align}
\mathcal{J}_{c}(p, q)
 	=  \min_{\hbar(\bm{x}, \breve{\bm{x}})} &\int_{\bm{x}} \int_{\breve{\bm{x}}}\hbar(\bm{x}, \breve{\bm{x}}) c(\bm{x}, \breve{\bm{x}})\\ \mbox{~~~~ s.t. ~~~} 
    &\int_{\bm{x}} \hbar(\bm{x}, \breve{\bm{x}}) = q(\breve{\bm{x}}),\\
    &\int_{\breve{\bm{x}}} \hbar(\bm{x}, \breve{\bm{x}}) = p(\bm{x}),\\
    &\hbar(\bm{x}, \breve{\bm{x}}) \geq 0, \quad \forall \bm{x}, \breve{\bm{x}}.
\end{align}
This defines a distance between the two distributions \( p(\bm{x}) \) and \( q(\breve{\bm{x}}) \). In practice, the marginal distributions are typically characterized directly from the data. Given two datasets \( \{\bm{x}_{k}, \forall k=1,2,\ldots, K\} \) and \( \{\breve{x}_{n}, \forall n=1,2,\ldots, N\} \), we can define matrices \( \bm{H} \) and \( \bm{C} \) of size \( K \times N \), along with vectors \( \bm{p} \) and \( \bm{q} \) of lengths \( K \)and \( N \); their specific definitions are as follows:
\begin{align}
    [\bm{H}]_{k,n} &= \hbar(\bm{x}_{k}, \breve{\bm{x}}_{n}), \\
    [\bm{C}]_{k,n} &= c(\bm{x}_{k}, \breve{\bm{x}}_{n}), \\
    [\bm{p}]_{k} &= p(\bm{x}_{k}), \\
    [\bm{q}]_{n} &= q(\breve{x}_{n}).
\end{align}
With these variable definitions, the average distance under the optimal joint distribution can be expressed as:
\begin{align}
    \mathcal{J}_{c}(p, q) = \min_{\bm{H}} \mathrm{tr} \left(\bm{H}^{T} \bm{C}\right) 
    \quad \text{s.t.} \quad 
    & \bm{H} \bm{1}_{N} = \bm{p}, \\
    & \bm{H}^{T} \bm{1}_{K} = \bm{q}, \\
    & [\bm{H}]_{k,n} \geq 0, \quad \forall k, n.
\end{align}
For convenience, let the optimal \( \bm{H} \) be denoted as \( \bm{H}_{o} \). The average distance under the optimal joint distribution can be simply represented as \( \mathcal{J}_{c}(p, q) = \mathrm{tr} \left(\bm{H}_{o}^{T} \bm{C}\right) \). The optimal distribution \( \bm{H}_{o} \) encodes all useful information about the distance matrix \( \bm{C} \) and the marginal distribution vectors \( \bm{p} \) and \( \bm{q} \).

Given a neural network \( \bm{x} = f(\bm{z}) \) to be trained, the objective is to make the distribution of \( \bm{x} \) closely approximate the distribution of \( \breve{\bm{x}} \). The distribution of \( \breve{\bm{x}} \) is characterized by the dataset \( \{\breve{x}_{n}, \forall n=1,2,\ldots, N\} \), while the distribution of \( \bm{x} \) is characterized by the dataset \( \{\bm{x}_{k}, \forall k=1,2,\ldots, K\} \) or \( \{f(\bm{z}_{k}), \forall k=1,2,\ldots, K\} \). To complete the training of \( f(\bm{z}) \), the goal is to compute the gradient of the target function with respect to the network output \( \bm{x} \), determining the ``direction" for the network output.

To seek the direction of network training when given \( \bm{x}_{k} \), we first need to compute the term related to \( \bm{x}_{k} \) in the average distance of the optimal joint distribution, expressed as:
\begin{align}
    \sum_{n} \left[\bm{H}_{o}\right]_{k,n} c(\bm{x}_{k}, \breve{\bm{x}}_{n}).
\end{align}
This means that the target function for each sample \( \bm{x}_{k} \) is constructed from all samples in the target dataset \( \{\breve{\bm{x}}_{n}\} \). Assuming the neural network \( f(\bm{z}) \) has \( M \) layers, given the input \( \bm{z}_{k} \), the gradient of the target function with respect to the network output \( \bm{x}_{k} \) can be computed as:
\begin{align}
    e_{k}^{(M)} = \sum_{n} \left[\bm{H}_{o}\right]_{k,n} \frac{\partial c(\bm{x}_{k}, \breve{\bm{x}}_{n})}{\partial \bm{x}_{k}}. \label{t-distance}
\end{align}
Given the distance function \( c(\bm{x}_{k}, \breve{\bm{x}}_{n}) \), we can compute the gradient of the target function with respect to the network output \( \bm{x}_{k} \); using backpropagation, we can complete the training of the neural network.

Consider a special case where \( c(\bm{x}_{k}, \breve{\bm{x}}_{n}) = \left\|\bm{x}_{k} - \breve{\bm{x}}_{n}\right\|_{2}^{2} \); according to (\ref{t-distance}), we can derive the gradient of the target function with respect to the network output as:
\begin{align}
    e_{k}^{(M)} = 2 \sum_{n} \left[\bm{H}_{o}\right]_{k,n} (\bm{x}_{k} - \breve{\bm{x}}_{n}) \\
    = 2 p(\bm{x}_{k}) \left(\bm{x}_{k} - \sum_{n} \alpha_{k,n} \breve{\bm{x}}_{n}\right),
\end{align}
where
\begin{align}
    \alpha_{k,n} = \frac{\left[\bm{H}_{o}\right]_{k,n}}{\sum_{n}\left[\bm{H}_{o}\right]_{k,n}} = \frac{\left[\bm{H}_{o}\right]_{k,n}}{p(\bm{x}_{k})}.
\end{align}
By definition, the coefficients \( \alpha_{k,n} \) satisfy \( 0 \leq \alpha_{k,n} \leq 1 \) and \( \sum_{n} \alpha_{k,n} = 1 \). In other words, by performing a weighted sum of the samples from the dataset \( \{\breve{\bm{x}}_{n}\} \), we can obtain the label for \( \bm{x}_{k} \), given by \( \sum_{n} \alpha_{k,n} \breve{\bm{x}}_{n} \), where the weights \( \alpha_{k,n} \) are optimized using optimal transport theory. This addresses a key question in network training: in which direction should each sample \( \bm{x}_{k} \) move to ensure that the overall distribution approaches the distribution of the given dataset \( \{\breve{\bm{x}}_{n}\} \).

The term \emph{optimal transport} refers to a classic mathematical problem\cite{villani2009optimal,lei2019geometric}. This concept can be illustrated with the transportation of goods between cities. Suppose there are two cities, A and B, with \( K \) shops in city A and \( N \) shops in city B. The goal is to transport goods from city A to city B in the most efficient way, embodying the idea of ``optimal transport." The distance from shop \( k \) in city A to shop \( n \) in city B is defined as \( c(\bm{x}_{k}, \breve{\bm{x}}_{n}) \), with their locations given by \( \bm{x}_{k} \) and \( \breve{\bm{x}}_{n} \). Additionally, the quantity of goods transported from shop \( k \) in city A to shop \( n \) in city B is \( \hbar(\bm{x}_{k}, \breve{\bm{x}}_{n}) \). The cost of sending these goods can be expressed as \( \sum_{k} \sum_{n} \hbar(\bm{x}_{k}, \breve{\bm{x}}_{n})c(\bm{x}_{k}, \breve{\bm{x}}_{n}) \). Given the predetermined quantity of goods at each shop in city A, denoted as \( p(\bm{x}_{k}) \), we have \( \sum_{n} \hbar(\bm{x}_{k}, \breve{\bm{x}}_{n}) = p(\bm{x}_{k}) \). Similarly, we want to ensure that each shop in city B receives the expected quantity of goods \( q(\breve{\bm{x}}_{n}) \), leading to \( \sum_{k} \hbar(\bm{x}_{k}, \breve{\bm{x}}_{n}) = q(\breve{\bm{x}}_{n}) \). This simple arrangement reveals that the optimization problem aligns with the previously defined optimization problem for the optimal joint distribution.

The concept of \emph{Earth Mover's Distance}\cite{rubner2000earth,arjovsky2017wasserstein} can be understood by considering the goods transported between cities as ``earth," and the machine used for transportation as a bulldozer (earth mover). In this scenario, the shops within the same city are closely situated (imagine a pile of earth). The problem of distributing goods in the optimal transport scenario transforms into the task of moving earth: given a desired pile of earth \( B \) with a specific shape and a ready pile \( A \), how can we optimally excavate earth from each grid point in pile \( A \) and allocate it to grid points in pile \( B \) to form the desired pile? This optimal method, combined with the distances between the grid points of the two piles, allows us to compute the ``average distance under the optimal joint distribution," which is the Earth Mover's Distance. The practical concern is not about the optimal transportation strategy itself, but rather the distance defined by this optimal strategy between the two piles (datasets), optimizing this distance to achieve distribution alignment and complete the training of the network.

\subsection{Area Under the ROC Curve: The Optimal Optimization Target in Detection Problems}

Detection algorithms often involve a trade-off between false positive rates and recall. For a given algorithm, a high recall can lead to a high risk of false positives, while a low false positive rate may incur a risk of missing detections. Evaluating a single metric alone does not adequately reflect the performance differences between algorithms. The ROC curve\footnote{Receiver Operating Characteristic.} provides a solution: if the ROC curve of one method completely encompasses that of another, we can conclude that the first method is superior. Furthermore, the area under the ROC curve (AUC) can serve as an optimization objective for training neural networks. We will elaborate on these points.

Consider a binary classification problem, specifically the classic signal detection issue: if an event occurs, the function outputs 1; if the event does not occur, it outputs 0. Given a training set \( \{(\bm{x}_{n}, y_{n}), \forall n=1,2,\ldots,N\} \), where \( y_{n}=1 \) indicates that the event occurs and \( y_{n}=0 \) indicates that it does not, the positive samples correspond to \( y_{n}=1 \), and the negative samples correspond to \( y_{n}=0 \). For convenience, we can define two indexed datasets: \( \mathcal{I}^{+} = \{n | y_{n}=1\} \) and \( \mathcal{I}^{-} = \{n | y_{n}=0\} \); the two datasets define the indices of positive and negative samples, containing \( N^{+} \) and \( N^{-} \) samples, respectively.

For detection problems, we typically design a function \( \hbar(\cdot) \), which can output a feature vector or embedding vector \( \bm{z}_{n} = \hbar(\bm{x}_{n}) \) for each sample \( \bm{x}_{n} \). Based on \( \bm{z}_{n} \), we can compute a score \( s_{n} \) as follows:
\begin{align}
    s_{n} = \frac{1}{1 + \expd{-\bm{w}^{T} \bm{z}_{n} + b}},
\end{align}
or
\begin{align}
    s_{n} = \frac{\bm{w}^{T} \bm{x}_{n}}{\left\|\bm{w}\right\| \cdot \left\|\bm{z}\right\|}.
\end{align}
In general, this score satisfies \( s_{n} < 1 \); a higher score indicates a greater probability of the event occurring. In practice, by setting a threshold \( \epsilon \), we can determine that if \( s_{n} \) exceeds this threshold, the sample \( \bm{x}_{n} \) is classified as containing the event; otherwise, it is classified as not occurring.

\subsubsection{Recall and False Positive Rate}

For a given threshold \( \epsilon \), recall is defined as the proportion of positive samples that have scores exceeding this threshold. In contrast, the false positive rate indicates how many negative samples have scores above this threshold, i.e., how many negative samples are misclassified as positive. By definition, recall \( t(\epsilon) \) and false positive rate \( f(\epsilon) \) can be calculated as follows:
\begin{align}
    t(\epsilon) &= \frac{1}{N^{+}} \sum_{n\in \mathcal{I}^{+}} \kappa\left(s_{n}\geq \epsilon\right), \\
    f(\epsilon) &= \frac{1}{N^{-}}\sum_{m \in \mathcal{I}^{-}} \kappa\left(s_{m} \geq \epsilon\right),
\end{align}
where \( \kappa(\mathrm{condition})=1 \) if the condition is true; otherwise, it is zero.

In practice, by lowering the threshold \( \epsilon \), we can often increase the recall; however, doing so typically raises the risk of a higher false positive rate. Conversely, setting a high threshold \( \epsilon \) effectively reduces the false positive rate but carries the risk of low recall; that is, if the event is present, we may fail to detect it under the high threshold assumption. Thus, discussing false positive rates without considering recall or vice versa is unscientific.

\subsubsection{Area Under the ROC Curve and Optimization Problems}

By definition, the area under the ROC curve (AUC) can be calculated as follows\cite{zhou2021machine,spackman1989signal}:
\begin{align}
    \mathrm{AUC} &= \int_{0}^{1} t(\epsilon) d f(\epsilon) \\
    &\approx \sum_{i=1}^{L-1} \left[f(\epsilon_{i+1}) - f(\epsilon_{i})\right] \cdot \frac{1}{2} \left[t(\epsilon_{i+1}) + t(\epsilon_{i})\right], \label{AUC-1}
\end{align}
where \( 1 \geq \epsilon_{0} > \epsilon_{1} > \cdots > \epsilon_{L-1} > \epsilon_{L} \geq 0 \). This means that, on the ROC curve, we sample \( L \) points based on different thresholds \( \epsilon \), calculate \( t(\epsilon_{i}) \) and \( f(\epsilon_{i}) \) for each, and then use a discrete approximation to compute the AUC.

In practice, how to select \( \epsilon_{i} \) and the total number of discrete points \( L \) is critical. An effective approach is to design based on the score distribution of negative samples. First, sort the scores of the negative sample set to obtain \( s_{1}^{\prime} > s_{2}^{\prime} > \cdots > s_{N^{-}-1}^{\prime} > s_{N^{-}}^{\prime} \), where \( s_{i}^{\prime} \) corresponds to the score of a sample in the negative sample set. By setting \( \epsilon_{i}=s_{i}^{\prime} \) and \( L = N^{-} \), we can verify:
\begin{align}
    f(\epsilon_{i}) = \frac{i}{N^{-}}.
\end{align}
Substituting this result into (\ref{AUC-1}), we can derive:
\begin{align}
    \mathrm{AUC} &\approx \frac{1}{N^{-}} \sum_{i=1}^{N^{-}-1} \frac{1}{2} \left[t(s^{\prime}_{i+1}) + t(s^{\prime}_{i})\right] \nonumber \\
    &= \frac{1}{N^{-} N^{+}} \sum_{i=1}^{N^{-}-1} \frac{1}{2} \left[\sum_{n\in \mathcal{I}^{+}} \kappa\left(s_{n}\geq s^{\prime}_{i+1}\right) + \sum_{n\in \mathcal{I}^{+}} \kappa\left(s_{n}\geq s^{\prime}_{i}\right)\right] \nonumber \\
    &= \frac{1}{N^{-} N^{+}} \sum_{n\in \mathcal{I}^{+}} \sum_{m\in \mathcal{I}^{-}} \kappa\left(s_{n} \geq s_{m}\right) \label{AUC-org} \\
    &= 1 - \frac{1}{N^{-} N^{+}} \sum_{n\in \mathcal{I}^{+}} \sum_{m\in \mathcal{I}^{-}} \kappa\left(s_{n}< s_{m}\right). \label{AUC-org-2}
\end{align}
Considering that
\begin{align}
    \frac{1}{N^{+}} \sum_{n\in \mathcal{I}^{+}} \kappa\left(s_{n}< s_{m}\right) = 1-\frac{1}{N^{+}} \sum_{n\in \mathcal{I}^{+}} \kappa\left(s_{m}\geq s_{n}\right),
\end{align}
it is straightforward to prove the transition from (\ref{AUC-org}) to (\ref{AUC-org-2}).

For a given sample set \( \{\bm{x}_{n}\} \), we can calculate the feature vector for each sample using \( \bm{z}_{n} = \hbar(\bm{x}_{n}) \), and based on the feature vector, compute the score set \( \{s_{n}\} \). By dividing the score set into two parts based on positive and negative samples \( \{s_{n}, \forall n\in \mathcal{I}^{+}\} \) and \( \{s_{m}, \forall m\in \mathcal{I}^{-}\} \), we can substitute these sets into (\ref{AUC-org}) to calculate the AUC value corresponding to the current detection algorithm. The closer this value is to 1, the better the performance of the detection algorithm.

The AUC calculation formula in (\ref{AUC-org}) cannot be computed for a single sample, nor can we obtain the gradient of AUC with respect to the feature vector \( \bm{z}_{n} \), making it challenging to use error backpropagation for parameter updates. To address this, we need to reformulate the AUC in (\ref{AUC-org}) and apply appropriate scaling.

Consider the following three inequalities:
\begin{align}
    \kappa(x< 0) &\leq \max(0, 1-x) = \kappa(x< 0) \cdot (1-x), \\
    -x &\leq -\ln\left(\frac{1}{1+\expd{-x}}\right), \\
    x &\leq -\ln \left(\frac{1}{1+\expd{x}}\right) = -\ln\left(1-\frac{1}{1+\expd{-x}}\right).
\end{align}
From this, we can derive:
\begin{align}
    \mathcal{J} &= 1 - \mathrm{AUC} \\
    &= \frac{1}{N^{-} N^{+}} \sum_{n\in \mathcal{I}^{+}} \sum_{m\in \mathcal{I}^{-}} \kappa\left(s_{n}< s_{m}\right) \\
    &= 1 - \frac{1}{N^{-} N^{+}} \sum_{n\in \mathcal{I}^{+}} \sum_{m\in \mathcal{I}^{-}} \kappa\left(s_{n}-s_{m}< 0\right) \\
    &\leq \frac{1}{N^{-} N^{+}} \sum_{n\in \mathcal{I}^{+}} \sum_{m\in \mathcal{I}^{-}} \kappa\left(s_{n}< s_{m}\right) \cdot \left(1-s_{n} + s_{m}\right) \\
    &=-\frac{1}{N^{+}} \sum_{n\in \mathcal{I}^{+}} \varpi_{n} \cdot (s_{n}-1) + \frac{1}{N^{-}} \sum_{m \in \mathcal{I}^{+}} \varpi_{m}^{\prime} \cdot s_{m} \\
    &\leq -\frac{1}{N^{+}} \sum_{n\in \mathcal{I}^{+}} \varpi_{n} \cdot \ln \left(\frac{1}{1+\expd{-s_{n}+1}}\right) - \nonumber\\
    &~~~~~~\frac{1}{N^{-}} \sum_{m \in \mathcal{I}^{+}} \varpi_{m}^{\prime} \cdot \ln\left(1-\frac{1}{1+\expd{-s_{m}}}\right), \label{J-ROC-1}
\end{align}
where
\begin{align}
    \varpi_{n} &= \frac{1}{N^{-}} \sum_{m \in \mathcal{I}^{+}}\kappa\left(s_{n}< s_{m}\right) \\
    &= \frac{1}{N^{-}} \sum_{m \in \mathcal{I}^{+}}\kappa\left(s_{m}> s_{n}\right) \\
    &\approx f(s_{n}), \label{f-sn-1} \\
    \varpi_{m}^{\prime} &= \frac{1}{N^{+}} \sum_{n\in \mathcal{I}^{+}}\kappa\left(s_{n}< s_{m}\right) \\
    &= 1 - \frac{1}{N^{+}} \sum_{n\in \mathcal{I}^{+}}\kappa\left(s_{m}\geq s_{n}\right) \\
    &= 1 - t(s_{m}), \label{t-sm-1}
\end{align}
Notably, \( f(s_{n}) \) represents the false positive rate when the threshold is \( \epsilon=s_{n} \); \( t(s_{m}) \) represents the recall when the threshold is \( \epsilon=s_{m} \). Both the false positive rate and recall can be computed during forward propagation for a given batch of data, and during backpropagation, they can be treated as fixed values.

Combining (\ref{J-ROC-1}), (\ref{f-sn-1}), and (\ref{t-sm-1}), we can express the cost function related to the detector as follows:
\begin{align}
    \mathcal{J}_{1}(\hbar) &= -\frac{1}{N^{+}} \sum_{n\in \mathcal{I}^{+}} f(s_{n}) \cdot \ln \left(\frac{1}{1+\expd{-s_{n}+1}}\right) \nonumber \\
    &\quad - \frac{1}{N^{-}} \sum_{m \in \mathcal{I}^{+}} \left[1 - t(s_{m})\right] \cdot \ln\left(1 - S_{d} \circ d(\hat{\bm{x}})\right).
\end{align}
Thus, we obtain a differentiable cost function concerning \( s_{n} \), \( s_{m} \), \( \bm{z}_{n} \), \( \bm{z}_{m} \), and the parameters of the detector \( \hbar(\cdot) \). This cost function can be directly used for training the neural network. In practice, \( f(s_{n}) \) and \( t_{s_{m}} \) can also be parameterized to yield more diverse cost functions.

\subsubsection{Discussions on the $F_{1}$-Score}

Previously, we mentioned that it is unreasonable to discuss false positive rates without considering recall or vice versa. In practice, when given a threshold \( \epsilon \), we often need a single value to reflect the quality of the method. On one hand, we can seek the optimal threshold value; on the other hand, we can directly perform preliminary comparisons of performance between methods. The value of the $F_{1}$-Score can serve this purpose.

Given a sample set \( \{\bm{x}_{n}\} \) and its positive and negative sample sets \( \{s_{n}, \forall n\in \mathcal{I}^{+}\} \) and \( \{s_{m}, \forall m\in \mathcal{I}^{-}\} \), for a given detector, let the number of Positive samples classified as Positive be denoted as \( \mathrm{PP} \), those classified as Negative as \( \mathrm{PN} \); simultaneously, let the number of Negative samples classified as Positive be \( \mathrm{NP} \), and those classified as Negative be \( \mathrm{NN} \). By definition, we have:
\begin{align}
    \mathrm{PP} + \mathrm{PN} &= N^{+}, \\
    \mathrm{NP} + \mathrm{NN} &= N^{-}, \\
    \text{Recall} &= \{R, \mathrm{TPR}, t(\epsilon)\} = \frac{\mathrm{PP}}{\mathrm{PP} + \mathrm{PN}} = \frac{\mathrm{PP}}{N^{+}}, \\
    \text{False Positive Rate} &= \{\mathrm{FPR}, f(\epsilon)\} = \frac{\mathrm{NP}}{\mathrm{NP} + \mathrm{NN}} = \frac{\mathrm{NP}}{N^{-}}, \\
    \text{Precision} &= \{P\} = \frac{\mathrm{PP}}{\mathrm{PP} + \mathrm{NP}}.
\end{align}
Here, ``Recall" is typically denoted as \( R \) (for Recall) or \( \mathrm{TPR} \) (for True Positive Rate), indicating the probability of detecting the event among all positive samples. ``False Positive Rate" is generally denoted as \( \mathrm{FPR} \), indicating the probability of misclassifying negative samples as positive. ``Precision" is typically denoted as \( P \) (for Precision), which indicates the probability that a sample classified as positive is actually positive.

Recall and false positive rate are a pair of contradictory relationships: high recall carries the risk of high false positive rates, which is described by the ROC curve; recall and precision also have a conflicting relationship: high recall risks low precision, which is described by the P-R curve.

Given a detector and a threshold, the value of the \( F_{1} \)-Score can be expressed as:
\begin{align}
    F_{1} &= 2 \times \frac{\text{Precision} \times \text{Recall}}{\text{Precision} + \text{Recall}} \\
    &= 2 \times \frac{P \times R}{P + R} \\
    &= 2 \times \frac{t(\epsilon)}{t(\epsilon) + \alpha f(\epsilon) + 1}. \label{F1-f-t}
\end{align}
Here, \( \alpha = \frac{N^{-}}{N^{+}} \) is the ratio of the number of negative samples to positive samples in the dataset. According to (\ref{F1-f-t}), when \( \alpha \) is small, indicating that negative samples are relatively few, the \( F_{1} \)-Score mainly depends on recall. Conversely, when \( \alpha \) is large, implying that positive samples are scarce, the focus shifts to the ratio of recall to false positive rate. When \( \alpha = 1 \), meaning the number of positive and negative samples is roughly equal, the \( F_{1} \)-Score considers both false positive rate and recall, providing a comprehensive evaluation.

\subsection{Diffusion Models: From Signal to Noise, From Noise to Signal}

Given data \( \bm{x} \), we assume it can gradually degrade into a Gaussian distribution, and can also be generated from that Gaussian distribution step by step. During the degradation process, let the result at step \( t \) be denoted as \( \bm{x}^{(t)} \); there are a total of \( T \) steps, with the final signal denoted as \( \bm{x}^{(T)} \). By assumption, \( \bm{x}^{(T)} \) follows a Gaussian distribution, i.e., \( \bm{x}^{(T)} \sim \mathcal{N}\left(\bm{x}^{(T)}; \bm{0}, \bm{I}\right) \). For convenience, we denote the degradation process as:
\begin{align}
    &\bm{x} = \nonumber\\
    &\bm{x}^{(0)} \rightarrowtail \bm{x}^{(1)} \rightarrowtail \cdots \rightarrowtail \bm{x}^{(t-1)} \rightarrowtail \bm{x}^{(t)} \rightarrowtail \bm{x}^{(t+1)} \rightarrowtail \cdots \rightarrowtail \bm{x}^{(T)}.
\end{align}
Here, \( \rightarrowtail \) indicates the degradation process, while \( \leftarrow \) represents the generative process. If the degradation process \( q\left(\bm{x}^{(t)}| \bm{x}^{(t-1)}\right) \) is known, we can derive the conditional probability using Bayes' theorem \cite{luo2022understanding}:
\begin{align}
    q\left[\bm{x}^{(t-1)}| \bm{x}^{(t)}, \bm{x}^{(0)}\right] 
    &= \frac{q\left[\bm{x}^{(t)}| \bm{x}^{(t-1)}, \bm{x}^{(0)}\right] q\left[\bm{x}^{(t-1)}| \bm{x}^{(0)}\right]}{q\left(\bm{x}^{(t)}| \bm{x}^{(0)}\right)}
   \nonumber \\
	&=\dfrac{q\left[\bm{x}^{(t)}| \bm{x}^{(t-1)}\right] q\left[\bm{x}^{(t-1)}|  \bm{x}^{(0)}\right]}{q\left[\bm{x}^{(t)}|  \bm{x}^{(0)}\right]}. \label{q-xt1-xt-x0}
\end{align}
In a variational diffusion model, \( q\left[\bm{x}^{(t)}| \bm{x}^{(t-1)}\right] \) follows a Gaussian distribution, defined as:
\begin{align}
    q\left[\bm{x}^{(t)}| \bm{x}^{(t-1)}\right] = \mathcal{N}\left[\bm{x}^{(t)}; \sqrt{\alpha_{t}} \bm{x}^{(t-1)}, (1-\alpha_{t})\bm{I}\right], \label{xt-sample-vdm}
\end{align}
where \( \alpha_{t} \in [0, 1] \) is a value that varies with \( t \). Under the assumption of Markov conditions, it can be derived that:
\begin{align}
    q\left(\bm{x}^{(t)}| \bm{x}^{(0)}\right) = \mathcal{N}\left[\bm{x}^{(t)}; \sqrt{\breve{\alpha}_{t}} \bm{x}^{(0)}, (1-\breve{\alpha}_{t})\bm{I}\right], \label{xt-x0-sample-vdm}
\end{align}
where \( \breve{\alpha}_{t} = \prod_{i=1}^{t} \alpha_{i} \). Substituting (\ref{xt-sample-vdm}) and (\ref{xt-x0-sample-vdm}) into (\ref{q-xt1-xt-x0}) yields \( q\left[\bm{x}^{(t-1)}| \bm{x}^{(t)}, \bm{x}^{(0)}\right] \), which also follows a Gaussian distribution. The specific form of the function \( q\left[\bm{x}^{(t-1)}| \bm{x}^{(t)}, \bm{x}^{(0)}\right] \) will be discussed later.
\begin{itemize}
    \item Equation (\ref{xt-sample-vdm}) describes the process of gradual degradation of samples. In fact, given \( \bm{x}^{(t-1)} \), \( \bm{x}^{(t)} \) can be sampled using the formula:
    \begin{align}
        \bm{x}^{(t)} = \sqrt{\alpha_{t}} \bm{x}^{(t-1)} + \sqrt{(1-\alpha_{t})} \bm{\epsilon},
    \end{align}
    where \( \bm{\epsilon} \sim \mathcal{N}\left(\bm{\epsilon}; \bm{0}, \bm{I}\right) \) follows a standard normal distribution. This means that by continually introducing Gaussian noise to the degraded sample from the previous time step, we can complete the degradation of samples.
    \item For any given \( t \), \( \bm{x}^{(t-1)} \) is closer to the real sample than \( \bm{x}^{(t)} \).
    \item As time \( t \) increases, \( \breve{\alpha}_{t} \) approaches zero; according to (\ref{xt-x0-sample-vdm}), sampling for \( \bm{x}^{(t)} \) results in:
    \begin{align}
        \bm{x}^{(t)} = \sqrt{\breve{\alpha}_{t}} \bm{x}^{(0)} + \sqrt{(1-\breve{\alpha}_{t})} \bm{\epsilon},
    \end{align}
    which tends toward a Gaussian distribution.
\end{itemize}
Unlike (\ref{xt-sample-vdm}) and (\ref{xt-x0-sample-vdm}), equation (\ref{q-xt1-xt-x0}) describes a signal recovery process. Given a degraded sample \( \bm{x}^{(t)} \) and the original sample \( \bm{x}^{(0)} \), we can use the function \( q\left[\bm{x}^{(t-1)}| \bm{x}^{(t)}, \bm{x}^{(0)}\right] \) to obtain a sample that is closer to the original sample \( \bm{x}^{(0)} \) than \( \bm{x}^{(t)} \). However, this recovery process requires the original sample \( \bm{x}^{(0)} \) as input, which, therefore, cannot be directly used in practice to generate high-quality sample data \( \bm{x} \), i.e., \( \bm{x}^{(0)} \).

\subsubsection{Data Generation with Variational Diffusion Models}

For data generation tasks, we often wish to generate a set of high-quality samples that closely resemble real data from a given random variable. In the variational diffusion model approach, this process is gradually completed through multiple steps:
\begin{align}
    &\bm{x} = \nonumber \\
    &\bm{x}^{(0)} \leftarrow \bm{x}^{(1)} \leftarrow \cdots \leftarrow \bm{x}^{(t-1)} \leftarrow \bm{x}^{(t)} \leftarrow \bm{x}^{(t+1)} \leftarrow \cdots \leftarrow \bm{x}^{(T)} \nonumber\\
    &= \bm{\epsilon},
\end{align}
where \( \leftarrow \) indicates the generative process, and \( \bm{x}^{(T)} = \bm{\epsilon} \), with \( \bm{\epsilon} \sim \mathcal{N}\left(\bm{\epsilon}; \bm{0}, \bm{I}\right) \) being a sample from a zero-mean Gaussian random process. Analogous to the modeling of the degradation process \( q\left[\bm{x}^{(t)}| \bm{x}^{(t-1)}\right] \), the sample recovery/generation process aims to model a function \( p\left[\bm{x}^{(t-1)}| \bm{x}^{(t)}\right] \) that, given any \( \bm{x}^{(t)} \), can generate a sample \( \bm{x}^{(t-1)} \) that is closer to the true sample.

Recalling the definition of \( q\left[\bm{x}^{(t-1)}| \bm{x}^{(t)}, \bm{x}^{(0)}\right] \), it describes a data recovery/generation process. Therefore, the function \( p\left[\bm{x}^{(t-1)}| \bm{x}^{(t)}\right] \) can be directly modeled after \( q\left[\bm{x}^{(t-1)}| \bm{x}^{(t)}, \bm{x}^{(0)}\right] \). However, a critical issue must be resolved: \( q\left[\bm{x}^{(t-1)}| \bm{x}^{(t)}, \bm{x}^{(0)}\right] \) requires the true sample \( \bm{x}^{(0)} \) as input, while the modeled function \( p\left[\bm{x}^{(t-1)}| \bm{x}^{(t)}\right] \) cannot (nor should it) use the true sample as input. To address this issue, the variational diffusion model trains a neural network to produce an estimate of the true sample \( \hat{\bm{x}}^{(0)} \), which is then substituted into \( q\left[\bm{x}^{(t-1)}| \bm{x}^{(t)}, \bm{x}^{(0)}\right] \) to enable modeling of \( p\left[\bm{x}^{(t-1)}| \bm{x}^{(t)}\right] \):
\begin{align}
    \hat{\bm{x}}^{(0)} = f(\bm{x}^{(t)}, t).
\end{align}
Here, \( f(\cdot, \cdot) \) represents a neural network, with its optimizable parameter set defined as \( \theta_{f} \). Under the condition of the function \( f(\bm{x}^{(t)}, t) \), the function \( p\left[\bm{x}^{(t-1)}| \bm{x}^{(t)}\right] \) can thus be modeled as:
\begin{align}
    p\left[\bm{x}^{(t-1)}| \bm{x}^{(t)}\right] &= q\left[\bm{x}^{(t-1)}| \bm{x}^{(t)}, \hat{\bm{x}}^{(0)}\right] \\
    &= q\left[\bm{x}^{(t-1)}| \bm{x}^{(t)}, f(\bm{x}^{(t)}, t)\right]. \label{p-xt1-xt-vdm}
\end{align}
According to \cite{luo2022understanding}, \( q\left[\bm{x}^{(t-1)}| \bm{x}^{(t)}, \bm{x}^{(0)}\right] \) presented by (\ref{q-xt1-xt-x0})  also follows a Gaussian distribution \cite{luo2022understanding}:
\begin{align}
    & q\left[\bm{x}^{(t-1)}| \bm{x}^{(t)}, \bm{x}^{(0)}\right]\nonumber\\
    & = \mathcal{N}\left[\bm{x}^{(t-1)}; \frac{\sqrt{\alpha_{t}} (1-\breve{\alpha}_{t-1}) \bm{x}^{(t)} + \sqrt{\breve{\alpha}_{t-1}} (1-\alpha_{t}) \bm{x}^{(0)}}{1-\breve{\alpha}_{t}}, \sigma_{t}^{2} \bm{I}\right],
\end{align}
where 
\begin{align}
    \sigma_{t}^{2} = \frac{(1-\alpha_{t})(1-\breve{\alpha}_{t-1})}{1-\breve{\alpha}_{t}}.
\end{align}
According to equation (\ref{p-xt1-xt-vdm}), the sample recovery/generation process can be described as:
\begin{align}
    \bm{x}^{(t-1)} &= \frac{1}{1-\breve{\alpha}_{t}}\left[\sqrt{\alpha_{t}} (1-\breve{\alpha}_{t-1}) \bm{x}^{(t)} + \sqrt{\breve{\alpha}_{t-1}} (1-\alpha_{t}) \hat{\bm{x}}^{(0)}\right] + \nonumber\\
    &~~~~ \sigma_{t} \bm{\epsilon}.
\end{align}
Here, \( \bm{\epsilon} \sim \mathcal{N}\left(\bm{\epsilon}; \bm{0}, \bm{I}\right) \) is a sample from a zero-mean Gaussian random process.

\subsubsection{Learning Network Parameters}

The neural network \( f(\bm{x}^{(t)}, t) \) for sample recovery/generation needs to be optimized in conjunction with the sample degradation process. To find the optimal parameters \( \theta_{f} \), we first need to define the optimization problem regarding this network. An effective way to do this is to maximize the evidence lower bound (ELBO):
\begin{align}
    &\ln p(\bm{x}) \nonumber \\
    &= \ln \int_{\bm{x}^{(1:T)}} p\left[\bm{x}^{(0)}, \bm{x}^{(1:T)}\right] \\
    &\geq \mathbb{E}_{q\left[ \bm{x}^{(1:T)} | \bm{x}^{(0)}\right]} \ln \frac{p\left[\bm{x}^{(0)}, \bm{x}^{(1:T)}\right]}{q\left[ \bm{x}^{(1:T)} | \bm{x}^{(0)}\right]} \\
    &= \mathbb{E}_{q\left[ \bm{x}^{(1)} | \bm{x}^{(0)}\right]} \ln p\left[\bm{x}^{(0)}| \bm{x}^{(1)}\right] - \mathrm{KL}\left\{q\left[ \bm{x}^{(T)} | \bm{x}^{(0)}\right]\| p\left[\bm{x}^{(T)}\right]\right\} \nonumber \\
    &\quad + \sum_{t=2}^{T} \mathbb{E}_{q\left[ \bm{x}^{(t)} | \bm{x}^{(0)}\right]} \mathrm{KL}\left\{q\left[ \bm{x}^{(t-1)} | \bm{x}^{(t)}, \bm{x}^{(0)}\right]\| p\left[\bm{x}^{(t-1)} | \bm{x}^{(t)} \right]\right\}.\nonumber
\end{align}
For Gaussian distributions, the KL distance between them has a closed-form solution\footnote{KL distance is a measure of the distance between two distributions.}. By simplifying, the optimization problem for the variational diffusion network concerning the neural network can be described as:
\begin{align}
   & \min_{\theta_{f}} \sum_{t=2}^{T} \mathbb{E}_{q\left[ \bm{x}^{(t)} | \bm{x}^{(0)}\right]} \mathrm{KL}\left\{q\left[ \bm{x}^{(t-1)} | \bm{x}^{(t)}, \bm{x}^{(0)}\right]\| p\left[\bm{x}^{(t-1)} | \bm{x}^{(t)} \right]\right\}, \\
   & \iff \min_{\theta_{f}} \sum_{t=2}^{T} \rho_{t} \cdot \mathbb{E}_{q\left[ \bm{x}^{(t)} | \bm{x}^{(0)}\right]} \left\| f(\bm{x}^{(t)}, t) - \bm{x}^{(0)}\right\|_{2}^{2}, \\
   & \iff \min_{\theta_{f}} \sum_{t=2}^{T} \rho_{t} \cdot \sum_{i} \left\| f(\bm{x}_{i}^{(t)}, t) - \bm{x}^{(0)}\right\|_{2}^{2},
\end{align}
where
\begin{align}
    \rho_{t} &= \frac{1}{2\sigma_{t}^{2}} \frac{\breve{\alpha}_{t-1}(1-\alpha_{t})^{2}}{(1-\breve{\alpha}_{t})^{2}}, \\
    \bm{x}_{i}^{(t)} &= \sqrt{\breve{\alpha}_{t}} \bm{x}^{(0)} + \sqrt{(1-\breve{\alpha}_{t})} \bm{\epsilon}_{i}, \quad \forall i.
\end{align}
Here, \( \bm{\epsilon}_{i} \) is the \( i \)-th sample from a standard Gaussian random process. According to studies on self-paced learning, \( \rho_{t} \) can be seen as a weight measuring the importance of the samples.

In neural network training, samples are typically fed into the trainer in batches for parameter learning. Given a sample \( \bm{x}_{n} \), the objective function for network training can be defined as:
\begin{align}
    \mathcal{J} = \sum_{n} \sum_{t} \sum_{i} \rho_{t} \cdot \left\| f(\bm{x}_{n, i}^{(t)}, t) - \bm{x}_{n}^{(0)}\right\|_{2}^{2},
\end{align}
where \( \bm{x}_{n, i}^{(t)} \) represents the \( i \)-th sample of the \( n \)-th sample degraded to step \( t \), and \( \bm{x}_{n}^{(0)} = \bm{x}_{n} \) is the original value of the \( n \)-th sample. According to the network's expression, we can also randomly sample \( n, t, i \) to construct an appropriate objective function. From the above expression, it is evident that the core of the diffusion network lies in the training issue of the denoising network in a Gaussian noise environment, where different samples are assigned different weights during training. These weights are related to the iteration count, which reflects the amount of noise or signal-to-noise ratio.

In summary, the variational diffusion model provides a framework for both signal degradation and recovery. By leveraging the relationships between these processes, we can effectively learn to generate high-quality samples that closely resemble real data. The optimization of network parameters through evidence lower bounds allows for flexible adaptations to different data distributions, enhancing the model's performance. Ultimately, this approach facilitates the generation of samples from noise, improving the overall quality and applicability of machine learning models in various domains.

\subsection{Data Visualization: A Special Data Transformation Method}

In high-dimensional space, distances between data samples can be measured using some metric. However, when given a batch of data, the relationships among them cannot be intuitively perceived through pairwise distances. Data visualization techniques address this issue. The basic idea is to constrain samples in a low-dimensional space such that their distances match the distances in high-dimensional space in some manner. We briefly introduce three methods: 1) Multidimensional Scaling (MDS) \cite{torgerson1952multidimensional}; 2) Stochastic Neighbor Embedding (SNE) \cite{hinton2002stochastic,van2008visualizing}; 3) Locally Linear Embedding (LLE) \cite{roweis2000nonlinear}.

Given a dataset \( \{\bm{x}_{n}, \forall n=1,2,\ldots,N\} \), with the distance between samples \( \bm{x}_{k} \) and \( \bm{x}_{n} \) denoted as \( \hat{d}_{k,n} \), the challenge is to find a set of data samples \( \bm{y}_{n} \) in a low-dimensional space such that the distance relationships among \( \bm{y}_{n} \) match those among \( \bm{x}_{n} \).

In the case that we have the low-dimensional dataset \( \{\bm{y}_{n}, n=1,2,\ldots,N\} \), we can visualize the \( \bm{y}_{n} \) points as scatter plots in a two-dimensional plane. Unless specified otherwise, the distance \( \hat{d}_{k,n} \) refers to the Euclidean distance between samples \( \bm{x}_{k} \) and \( \bm{x}_{n} \):
\begin{align}
    \hat{d}_{k,n} = \left\|\bm{x}_{k} - \bm{x}_{n}\right\|_{2}.
\end{align}

\subsubsection{Multidimensional Scaling}

In the low-dimensional space, the distance between samples \( \bm{y}_{k} \) and \( \bm{y}_{n} \) can be defined as \( \left\|\bm{y}_{k} - \bm{y}_{n}\right\|_{2} \). The most intuitive way to establish an optimization problem that constrains these pairwise distances to match is as follows:
\begin{align}
    \left\|\bm{y}_{k} - \bm{y}_{n}\right\|_{2} = \hat{d}_{k,n}, \quad \forall k,n=1,2,\ldots,N. \label{opt-MDS}
\end{align}
For convenience, we introduce the centering matrix \( \bm{Q} = \mathbf{I}_{N} - \frac{1}{N} \bm{1}_{N} \bm{1}_{N}^{T} \), where \( \bm{I}_{N} \) is the \( N \times N \) identity matrix and \( \bm{1}_{N} \) is a column vector of length \( N \) with all elements equal to 1.

Next, we define the \( L \times N \) matrix \( \bm{Y} \) as:
\begin{align}
    \bm{Y} = \left[\begin{array}{cccc}
        \bm{y}_{1} & \bm{y}_{2} & \cdots & \bm{y}_{N}
    \end{array}\right].
\end{align}
Simultaneously, define the \( N \times N \) matrix \( \bm{D} \) as:
\begin{align}
    \left[\bm{D}\right]_{k,n} = \hat{d}_{k,n}^{2},
\end{align}
where \( [\cdot]_{k,n} \) indicates the \( (k,n) \)-th element of the matrix. Under the centering constraint on \( \bm{Y} \), equation (\ref{opt-MDS}) can be  expressed as:
\begin{align}
    \bm{Y}^{T}\bm{Y} = \bm{Q} \bm{D} \bm{Q}^{T}. \label{mds-yy-qdq}
\end{align}
For data visualization, we have \( N \gg  L \); the matrix \( \bm{Y}^{T} \) is typically a ``tall" matrix, and the rank of the left matrix \( \bm{Y}^{T}\bm{Y} \) is far less than the dimensionality \( N \). Considering the eigen-decomposition of the matrix \( \bm{Q} \bm{D} \bm{Q}^{T} \):
\begin{align}
    \bm{Q} \bm{D} \bm{Q}^{T} = \sum_{n=1}^{N} \lambda_{n} \bm{u}_{n} \bm{u}_{n}^{T},
\end{align}
where \( \lambda_{1} \geq \lambda_{2} \geq \ldots \geq \lambda_{N} \) are the eigenvalues of the matrix, and \( \bm{u}_{n} \) are the corresponding eigenvectors. Selecting the largest \( L \) eigenvalues and their corresponding eigenvectors constructs the following two matrices:
\begin{align}
    \bm{U}_{L} = \left[\begin{array}{cccc}
        \bm{u}_{1} & \bm{u}_{2} & \cdots & \bm{u}_{L}
    \end{array}\right], \\
    \bm{\Lambda}_{L} = \mathrm{diag}\left[\begin{array}{cccc}
        \lambda_{1}, & \lambda_{2}, & \ldots, & \lambda_{L}
    \end{array}\right].
\end{align}
Under the least squares criterion, the optimal \( \bm{Y} \) corresponding to equation (\ref{mds-yy-qdq}) satisfies:
\begin{align}
    \bm{Y} = \sqrt{\bm{\Lambda}_{L}} \bm{U}_{L}^{T},
\end{align}
where \( \sqrt{(\cdot)} \) indicates taking the square root of each element of the matrix. By extracting the \( n \)-th column of matrix \( \bm{Y} \), we obtain the low-dimensional coordinates \( \bm{y}_{n} \) corresponding to sample \( \bm{x}_{n} \). Thus, we complete the mapping of data from high-dimensional to low-dimensional space.

The multidimensional scaling method can be extended in two ways: 1) Defining different distance metrics, such as proposing methods better suited for the task to compute distances \( \hat{d}_{k,n} \) between data \( \bm{x}_{k} \) and \( \bm{x}_{n} \), as in the Isomap algorithm; 2) Establishing better optimization problems to solve for data coordinates in low-dimensional space:
\begin{align}
    \min_{\{\bm{y}_{n}, \forall n\}} \sum_{k} \sum_{n} \mathcal{J}\left[\bm{y}_{k}, \bm{y}_{n}, \hat{d}_{k,n}\right].
\end{align}
Here, \( \mathcal{J}(\cdot) \) defines the distance between pairs of data in low-dimensional space, and the cost function between this distance and the expected distance, such as \( \mathcal{J}\left[\bm{y}_{k}, \bm{y}_{n}, \hat{d}_{k,n}\right] = \left|\left\|\bm{y}_{k}-\bm{y}_{n}\right\|_{2} - \hat{d}_{k,n}\right| \) or \( \mathcal{J}\left[\bm{y}_{k}, \bm{y}_{n}, \hat{d}_{k,n}\right] = \left(\left\|\bm{y}_{k}-\bm{y}_{n}\right\|_{2} - \hat{d}_{k,n}\right)^{2} \), etc.

\subsubsection{Locally Linear Embedding Method}

The locally linear embedding (LLE) method  employs an encoding-decoding approach to map data from high-dimensional space to low-dimensional coordinates. The fundamental idea is to establish relationships between data samples and their neighbors using different coefficients. Subsequently, these relationships are constrained in the low-dimensional space to complete the data reduction.

LLE does not require direct distance specifications between pairs of samples; however, it does require the specification of neighboring samples for each sample. Specifically, for each sample \( \bm{x}_{n} \), its neighboring samples \( \bm{x}_{n_{i}} \) must be defined, where \( i = 1, 2, \ldots, K \) and \( n_{i} \) is an index ranging from 1 to \( N \). For convenience, we define the matrix of neighboring samples as:
\begin{align}
    \breve{\bm{X}}_{n} = \left[\begin{array}{cccc}
    \bm{x}_{n_{1}} & \bm{x}_{n_{2}} & \cdots & \bm{x}_{n_{K}}
    \end{array}\right].
\end{align}
To simplify the description, we assume each sample is a column vector of length \( M \). Thus, the dimension of the matrix \( \breve{\bm{X}}_{n} \) is \( M \times K \). For each sample \( \bm{x}_{n} \), we can find a length-\( K \) vector \( \bm{h}_{n} \) such that:
\begin{align}
    \min_{\bm{h}_{n}} \left\|\bm{x}_{n} - \breve{\bm{X}}_{n} \bm{h}_{n}\right\|_{2}^{2} + \epsilon \left\|\bm{h}_{n}\right\|_{2},
\end{align}
where \( \epsilon \geq 0 \) is a regularization parameter. Solving this problem yields the optimal coefficients:
\begin{align}
    \bm{h}_{n} = \left(\breve{\bm{X}}_{n}^{T} \breve{\bm{X}}_{n} + \epsilon \bm{I}\right)^{-1} \breve{\bm{X}}_{n}^{T} \bm{x}_{n}.
\end{align}
This vector encodes the relationship between sample \( \bm{x}_{n} \) and its neighboring samples. To complete the dimensionality reduction, we need to find low-dimensional data samples that satisfy the same relationships.

Given the low-dimensional data samples \( \bm{y}_{n} \), the optimization problem to constrain their relationships can be expressed as:
\begin{align}
    \min_{\{\bm{y}_{n}, \forall n\}} \left\|\bm{y}_{n} - \breve{\bm{Y}}_{n} \bm{h}_{n}\right\|_{2}^{2}.
\end{align}
By extending \( \bm{h}_{n} \) into a length-\( N \) vector \( \tilde{\bm{h}}_{n} \), where coefficients corresponding to neighboring samples are the same and all other coefficients are zero, the problem can be formulated as:
\begin{align}
    \min_{\bm{Y}} \sum_{n} \left\|\bm{Y} \bm{i}_{n} - \bm{Y} \tilde{\bm{h}}_{n}\right\|_{2}^{2}.
\end{align}
This leads to:
\begin{align}
    \min_{\bm{Y}} \mathrm{tr}\left\{\bm{Y}(\bm{I} - \bm{H})(\bm{I} - \bm{H})^{T} \bm{Y}^{T}\right\},
\end{align}
where
\begin{align}
    \bm{H} = \left[\begin{array}{cccc}
    \tilde{\bm{h}}_{1} & \tilde{\bm{h}}_{2} & \cdots & \tilde{\bm{h}}_{N}
    \end{array}\right]
\end{align}
is an \( N \times N \) matrix. Solving this optimization problem requires additional constraints; otherwise, one could trivially minimize the objective function by setting \( \bm{Y} = \bm{0} \). To address this issue, we can reformulate the original problem as:
\begin{align}
    \min_{\bm{Y}} \frac{\mathrm{tr}\left\{\bm{Y}(\bm{I} - \bm{H})(\bm{I} - \bm{H})^{T} \bm{Y}^{T}\right\}}{\mathrm{tr}\left\{\bm{Y} \bm{Y}^{T}\right\}}.
\end{align}
Using eigen-decomposition, the matrix \( (\bm{I} - \bm{H})(\bm{I} - \bm{H})^{T} \) can be expressed as:
\begin{align}
    (\bm{I} - \bm{H})(\bm{I} - \bm{H})^{T} = \sum_{n=1}^{N} \lambda_{n} \bm{u}_{n} \bm{u}_{n}^{T},
\end{align}
where \( \lambda_{1} \leq \lambda_{2} \leq \cdots \leq \lambda_{N} \) are the eigenvalues of the matrix, and \( \bm{u}_{n} \) are the corresponding eigenvectors. The optimal \( \bm{Y} \) is formed by selecting the smallest \( L \) eigenvalues and their corresponding eigenvectors:
\begin{align}
    \bm{Y} = \left[\begin{array}{cccc}
    \bm{u}_{1} & \bm{u}_{2} & \cdots & \bm{u}_{L}
    \end{array}\right]^{T}.
\end{align}
Thus, the introduction of the locally linear embedding method is complete. This method can also be extended in various ways, allowing for the design of more suitable encoding-decoding methods tailored to different tasks. For instance, by introducing a constraint such as \( \bm{h}_{n}^{T} \bm{1} = 1 \) in the solution for \( \bm{h}_{n} \), we can obtain \( \bm{h}_{n} = \bm{\Phi}_{n}^{-1} \bm{1}/(\bm{1}^{T} \bm{\Phi}_{n}^{-1} \bm{1}) \), where \( \bm{\Phi}_{n} = (\bm{x}_{n} \bm{1}^{T} - \breve{\bm{X}}_{n})(\bm{x}_{n} \bm{1}^{T} - \breve{\bm{X}}_{n})^{T} \). Of course, readers can also devise specific encoding problems based on their task requirements to achieve more efficient local embedding data visualization methods.

\subsubsection{Stochastic Neighbor Embedding Method}

The core idea of the Stochastic Neighbor Embedding (SNE) method is to describe the relationships between samples using conditional probabilities and then constrain the relationships in low-dimensional coordinates to approximate those in high-dimensional space. By optimizing the distance between probability distributions, we can obtain the coordinates of samples in low-dimensional space.

Given a dataset \( \{\bm{x}_{n}, \forall n=1,2,\ldots, N\} \), the conditional probability of \( \bm{x}_{k} \) given sample \( \bm{x}_{n} \) can be described as:
\begin{align}
    p_{k| n} = \frac{\expd{-\frac{\left\|\bm{x}_{k} - \bm{x}_{n}\right\|_{2}^{2}}{2\sigma_{n}^{2}}}}{\sum_{i\neq n} \expd{-\frac{\left\|\bm{x}_{i} - \bm{x}_{n}\right\|_{2}^{2}}{2\sigma_{n}^{2}}}} = \frac{\expd{-\frac{\hat{d}_{k,n}^{2}}{2\sigma_{n}^{2}}}}{\sum_{i\neq n} \expd{-\frac{\hat{d}_{i,n}^{2}}{2\sigma_{n}^{2}}}}.
\end{align}
In the low-dimensional space, using a similar approach, we can also define the conditional probability of \( \bm{y}_{k} \) given \( \bm{y}_{n} \). For convenience, we denote it as \( q_{k| n} \). The most common distance for measuring the distance between probability distributions is the KL divergence. Under the criterion of minimizing KL divergence, the optimization problem regarding low-dimensional coordinates \( \bm{y}_{n} \) can be described as:
\begin{align}
    \min_{\{\bm{y}_{n}, \forall n\}} \sum_{k} \sum_{n} p_{k| n} \ln \frac{p_{k| n}}{q_{k| n}}.
\end{align}
This can be rewritten as:
\begin{align}
    \min_{\{\bm{y}_{n}, \forall n\}} - \sum_{k} \sum_{n} p_{k| n} \ln q_{k| n}.
\end{align}
For each \( \bm{y}_{m} \), its objective function can be expressed as:
\begin{align}
    \mathcal{J}(\bm{y}_{m}) = -\sum_{n} p_{m| n} \ln q_{m| n} - \sum_{k} p_{k| m} \ln q_{k| m}.
\end{align}
In this equation, \( q_{m| n} \) and \( q_{k| m} \) are functions of the low-dimensional coordinates \( \bm{y}_{m} \). Under the condition of Euclidean distance, we denote the distance between \( \bm{y}_{k} \) and \( \bm{y}_{n} \) as:
\begin{align}
    \hat{t}_{k,n} = \left\|\bm{y}_{k} - \bm{y}_{n}\right\|_{2}.
\end{align}
Thus, we can derive:
\begin{align}
    \nabla\mathcal{J}(\bm{y}_{m}) &= \frac{\partial }{\partial \bm{y}_{m}}\mathcal{J}(\bm{y}_{m}) \\
    &= -2\sum_{n} \frac{ p_{m| n}}{q_{m| n}} \cdot \frac{\partial q_{m| n}}{\partial \hat{t}_{m,n}^{2}} \cdot (\bm{y}_{m}-\bm{y}_{n}) - \nonumber\\
    &~~~~~~~~ 2\sum_{k} \frac{p_{k| m}}{q_{k| m}} \cdot \frac{\partial q_{k| m}}{\partial \hat{t}_{k,m}^{2}} \cdot (\bm{y}_{m}-\bm{y}_{k}) \\
    &= -2\sum_{n} \left[ \frac{ p_{m| n}}{q_{m| n}} \cdot \frac{\partial q_{m| n}}{\partial  \hat{t}_{m,n}^{2}} +  \frac{p_{n| m}}{q_{n| m}} \cdot \frac{\partial q_{n| m}}{\partial \hat{t}_{n,m}^{2}} \right]  (\bm{y}_{m}-\bm{y}_{n}).
\end{align}
Given the function \( q_{k| n} \), we can compute the values of \( \frac{\partial q_{m| n}}{\partial  \hat{t}_{m,n}^{2}} \) and \( \frac{\partial q_{n| m}}{\partial  \hat{t}_{n,m}^{2}} \). Using gradient descent algorithms, we can perform the search for low-dimensional coordinates, such as:
\begin{align}
    \bm{y}_{m}^{(i+1)} = \bm{y}_{m}^{(i)} - \alpha \nabla\mathcal{J}(\bm{y}_{m}^{(i)}) + \beta(\bm{y}_{m}^{(i)} - \bm{y}_{m}^{(i-1)}). \label{tSNE-update-ym}
\end{align}
Letting
\begin{align}
    q_{k| n} = \frac{\expd{-\hat{t}_{k,n}^{2}}}{\sum_{i\neq n} \expd{-\hat{t}_{i,n}^{2}}},
\end{align}
we can succinctly express \( \nabla\mathcal{J}(\bm{y}_{m}) \) as:
\begin{align}
    &\nabla\mathcal{J}(\bm{y}_{m}) \nonumber\\
    &= -2\sum_{n}(p_{m| n} + p_{n| m} - q_{m| n} - q_{n| m}) (\bm{y}_{m} - \bm{y}_{n}).
\end{align}
However, under this distribution, the method may not effectively distinguish in low-dimensional space. To address this issue, the t-SNE  assumes that the distances between low-dimensional data points follow a t-distribution:
\begin{align}
    q_{k| n} = \frac{\Gamma\left(\frac{\zeta+1}{2}\right)}{\sqrt{\pi \zeta} \Gamma\left(\frac{\zeta}{2}\right)} \cdot \left(1+\frac{\hat{t}_{k,n}^{2}}{\zeta}\right)^{-\frac{\zeta+1}{2}}.
\end{align}
It can be verified that \( q_{k| n} = q_{n|k} \), and:
\begin{align}
    \frac{ 1}{q_{m| n}} \cdot \frac{\partial q_{m| n}}{\partial  \hat{t}_{m,n}^{2}} &= \frac{1}{q_{n| m}} \cdot \frac{\partial q_{n| m}}{\partial \hat{t}_{n,m}^{2}} \\
    &= -\frac{2}{\zeta(\zeta+1)} \cdot \left(1+ \frac{\hat{t}_{m,n}^{2}}{\zeta}\right)^{-1}.
\end{align}
At this point, we can derive:
\begin{align}
    &\nabla\mathcal{J}(\bm{y}_{m})\nonumber \\ &= \frac{4}{\zeta(\zeta+1)} \sum_{n}(p_{m| n} + p_{n| m}) \left(1+ \frac{\hat{t}_{m,n}^{2}}{\zeta}\right)^{-1}(\bm{y}_{m} - \bm{y}_{n}). \label{t-SNE-updateJ}
\end{align}
Substituting (\ref{t-SNE-updateJ}) into (\ref{tSNE-update-ym}) allows for the updating of low-dimensional coordinates \( \bm{y}_{m} \). In practice, one can take \( \zeta=1 \) as a hyperparameter for the t-distribution, corresponding to the Cauchy distribution. Comparing the density functions of the Cauchy distribution and the Gaussian distribution reveals that replacing the Gaussian distribution with the Cauchy distribution can bring closer together originally adjacent samples in low-dimensional space while pushing farther apart those that are remote, thereby indirectly reducing the distances within classes and amplifying the distances between classes, achieving better data visualization results.
 
\section{Conclusions}

In conclusion, this paper has provided a comprehensive overview of the principles and methods involved in acoustic signal transformation, detection, and filtering. As the field continues to evolve, driven by technological advancements and increasing application demands, understanding these core techniques becomes essential for researchers and practitioners alike.
The diversity of acoustic signals, shaped by their varied applications, necessitates a nuanced approach to processing techniques. By categorizing methods into transformation, detection, and filtering, we can better appreciate their roles in enhancing signal clarity and extracting meaningful information.
Furthermore, the shift from knowledge-driven to data-driven methodologies, particularly with the rise of deep learning, has opened new avenues for innovation in acoustic signal processing. This transformation not only improves performance but also encourages the development of novel applications in fields such as speech communication, health monitoring, and industrial diagnostics.

Ultimately, this work aims to provide a simple summary that allows readers to draw insights from various fields. However, it does not claim to be exhaustive. Due to the author's limited capabilities, there may be inadvertent errors within the text. If there are areas for improvement, please do not hesitate to reach out to the author. Additionally, if readers notice any missing or improperly cited references, they are encouraged to contact the author via email. If the author believe that there are important methods that should be included, please also reach out, providing the title of the relevant paper, and the author will make continuous revisions in future versions.

\bibliographystyle{IEEEtran}

\end{document}